\documentclass[12pt]{article}
\usepackage{latexsym}
\usepackage{amsmath,amsfonts}
\usepackage{times}
\allowdisplaybreaks[4]

\catcode`\@=11

\newcount\hour
\newcount\minute
\newtoks\amorpm \hour=\time\divide\hour by 60\minute
=\time{\multiply\hour by 60 \global\advance\minute by-\hour}
\edef\standardtime{{\ifnum\hour<12 \global\amorpm={am}%
        \else\global\amorpm={pm}\advance\hour by-12 \fi
        \ifnum\hour=0 \hour=12 \fi
        \number\hour:\ifnum\minute<10
        0\fi\number\minute\the\amorpm}}
\edef\militarytime{\number\hour:\ifnum\minute<10
0\fi\number\minute}

\def\draftlabel#1{{\@bsphack\if@filesw {\let\thepage\relax
   \xdef\@gtempa{\write\@auxout{\string
      \newlabel{#1}{{\@currentlabel}{\thepage}}}}}\@gtempa
   \if@nobreak \ifvmode\nobreak\fi\fi\fi\@esphack}
        \gdef\@eqnlabel{#1}}
\def\@eqnlabel{}
\def\@vacuum{}
\def\marginnote#1{}
\def\draftmarginnote#1{\marginpar{\raggedright\scriptsize\tt#1}}
\def\draft{
        \pagestyle{plain}
        \overfullrule=2pt
        \oddsidemargin -.5truein
        \def\@oddhead{\sl \phantom{\today\quad\militarytime} \hfil
        \smash{\Large\sl DRAFT} \hfil \today\quad\militarytime}
        \let\@evenhead\@oddhead
        \let\label=\draftlabel
        \let\marginnote=\draftmarginnote
        \def\ps@empty{\let\@mkboth\@gobbletwo
        \def\@oddfoot{\hfil \smash{\Large\sl DRAFT} \hfil}
        \let\@evenfoot\@oddhead}
        \def\@eqnnum{(\theequation)\rlap{\kern\marginparsep\tt\@eqnlabel}%
        \global\let\@eqnlabel\@vacuum}  }

\newcommand{\rf}[1]{(\ref{#1})}
\renewcommand{\theequation}{\thesection.\arabic{equation}}
\renewcommand{\thefootnote}{\fnsymbol{footnote}}
\newcommand{\newsection}{    
\setcounter{equation}{0}\section}
\def\appendix#1{\addtocounter{section}{1}\setcounter{equation}{0}
\renewcommand{\thesection}{\Alph{section}}
\section*{Appendix \thesection\protect\indent \parbox[t]{11.15cm}{#1}}
\addcontentsline{toc}{section}{Appendix \thesection\ \ \ #1}}

\def\be{\begin{equation}}
\def\ee{\end{equation}}
\def\beq{\begin{eqnarray}}
\def\eeq{\end{eqnarray}}

\def\phik{|\phi\rangle}
\def\phibr{\langle\phi|}

\def\phibfk{|\phibf\rangle}
\def\varphibfk{|\varphibf\rangle}

\def\oplussm{{\scriptscriptstyle \oplus}}
\def\ominussm{{\scriptscriptstyle \ominus}}

\def\abf{{\bf a}}
\def\bbf{{\bf b}}
\def\cbf{{\bf c}}
\def\dbf{{\bf d}}
\def\hbf{{\bf h}}
\def\nbf{{\bf n}}

\def\ibf{{\bf i}}
\def\iibf{{\bf ii}}
\def\iiibf{{\bf iii}}
\def\ivbf{{\bf iv}}
\def\vbf{{\bf v}}
\def\vibf{{\bf vi}}

\def\field{{\rm field}}

\def\diff{{\rm diff}}
\def\cur{{\rm cur}}
\def\sh{{\rm sh}}
\def\eff{{\rm eff}}
\def\irm{{\rm i}}
\def\mrm{{\rm m}}
\def\intrm{{\rm int}}
\def\longrm{{\rm long}}

\def\half{{\frac{1}{2}}}

\def\CC{{\cal C}}
\def\OO{{\cal O}}
\def\LL{{\cal L}}
\def\MM{{M}}
\def\TT{{\cal T}}

\def\alphab{\bar\alpha}

\def\zetab{\bar\zeta}

\def\Ab{\bar{A}}
\def\Xb{\bar{X}}
\def\Yb{\bar{Y}}
\def\Wb{\bar{W}}

\def\xb{\bar{x}}
\def\yb{\bar{y}}
\def\vb{\bar{v}}
\def\wb{\bar{w}}

\def\Ksm{{\scriptscriptstyle K}}

\def\AdSsm{{\scriptscriptstyle AdS}}

\def\phibf{{\boldsymbol{\phi}}}
\def\varphibf{{\boldsymbol{\varphi}}}

\def\Mwt{{\widetilde M}}

\begin{document}


\begin{flushright}
FIAN-TD-2015-06 \hspace{1.2cm} {}~\\
arXiv: 1507.06584  [hep-th]
\end{flushright}
\vspace{1cm}

\begin{center}

{\Large \bf Light-cone AdS/CFT-adapted approach to

\bigskip
AdS fields/currents, shadows, and conformal fields}

\vspace{2.5cm}

R.R. Metsaev\footnote{ E-mail: metsaev@lpi.ru }

\vspace{1cm}

{\it Department of Theoretical Physics, P.N. Lebedev Physical
Institute, \\ Leninsky prospect 53,  Moscow 119991, Russia }

\vspace{3.5cm}

{\bf Abstract}

\end{center}

Light-cone gauge formulation of fields in AdS space and conformal field theory in flat space adapted for the study of AdS/CFT correspondence is developed. Arbitrary spin mixed-symmetry fields in AdS space and arbitrary spin mixed-symmetry currents, shadows, and conformal fields in flat space are considered on an equal footing. For the massless and massive fields in AdS and the conformal fields in flat space, simple light-cone gauge actions leading to decoupled equations of motion are found.
For the currents and shadows, simple expressions for all 2-point functions are also found.
We demonstrate that representation of conformal algebra generators on space of
currents, shadows, and conformal fields can be built in terms of spin operators
entering the light-cone gauge formulation of AdS fields. This considerably simplifies the study of AdS/CFT correspondence. Light-cone gauge actions for totally symmetric arbitrary spin long conformal fields in flat space are presented. We apply our approach to the study of totally antisymmetric (one-column) and mixed-symmetry (two-column) fields in AdS space and currents, shadows, and conformal fields in flat space.

\newpage
\renewcommand{\thefootnote}{\arabic{footnote}}
\setcounter{footnote}{0}

\section{Introduction}

In spite of its Lorentz noncovariance, the light-cone gauge approach to relativistic field and string dynamics \cite{Dirac:1949cp} simplifies considerably the study of various problems of quantum field and string theories. This is to say that a
number of important problems of modern quantum field and string theories have successfully
been solved by using the light-cone gauge formalism. Perhaps the most attractive example of application of this formalism is the building of the light-cone gauge formulation of superstring field theory \cite{Green:1984fu}.

Motivated by desire to provide the light-cone gauge framework for the study of the conjectured string/gauge duality \cite{Maldacena:1997re} and its cousins, the light-cone gauge formulation of AdS field theory in Ref.\cite{Metsaev:1999ui} and light-cone gauge formulation of AdS superstring theory in  Refs.\cite{Metsaev:2000yf,Metsaev:2000yu} were developed. Also, in Ref.\cite{Metsaev:1999ui}, light-cone gauge approach to CFT was developed.  AdS/CFT correspondence for arbitrary spin massless AdS fields was studied in Ref.\cite{Metsaev:1999ui} and it was demonstrated that the light-cone approach simplifies considerably the study of AdS/CFT correspondence.

To explain our terminology let us discuss briefly the group theoretical interpretation of fields in $AdS_{d+1}$ and related currents, shadows, and conformal fields in $R^{d-1,1}$ we are going to study in this paper. Massless and massive fields propagating in $AdS_{d+1}$ space are associated
with unitary positive-energy lowest weight irreducible representations of the $so(d,2)$ algebra.
We will denote such representations by
$D(E_0,\hbf)$, where the $E_0$ is the lowest eigenvalue of the energy
operator, while the $\hbf = (h_1,\ldots, h_r)$, $r=[\frac{d}{2}]$ stands for the highest
weight of the $so(d)$ algebra representation. For bosonic fields,
the highest weights $h_i$ are integers. As shown in Ref.\cite{Metsaev:1995re}, the $E_0$ and
$\hbf$ satisfy the following restriction:%
\footnote{ For $d=3,4$, the restriction \rf{05072015-man-01}
was obtained in Refs.\cite{Evans,Mack:1975je}. The study of
unitary representations of various superalgebras may be found,
e.g., in Refs.\cite{Dobrev:1985qv,Gunaydin:1998sw}.}
\be \label{05072015-man-01}
E_0 \geq h_k  - k  - 1 + d\,,
\ee
where $k$ is defined from the relation%
\footnote{
For $d$-even,  the weight $h_r$ in \rf{05072015-man-01}, \rf{05072015-man-02} should be replaced by $|h_r|$. Bosonic irrep of the $so(d)$ algebra having highest weight $\hbf$ can be described by tensor field whose $so(d)$ space tensor indices have the structure of the Young tableaux labelled by $\hbf$. In such Young tableaux, $h_i$ is equal to length of the $i$-th row.
}
\be \label{05072015-man-02}
h_1
=
\ldots
=
h_k
>
h_{k+1}
\ge
h_{k+2}
\ge
\ldots
\ge h_r
\geq 0\,.
\ee
Using the notation
\be \label{05072015-man-03}
E_0^{\mrm = 0} \equiv h_k  - k  - 1 + d \,,
\ee
we note that, for massless and massive fields, restriction in \rf{05072015-man-01}
takes the form%
\footnote{ As known, the case with $E_0 = E_0^{\mrm = 0}$, and $k=d/2$, $d$-even is not associated with local field propagating in $AdS_{d+1}$. This case will be ignored in this paper.
}
\beq
\label{27062015-man-05a} && E_0 = E_0^{\mrm = 0} \,, \hspace{4.6cm} \hbox{ for massless fields in } AdS_{d+1},
\\
\label{05072015-man-04} && E_0 > h_k  - k  - 1 + d\,, \hspace{3cm} \hbox{ for massive fields in } AdS_{d+1}.
\eeq

Conformal currents in $R^{d-1,1}$ are also associated with the representations $D(\Delta_\cur,\hbf)$ of the $so(d,2)$ algebra, where $\Delta_\cur$ stands for the conformal dimension of the currents. In this paper, conformal currents with $\Delta_\cur = E_0^{\mrm=0}$ and  $\Delta_\cur > E_0^{\mrm=0}$ will be referred to as short and long currents respectively. Conformal shadows in $R^{d-1,1}$ are associated with non-unitary representations of the $so(d,2)$ algebra labeled by $\Delta_\sh$, $\hbf$, where $\Delta_\sh = d-\Delta_\cur$. In this paper, conformal shadows having $\Delta_\sh = d- E_0^{\mrm=0}$ are referred to as short shadows, while conformal shadows having  $\Delta_\sh = d - E_0$ with $E_0$ as in \rf{05072015-man-04} are referred to as long shadows.%
\footnote{
In the literature, the short currents and shadows are referred to as canonical currents and shadows respectively, while the long currents and shadows are referred to as anomalous currents and shadows respectively. We recall that, in Lorentz covariant approach, the short currents with $\Delta_\cur= d-1$, $\hbf =(1,0,\ldots,0)$ and $\Delta_\cur = d$, $\hbf=(2,0,\ldots,0)$ are described by the respective  conserved vector current and conserved symmetric traceless rank-2 tensor field
(energy-momentum tensor). Discussion of the short currents bilinear in arbitrary spin massless fields may be found in Ref.\cite{Konstein:2000bi}.
}

The short shadows in $R^{d-1,1}$ can be used to build the conformal invariant Lagrangian dynamics of fields propagating in $R^{d-1,1}$. Such fields will be referred to as short conformal fields.
The long shadows with some particular values of $\Delta_\sh$ can also be used to build the conformal invariant Lagrangian dynamics of fields propagating in $R^{d-1,1}$. Such fields will be referred to as long conformal fields. Most of conformal fields enter higher-derivative Lagrangian dynamics
and are associated with non-unitary representations of the $so(d,2)$ algebra labeled by $\Delta$, $\hbf$. For the short conformal fields, $\Delta = d- E_0^{\mrm=0}$, while for the long conformal fields $\Delta = d- E_{0,\longrm}$, where $E_{0,\longrm}$ takes integer or half integer values which satisfy the restriction \rf{05072015-man-04}.%
\footnote{
In Lorentz covariant approach, the short conformal fields were studied in Refs.\cite{Fradkin:1985am,Segal:2002gd,Vasiliev:2009ck}.
}

Throughout this paper, fields with $\hbf = (h_1,0,\ldots,0)$ are referred to as totally symmetric fields, while fields with $\hbf = (h_1,\ldots, h_r)$, $h_1> h_2 \geq 1$ are referred to as mixed-symmetry fields.%
\footnote{
Fields with $\hbf =(1,\ldots,1,0,\ldots,0)$ are referred to as totally antisymmetric
fields.
}
The light-cone gauge description of arbitrary spin totally symmetric short currents (shadows) and long  currents (shadows) was developed in the respective Ref.\cite{Metsaev:1999ui} and Ref.\cite{Metsaev:2013kaa}. The general light-cone formalism for arbitrary spin mixed-symmetry short currents (shadows) and long  currents (shadows) was developed in Ref.\cite{Metsaev:2005ws}.%
\footnote{ We recall that, in the framework of AdS/CFT, the long currents (shadows) are dual to bulk massive AdS fields.
}
We note however that the light-cone gauge formulation of CFT in Ref.\cite{Metsaev:2005ws} has the following unsatisfactory feature. Representation for generators of the conformal symmetries obtained in Ref.\cite{Metsaev:2005ws} is non-local.%
\footnote{ Using the notation $\partial^i$ and $\partial^-$ for derivatives with respect to the transverse space coordinates $x^i$ and the light-cone time coordinate $x^+$, we note that
if a representation for generators of the conformal symmetries is non-polynomial in the $\partial^i$, $\partial^-$, then such representation is referred to as non-local. Accordingly,
if a representation for generators of the conformal symmetries is polynomial in the $\partial^i$, $\partial^-$, then such representation is referred to as local.
}

One of the aims of this paper is to obtain the local representation for generators of the conformal symmetries in CFT. Doing so, we obtain not only the desired light-cone gauge formulation of currents and shadow but also new light-cone gauge formulation of fields in AdS space and conformal fields in flat space. The new formulations of AdS field theory and CFT are based on the use of one and the same spin operators satisfying some equations which take the form of deformed commutation relations of $so(d)$ algebra. Finding a solution to these equations for the spin operators implies immediately the complete light-cone description of both the AdS field theory and CFT. Therefore, in the framework of our new light-cone gauge approach, the study of the  AdS/CFT correspondence is realized in a rather straightforward way. This is why we refer the new approach as AdS/CFT-adapted light-cone approach.

The reason why it is possible to develop the light-cone formulations of AdS field theory and CFT in terms of one and the same spin operators is related to the following fact. In the light-cone gauge formulations of AdS field theory and CFT, we deal with fields which are not subject to any differential constraints. For the light-cone gauge field theory in $AdS_{d+1}$, it is convenient to decompose light-cone gauge massless and massive fields into irreps of the $so(d-2)$ algebra.
For the light-cone gauge CFT in $R^{d-1,1}$, it is also convenient to decompose
light-cone gauge short currents (shadows) and long currents (shadows)
into irreps of the $so(d-2)$ algebra. In the framework of AdS/CFT correspondence, massless
AdS field is related to short current (shadow). Namely, considering massless field
in $AdS_{d+1}$ and the corresponding short current (shadow) in $R^{d-1,1}$, we note that
irreps of the $so(d-2)$ algebra which enter the light-cone gauge formulation of
the massless field in $AdS_{d+1}$ coincide with irreps of the $so(d-2)$ algebra which enter the light-cone gauge formulation of the short current (shadow) in $R^{d-1,1}$. In other words, there is precise matching between irreps of the $so(d-2)$ algebra involved in the light-cone gauge formulation of the massless field in $AdS_{d+1}$ and irreps of the $so(d-2)$ algebra involved in the light-cone gauge formulation of the corresponding short current (shadow) in $R^{d-1,1}$. The same matching happens for massive fields and the corresponding long currents (shadows). Namely, irreps of the $so(d-2)$ algebra which enter the light-cone gauge formulation of the massive field in $AdS_{d+1}$ coincide with irreps of the $so(d-2)$ algebra which enter the light-cone gauge formulation of the long current (shadow) in $R^{d-1,1}$. In fact, it is the remarkable matching of bulk and boundary irreps of the $so(d-2)$ algebra that explains why it is possible to develop the light-cone gauge formulations of fields in $AdS_{d+1}$  and currents (shadows) in $R^{d-1,1}$ in terms of one and the same spin operators.

For the reader convenience, let us mention briefly some results concerning the Lorentz covariant description of mixed-symmetry AdS fields. General mixed-symmetry {\it massless} fields in $AdS_{d+1}$, $d$-arbitrary, were studied in Refs.\cite{Metsaev:1995re}, \cite{Brink:2000ag}-\cite{Campoleoni:2012th}. Two-column mixed-symmetry {\it massless} fields in $AdS_{d+1}$ were considered in Ref.\cite{Alkalaev:2003hc}, while two-row mixed-symmetry {\it massless} fields in $AdS_5$ were investigated in Refs.\cite{Alkalaev:2006hq}. General mixed-symmetry {\it massive} fields in $AdS_{d+1}$ were studied in Refs.\cite{Alkalaev:2011zv,Burdik:2011cb}. Two-column and two-row mixed-symmetry {\it massive} fields in $AdS_{d+1}$ were investigated in Ref.\cite{deMedeiros:2003px} and Refs.\cite{Zinoviev:2008ve} respectively. Interacting hook mixed-symmetry massive fields in $AdS_{d+1}$ were considered in Refs.\cite{Zinoviev:2010av}, while interacting two-row and two-column  mixed-symmetry massless fields in $AdS_{d+1}$ were studied in Refs.\cite{Alkalaev:2010af} and Refs.\cite{Boulanger:2011se} respectively.%
\footnote{
Successful applications of light-cone gauge and BRST approaches in string theory give hope that these approaches will be useful for the study of interacting mixed-symmetry AdS fields.
In flat space, interacting mixed-symmetry fields were studied by the BRST method in Refs.\cite{Koh:1986vg}-\cite{Metsaev:2012uy}
and by the light-cone gauge method in Refs.\cite{Metsaev:1993mj}-\cite{Metsaev:2007rn}.
}

The paper is organized as follows. In Sec.\ref{section-02}, we start with brief review of light-cone approach to AdS fields developed in Ref.\cite{Metsaev:1999ui}. After that, we present our new light-cone formulation which is adapted for the study of AdS/CFT. We present light-cone gauge action which leads to decoupled equations of motion for light-cone gauge AdS fields. Realization of relativistic symmetries of light-cone gauge AdS fields is also discussed.  In Sec.\ref{section-03}, we discuss our new light-cone gauge formulation of currents and shadows in flat space. Two-point functions for light-cone gauge currents and shadows are presented.  Realization of conformal  symmetries on space of currents and shadows is discussed. Sec.\ref{section-4} is devoted to the study of AdS/CFT. We demonstrate that, in the framework of AdS/CFT, the normalizable modes of AdS fields are related to currents, while the non-normalizable modes of AdS fields are related to shadows. Also we show that light-cone gauge action of AdS field computed on solution of Dirichlet problem coincides with light-cone gauge two-point function of shadow.  In Sec.\ref{sec-05}, we study the light-cone gauge formulation of conformal fields in flat space. We present the action which leads to decoupled equations of motion for light-cone gauge conformal fields. Simple examples of short and long totally symmetric arbitrary spin conformal fields are discussed with some details.
In Secs.\ref{s6},\ref{sec-07}, to illustrate our approach,  we deal with totally antisymmetric and two-column fields which, in the earlier literature, have not been studied in the framework of light-cone gauge approach. For such fields, we demonstrate how finding a solution to  our defining equations for the spin operators leads immediately to the complete light-cone description of fields in AdS space and currents, shadows, and conformal fields in flat space. Various technical details are collected in three Appendices.

\newsection{Light-cone gauge formulation of fields in $AdS_{d+1}$ space}\label{section-02}

We begin with the discussion of light-cone gauge formulation of
field dynamics in AdS space. To this end we use Poincar\'e parametrization of $AdS_{d+1}$ space given by (for details of our notation, see Appendix A)
\be \label{18052015-man-01}
ds^2 = \frac{1}{z^2}(dx^a dx^a + dz dz)\,.
\ee
To simplify the presentation, we collect all scalar, vector, tensor, and arbitrary spin fields into a ket-vector $|\phi\rangle$. By definition, in light-cone gauge approach, all fields entering the ket-vector $\phik$ are not subject to any differential constraints. In terms of the ket-vector $|\phi\rangle$, light-cone action and Lagrangian can be cast into the following form \cite{Metsaev:1999ui}:
\beq
\label{18052015-man-02} S & = &  \int d^dx dz \, \LL \,,
\\
\label{18052015-man-03} && \LL  = \half \langle \phi|\bigl(\Box + \partial_z^2 -\frac{1}{z^2}A\bigr)|\phi\rangle\,, \qquad
\\
\label{18052015-man-04} &&  \Box = 2\partial^+\partial^- + \partial^i\partial^i\,,
\eeq
where $\Box$ in \rf{18052015-man-03},\rf{18052015-man-04} stands for the D'Alembertian operator in $R^{d-1,1}$. Operator $A$ appearing in \rf{18052015-man-03} is independent of space-time coordinates and their derivatives. This operator acts only on spin D.o.F of fields collected into the ket-vector $|\phi\rangle$.

\noindent {\bf Light-cone gauge realization of relativistic symmetries}. Relativistic symmetries of fields propagating in $AdS_{d+1}$ space are described by the $so(d,2)$ algebra. The Lorentz symmetries of fields in  $AdS_{d+1}$ space are described by the $so(d,1)$  algebra. The use of Poincar\'e parametrization \rf{18052015-man-01} spoils manifest $so(d,1)$  Lorentz algebra symmetries. Namely, only the $so(d-1,1)$ algebra symmetries are manifest when we use Poincar\'e parametrization \rf{18052015-man-01}. Note however that the choice of the light-cone gauge spoils the manifest $so(d-1,1)$ algebra symmetries. This is to say that, in the framework of the light-cone approach, in order to complete the description of relativistic dynamics of fields in $AdS_{d+1}$ space we have to work out an explicit realization of the $so(d-1,1)$ algebra symmetries and the remaining relativistic symmetries as well. To this end we now  discuss the $so(d,2)$ algebra symmetries of the light-cone gauge action given in \rf{18052015-man-02}.

In the light-cone formulation,  generators of the $so(d,2)$ algebra can be separated into the following two groups:
\beq
\label{18052015-man-01a} && P^i,\ \ P^+, \ \ J^{+i}, \ \ J^{+-}, \ \ J^{ij}, \ \  D, \ \ K^i,\ \ K^+,\hspace{1cm} \hbox{ kinematical generators};\qquad
\\
\label{18052015-man-02a} && P^-,\ \ J^{-i}\,, \ \ K^-\hspace{5.7cm} \hbox{ dynamical
generators},
\eeq
where the vector indices of the $so(d-2)$ algebra take values $i,j=1,\ldots,d-2$ (for details of our notation see Appendix A). For the case of free fields, the field theoretical realization of $so(d,2)$ algebra generators $G_\field$ in \rf{18052015-man-01a}, \rf{18052015-man-02a} takes the following form:
\be \label{19052015-man-31}
G_\field = \int dz dx^-d^{d-2} x\,\langle\partial^+\phi|G_\diff|\phi\rangle.
\ee
Quantity $G_\diff$ appearing on right hand side \rf{19052015-man-31} stands for the realization of  generators \rf{18052015-man-01a}, \rf{18052015-man-02a} in terms of differential operators defined on space of the ket-vector $\phik$. Light-cone gauge action \rf{18052015-man-02} is invariant under the transformation $\delta\phik= G_\diff \phik$. In Ref.\cite{Metsaev:1999ui}, we found the following explicit expressions for the differential operators $G_\diff$:

\noindent {\bf Kinematical generators},
\beq
\label{19052015-man-32} && P^+=\partial^+\,, \hspace{3.4cm} P^i=\partial^i\,,
\\
\label{19052015-man-33} && J^{+-} = x^+P^-  - x^-\partial^+\,, \hspace{1cm}
J^{+i}=  x^+\partial^i - x^i\partial^+\,,
\\
\label{19052015-man-35} && J^{ij} = x^i\partial^j-x^j\partial^i + M^{ij}\,,
\\
\label{19052015-man-36} && D = x^+P^- + x^-\partial^+ + x^i\partial^i + z\partial_z + \frac{d-1}{2}\,,
\\
\label{19052015-man-37} && K^+ = -\frac{1}{2}(2x^+x^- + x^ix^i + z^2)\partial^+ +x^+D \,,
\\
\label{19052015-man-38} && K^i = -\frac{1}{2}(2x^+x^- + x^jx^j + z^2 )\partial^i + x^i D +
M^{ij} x^j + M^{i-}x^+ + M^{\ominussm i} \,,
\eeq

\noindent {\bf Dynamical generators},
\beq
\label{19052015-man-39} && P^-=\frac{-\partial^i\partial^i + \MM^2}{2\partial^+}\,,
\\
\label{19052015-man-40} && J^{-i}=x^-\partial^i-x^i P^- + M^{-i}\,,
\\
\label{19052015-man-41} && K^- = -\frac{1}{2}(2x^+x^-  + x^jx^j + z^2) P^- + x^-D
+ x^i M^{-i} - M^{\ominussm i} \frac{\partial^i}{\partial^+} +  \frac{1}{\partial^+} B \,, \qquad
\\
\label{19052015-man-42} && \hspace{1cm}  \MM^2 \equiv -\partial_z^2 + \frac{1}{z^2} A\,,
\\
\label{19052015-man-43} && \hspace{1cm}  M^{-i} \equiv
M^{ij}\frac{\partial^j}{\partial^+}+\frac{1}{\partial^+}M^{\oplussm i}\,,
\\
\label{19052015-man-44} &&  \hspace{1cm} M^{\oplussm i}   =  - M^{zi}\partial_z - \frac{1}{2z} [M^{zi},A]\,, \hspace{1cm}  M^{\ominussm i}  =   - zM^{zi}\,,
\\
\label{19052015-man-46} && \hspace{1cm}  B = \half \Bigl( A -  \half M^{ij}M^{ij} - \langle \CC_{so(d,2)}\rangle  - \frac{d^2-1}{4}\Bigr)\,.
\eeq
In \rf{19052015-man-46} and below, the quantity $\langle \CC_{so(d,2)}\rangle$ stands for an eigenvalue of the second order Casimir operator of the $so(d,2)$ algebra. For the representation $D(E_0,\hbf)$, the $\langle \CC_{so(d,2)}\rangle$ is given by
\be \label{19062015-man-01}
\langle \CC_{so(d,2)}\rangle = E_0(E_0-d)+
\sum_{\sigma=1}^{[d/2]} h_\sigma (h_\sigma -2\sigma +d)\,.
\ee
The spin operators $M^{ij}$, $i,j=1,\ldots,d-2$, form commutation relations of the $so(d-2)$ algebra, while the spin operators $M^{ij}$ and $M^{zi}$ form commutation relations of the $so(d-1)$ algebra (see relations in \rf{31012015-03}-\rf{31012015-03b}). Hermitian conjugation rules are given by
\be
A^\dagger = A\,, \qquad M^{ij\dagger} = - M^{ij}\,, \qquad M^{zi\dagger} = - M^{zi}\,.
\ee

From \rf{19052015-man-32}-\rf{19052015-man-46}, we see that a knowledge of the operators $A$, $M^{ij}$, $M^{zi}$ allows us to fix the realization of the kinematical and dynamical generators on space of the ket-vector $\phik$. Thus problem of the light-cone gauge description of AdS fields amounts to finding an explicit expressions for the operators $A$, $M^{ij}$, $M^{zi}$ which we refer to as basic operators. Closed system of equations for these basic operators was found in Ref.\cite{Metsaev:1999ui} (see Eqs.\rf{31012015-01}-\rf{31012015-03b} in Appendix B in this paper).

The presentation above-given summarizes the light-cone formulation of AdS fields developed in Ref.\cite{Metsaev:1999ui}. We now proceed to the discussion of AdS/CFT-adapted light-cone formulation which is based on the use of new basic operators. Such new basic operators turn out to be very convenient for the study of AdS/CFT correspondence and conformal fields.

\noindent {\bf New basic operators}. In this paper, we introduce new basic operators denoted by $\nu$, $W^i$, $\Wb^i$. As before, the $so(d-2)$ algebra spin operator $M^{ij}$ also enters the game. In terms of the new operators and the operator $M^{ij}$, the operators $A$, $B$, $M^{zi}$ and $M^{\oplussm i}$ take the form
\beq
\label{18052015-man-06} && A  = \nu^2 - \frac{1}{4}\,,
\\
\label{19052015-man-47} && M^{zi} = W^i - \Wb^i\,,
\\
\label{19052015-man-48} && M^{\oplussm i} = \TT_{\nu+\half} \Wb^i - W^i \TT_{-\nu-\half}\,,
\\
\label{19052015-man-49} && B = \half \Bigl( \nu^2  - \langle \CC_{so(d,2)} \rangle - \half M^{ij}M^{ij} - \frac{d^2}{4} \Bigr)\,,
\\
\label{19052015-man-50} && \qquad \ \TT_\nu \equiv \partial_z + \frac{\nu}{z}\,.
\eeq
We note the following

\noindent {\bf Defining equations for the new basic operators $\nu$, $W^i$, $\Wb^i$, $M^{ij}$}:
\beq
\label{04032015-00} && [\nu , M^{ij}]=0\,,
\\
\label{04032015-00a} && [M^{ij},M^{kl}] = \delta^{jk}M^{il} + 3 \hbox{ terms} \,, \qquad M^{ij} = - M^{ji}\,,
\\
\label{04032015-01} && [\nu, W^i] = W^i\,,
\\
\label{04032015-02} && [\nu,\Wb^i]= - \Wb^i\,,
\\
\label{04032015-02a} && [W^i, M^{jk}]= \delta^{ij} W^k - \delta^{ik} W^j\,,
\\
&& [\Wb^i, M^{jk}]= \delta^{ij} \Wb^k - \delta^{ik} \Wb^j\,,
\\
\label{04032015-03} && [W^i,W^j]= 0 \,,
\\
&& [\Wb^i,\Wb^j] = 0 \,,
\\
\label{04032015-04} && W^i \Wb^j  -  W^j \Wb^i + \Wb^i W^j - \Wb^j W^i = M^{ij}\,,
\\
\label{04032015-05} && (\nu-1) \bigl(W^i \Wb^j + W^j \Wb^i \bigr) - (\nu+1) \bigl(\Wb^i W^j + \Wb^j W^i \bigr)   = \delta^{ij} B + \half \{ M^{il},M^{lj} \}\,, \qquad
\eeq
where the operator $B$ appearing in \rf{04032015-05} is defined by relations \rf{19052015-man-49}, \rf{19062015-man-01}. We assume the hermitian conjugation rules,
\be \label{04032015-05a}
\nu^\dagger = \nu\,, \qquad M^{ij\dagger} = - M^{ij}\,, \qquad W^{i\dagger} = \Wb^i\,.
\ee

Equations \rf{04032015-00}-\rf{04032015-05} together with relations in \rf{19052015-man-49}, \rf{19062015-man-01} constitute the defining equations of AdS/CFT-adapted light-cone gauge formulation of the relativistic dynamics in AdS space we suggest in this paper. These equations are equivalent to the equations for operators $A$, $M^{zi}$, $M^{ij}$ we obtained in Ref.\cite{Metsaev:1999ui}. Derivation of Eqs.\rf{04032015-00}-\rf{04032015-05} from the ones in Ref.\cite{Metsaev:1999ui} may be found in Appendix B.

The following remarks are in order.

\noindent
\ibf) For AdS field associated with the representation $D(E_0,\hbf)$, the operator $\nu$ can be presented as
\be \label{18052015-man-07}
\nu = \kappa + M^z\,,   \qquad \qquad  \kappa \equiv E_0 - \frac{d}{2}\,,
\ee
where we introduce a new operator $M^z$. The use of the representation for $\nu$ in \rf{18052015-man-07} is advantageous in view of the following reason. It turns out that eigenvalues of the operator $M^z$ on the space of ket-vector $\phik$ are integers. Namely, for massive field, eigenvalues of $M^z$ take values $-h_k,-h_k+1, \ldots, h_k$, while,
for massless field, eigenvalues of $M^z$ take values $-h_k,-h_k+1, \ldots, h_{k+1}$.

\noindent \iibf) The particular representation for the operator $B$ in \rf{19052015-man-49} is simply obtained from general representation for the operator $B$ in \rf{19052015-man-46} and the particular representation for the operator $A$ in \rf{18052015-man-06}. For AdS field associated with the representation $D(E_0,\hbf)$, we can use relations \rf{18052015-man-07} and \rf{19062015-man-01} to represent the operator $B$ \rf{19052015-man-49} in the following two forms:
\beq
\label{18052015-man-08a} B  & = &  \half\Bigl(\nu^2  - \kappa^2 - \langle \CC_{so(d)}\rangle - \half M^{ij}M^{ij}\Bigr)
\nonumber\\
& = & \kappa M^z + \half\Bigl( M^z M^z -  \half M^{ij}M^{ij} - \langle \CC_{so(d)}\rangle \Bigr)\,,
\eeq
where quantity $\langle \CC_{so(d)}\rangle$ appearing in \rf{18052015-man-08a} stands for eigenvalue of the second order Casimir operator of the $so(d)$ algebra. For the representation labelled by the highest weight $\hbf=(h_1,\ldots, h_r)$, $r=[\frac{d}{2}]$, the $\langle \CC_{so(d)}\rangle$  is given by the well known expression,
\be
\langle \CC_{so(d)}\rangle = \sum_{\sigma=1}^{[d/2]} h_\sigma (h_\sigma -2\sigma +d)\,.
\ee

\noindent
\iiibf) The particular representation for the operator $M^{\oplussm i}$ given in \rf{19052015-man-48} is obtained from general representation in \rf{19052015-man-44} by using \rf{18052015-man-06}, \rf{19052015-man-47} and commutators in \rf{04032015-01}, \rf{04032015-02}. The operators $\TT_\nu$  \rf{19052015-man-50} and $M^{\oplussm i}$ \rf{19052015-man-48} satisfy the following interesting relations
\be
-\TT_{\nu+\half} \TT_{-\nu-\half} = \MM^2\,, \qquad [M^{\oplussm i}, M^{\oplussm j}] = \MM^2 M^{ij}\,, \qquad \MM^2 \equiv  -\partial_z^2 + \frac{1}{z^2}(\nu^2-\frac{1}{4})\,.
\ee

\noindent \ivbf) Using \rf{04032015-01}, \rf{04032015-02}, we can represent Eqs.\rf{04032015-05} in the following concise form
\be
W^i \nu \Wb^j + W^j \nu \Wb^i - \Wb^i \nu W^j - \Wb^j \nu W^i  = \delta^{ij} B + \half \{ M^{il},M^{lj} \}\,. \qquad
\ee

\noindent \vbf) The defining equations \rf{04032015-00}-\rf{04032015-05} take the form of some special deformation of commutation relations of the $so(d)$ algebra. In the flat space limit which is realized as $R\rightarrow \infty$, where $R$ is radius of AdS space, the defining equations \rf{04032015-00}-\rf{04032015-05} become the usual commutators of the $so(d)$ algebra. To see this we note the following relations in the limit as $R\rightarrow \infty$:
\beq
\label{09072015-man-01} && \kappa|_{_{R \rightarrow \infty}} \rightarrow  R \mrm \,, \qquad W^i |_{_{R \rightarrow \infty}} \rightarrow W_0^i\,, \qquad \Wb^i |_{_{R \rightarrow \infty}} \rightarrow \Wb_0^i\,, %
\\
\label{09072015-man-02} && M^{ij} = M_0^{ij}\,, \qquad  \ \ \ M^z = M_0^z\,,
\eeq
where $\mrm$ in \rf{09072015-man-01} stands for mass parameter of a massive field.%
\footnote{
The asymptotic behaviour of $\kappa$ in \rf{09072015-man-01} can easily been obtained from relation (5.62) in Ref.\cite{Metsaev:2003cu}. Note that, in Ref.\cite{Metsaev:2003cu}, we set the AdS radius $R=1$. To restore the dependence on the $R$, we should make the re-scaling $ m\rightarrow m R$ in relation (5.62) in Ref.\cite{Metsaev:2003cu}.
}
Using \rf{09072015-man-01}, \rf{09072015-man-02}, we see that, in the flat limit, commutators in \rf{04032015-00}, \rf{04032015-01}, \rf{04032015-02}, \rf{04032015-05} take the form
\beq
\label{09072015-man-03} && [M_0^z,M_0^{ij}] = 0 \,, \qquad [M_0^z, W_0^i] = W_0^i\,, \qquad  [M_0^z,\Wb_0^i]= - \Wb_0^i\,,
\\
\label{09072015-man-04} && W_0^i \Wb_0^j + W_0^j \Wb_0^i   - \Wb_0^i W_0^j - \Wb_0^j W_0^i   = \delta^{ij} M_0^z\,.
\eeq
Obviously, commutators \rf{09072015-man-03}, \rf{09072015-man-04} together with the ones in \rf{04032015-00a}, \rf{04032015-02a}-\rf{04032015-04} form the commutators of the $so(d)$ algebra.

\noindent \vibf)
The explicit form of the operator $\nu$ is known for
the following dynamical systems:

  \abf) totally symmetric arbitrary spin  massless and massive fields in $AdS_{d+1}$, $d\geq 3$, \cite{Metsaev:1999ui,Metsaev:2013kaa};

 \bbf) mixed symmetry arbitrary spin massless and self-dual massive
fields in $AdS_5$ \cite{Metsaev:2002vr};

  \cbf) mixed symmetry arbitrary spin massive fields in $AdS_5$ \cite{Metsaev:2014sfa};

 \dbf) type IIB supergravity in $AdS_5 \times S^5$ and $AdS_3 \times
S^3$ backgrounds \cite{Metsaev:1999gz,Metsaev:2000mv}.

\medskip
Just to illustrate our approach we consider arbitrary spin massless fields in $AdS_4$.

{\bf Massless arbitrary spin field in $AdS_4$}.  For $AdS_4$, one gets $d=3$ and hence the vector index of the $so(d-2)$ algebra $i$ takes only one value, $i=1$.
In order to describe field content entering dynamics of massless field in $AdS_4$ we use oscillators $\alpha^1$, $\alpha^z$. Commutation relations for the oscillators, the vacuum $|0\rangle$, and hermitian conjugation rules are defined as
\be
[\alphab^1,\alpha^1] = 1\,, \qquad [\alphab^z,\alpha^z]= 1\,, \qquad \alphab^1 |0\rangle = 0 \,, \qquad \alphab^z|0\rangle = 0\,, \qquad \alpha^{1\dagger} = \alphab^1\,, \qquad \alpha^{z\dagger} = \alphab^z\,.
\ee
Using such oscillators, we introduce the ket-vector $\phik$ to discuss massless spin-$s$ field,
\be
\label{18062015-man-01}  \phik = |\phi_s\rangle + \alpha^1 |\phi_{s-1}\rangle\,,
\qquad  |\phi_s\rangle \equiv \frac{\alpha_z^s}{\sqrt{s!}} \phi_s|0\rangle\,, \qquad |\phi_{s-1}\rangle \equiv \frac{\alpha_z^{s-1}}{\sqrt{(s-1)!}} \phi_{s-1}|0\rangle\,, \qquad
\ee
$\alpha_z =\alpha^z$, where two real-valued fields $\phi_s=\phi_s(x,z)$ and $\phi_{s-1}=\phi_{s-1}(x,z)$ appearing in \rf{18062015-man-01} stand for two propagating physical D.o.F which describe spin-$s$ massless field in $AdS_4$. Operators $A$, $\nu$, $W^1$, $\Wb^1$, $B$ can be read from results in Ref.\cite{Metsaev:2013kaa},
\beq
\label{18062015-man-06} && A = 0\,, \qquad  \nu = s - \half -  N_z\,, \qquad  B = - s^2\,,
\\
\label{18062015-man-04} && W^1 = - \sqrt{s} \alpha^1(1-N_\alpha)\bar\alpha^z\,, \hspace{1.3cm}  \Wb^1 = - \sqrt{s} \alpha^z \bar\alpha^1\,,
\\
&& N_\alpha\equiv \alpha^1\alphab^1\qquad N_z\equiv \alpha^z\alphab^z\,.
\eeq
The relation $A=0$ in \rf{18062015-man-06} is obtained by using general formula \rf{18052015-man-06} and noticing that $\nu$ given in \rf{18062015-man-06} takes values $\pm 1/2$ on space of ket-vector \rf{18062015-man-01}. Using \rf{18062015-man-06} in \rf{18052015-man-03}, we get Lagrangian of massless arbitrary spin field in $AdS_4$,
\be \label{18062015-man-07a}
\LL =  \half \phibr (\Box + \partial_z^2)\phik\,, \hspace{1cm}  \hbox{ for massless field in } AdS_4\,.
\ee

A a side remark we note that all unitary representations of the $so(d+1,2)$ algebra which are associated with fields propagating in $R^{d,1}$ and $AdS_{d+1}$ were found in Refs.\cite{Siegel:1988gd,Metsaev:1995jp}. For fields in $AdS_{d+1}$ which respect not only the relativistic $so(d,2)$ symmetries but also the conformal $so(d+1,2)$ symmetries, the operator $A$ is equal to zero and the light-cone gauge action \rf{18052015-man-02} is invariant under transformations of the $so(d+1,2)$ algebra.%
\footnote{ As shown in Ref.\cite{Metsaev:1995jp}, for fields in $AdS_{d+1}$ which
respect the conformal $so(d+1,2)$ symmetries, one has the following relations and restrictions for $E_0$, $h_i$, and $d$: $h_1= h_2 =\ldots = h_r$, $r\equiv \frac{d-1}{2}$, $E_0 = h_1 + \frac{d-1}{2}$, $d$-odd. Recent interesting notes on conformal symmetries of massless and partial-massless fields may be found in Refs.\cite{Beccaria:2015vaa}.
}
We note then, that, for arbitrary spin massless fields in $AdS_4$, the light-cone gauge action \rf{18052015-man-02},\rf{18062015-man-07a} is invariant under the conformal $so(4,2)$ symmetries.

\newsection{Light-cone gauge approach to currents and shadows in $R^{d-1,1}$}\label{section-03}

Light-cone gauge formulation of currents (shadows) can be developed by starting with Lorentz covariant approach to currents (shadows). Doing so, we use differential constraints appearing in Lorentz covariant formulation of currents (shadows) to solve all fields entering Lorentz covariant approach in terms of unconstrained fields.%
\footnote{ Study of differential constraints for short currents may be found in Refs.\cite{Shaynkman:2004vu}-\cite{Costa:2014rya}.
}
It is such unconstrained fields that constitute a field content entering the light-cone gauge formulation of currents (shadows). For the case of totally symmetric fields, the  demonstration of how the Lorentz covariant approach can be used for the derivation of the light-cone gauge formulation of currents (shadows) in $R^{d-1,1}$, $d$-arbitrary, may be found in Ref.\cite{Metsaev:1999ui}.%
\footnote{
The use of Lorentz covariant approach for the derivation of canonical formulation of totally symmetric arbitrary spin  currents in $R^{3,1}$ may be found in Ref.\cite{Koch:2014aqa}. Canonical formulation of totally symmetric arbitrary spin massless and massive fields in $AdS_{d+1}$, $d$-arbitrary, may be found in Ref.\cite{Metsaev:2011iz}. In the framework of canonical formulation, interesting recent discussion of AdS particle/string may be found in Ref.\cite{Jorjadze:2012jk}.
}
However this strategy is difficult to realize in the case of mixed-symmetry currents (shadows) because, in general, Lorentz covariant formulations of mixed-symmetry currents (shadows) are complicated or not available. One of the attractive features of the light-cone approach is that this approach admits to develop light-cone gauge formulation of currents and shadows without the use of Lorentz covariant formulation. Namely, in Ref.\cite{Metsaev:2005ws}, we developed the general light-cone formulation of arbitrary spin mixed-symmetry currents and shadows in $R^{d-1,1}$.

The light-cone gauge formulation of CFT worked out in Ref.\cite{Metsaev:2005ws} has the following unsatisfactory feature. Representation for generators of the $so(d,2)$ algebra obtained in Ref.\cite{Metsaev:2005ws} is non-local.%
\footnote{ We recall that if a representation for generators of the conformal symmetries is non-polynomial in the derivatives $\partial^i$, $\partial^-$, then such representation is referred to as non-local. Accordingly,
if a representation for the generators is polynomial in the $\partial^i$, $\partial^-$, then such representation is referred to as local.
We recall also that appearance of terms non-polynomial in the derivative $\partial^+$ is unavoidable
in light-cone gauge formulation.
}
In Ref.\cite{Metsaev:2005ws}, we noticed that the local representation can be obtained by choosing a suitable basis of currents (shadows). However an explicit local realization of the $so(d,2)$ algebra generators on such suitable basis of currents (shadows) has not been worked out in Ref.\cite{Metsaev:2005ws}. In this section, we fill this gap and find the following interesting result. It turns out that the local representation of the $so(d,2)$ algebra generators is entirely formulated in terms of basic operators $\nu$, $W^i$, $\Wb^i$ which we used to develop the light-cone gauge formulation of arbitrary spin AdS fields in Sec.\ref{section-02}.
In other words, it is the basic operators $\nu$, $W^i$, $\Wb^i$ discussed in Sec.\ref{section-02} that allows us to develop the local representation of the $so(d,2)$ algebra generators for currents and shadows. We start our presentation of new light-cone gauge approach to CFT with the discussion of  light-cone gauge 2-point vertices for currents and shadows.

\noindent {\bf Light-cone gauge 2-point vertices of currents and shadows}. To simplify the presentation we collect all scalar, vector, tensor, and arbitrary spin fields entering the light-cone gauge formulation of currents and shadows into the respective ket-vectors $|\phi_\cur\rangle$ and $|\phi_\sh\rangle$. By definition, in light-cone gauge approach, all fields entering the ket-vector $|\phi_\cur\rangle$, $|\phi_\sh\rangle$ are not subject to any differential constraints.

For currents and shadows, one can construct three 2-point vertices. The first 2-point vertex, which we denote by $\Gamma^{\cur-\sh}$, is a local functional of current and shadow. We now note the following expression for the local vertex:
\be \label{19052015-man-01}
\Gamma^{\rm cur-\sh} = \int d^d x \, \LL^{\rm cur-sh}\,,  \hspace{1.3cm} \LL^{\rm cur-sh} = \langle \phi_\cur| |
\phi_\sh\rangle \,.
\ee

The second 2-point function denoted by $\Gamma^{\rm \sh-\sh}$ is a nonlocal functional of two shadows, while the third 2-point function, denoted by $\Gamma^{\rm \cur-\cur}$, is a nonlocal functional of two currents. The 2-point functions $\Gamma^{\rm \sh-\sh}$ and $\Gamma^{\rm \cur-\cur}$ are given by
\beq
\label{19052015-man-03} && \Gamma^{\rm sh-sh}= \int d^dx_1 d^dx_2\, \LL_{12}^{\rm sh-sh} \,,
\\
\label{19052015-man-04} && \hspace{1.3cm} \LL_{12}^{\rm sh-sh} \equiv \half \langle\phi_\sh(x_1)|
\frac{f_\nu^\sh}{ |x_{12}|^{2\nu + d }} |\phi_\sh (x_2)\rangle \,,
\\
\label{19052015-man-05} && \hspace{1.3cm} f_\nu^\sh \equiv \frac{4^\nu \Gamma(\nu + \frac{d}{2})\Gamma(\nu + 1)}{4^\kappa
\Gamma(\kappa + \frac{d}{2})\Gamma(\kappa + 1)} \,,
\\[5pt]
\label{19052015-man-06} && \Gamma^{\rm cur-cur}= \int d^dx_1 d^dx_2\, \LL_{12}^{\rm cur-cur} \,,
\\
\label{19052015-man-07} && \hspace{1.3cm} \LL_{12}^{\rm cur-cur} \equiv \half \langle\phi_\cur(x_1)|
f_\nu^\cur |x_{12}|^{2\nu - d } |\phi_\cur (x_2)\rangle \,,
\\
\label{19052015-man-08} && \hspace{1.3cm} f_\nu^\cur \equiv \frac{4^\kappa
\Gamma(\kappa + 1 - \frac{d}{2})\Gamma(\kappa)}{4^\nu \Gamma(\nu + 1 - \frac{d}{2})\Gamma(\nu)} \,,
\\
&& \hspace{2cm} |x_{12}|^2 \equiv x_{12}^a x_{12}^a\,, \qquad x_{12}^a = x_1^a - x_2^a\,,
\eeq
where expression for the operator $\nu$ appearing in \rf{19052015-man-05}, \rf{19052015-man-08} is given in \rf{18052015-man-07}.

It the literature, sometimes one prefers to present the 2-point vertices in the momentum space.
Therefore, for the reader convenience, we represent our above-given expressions for  $\Gamma^{\rm sh-sh}$,$\Gamma^{\rm cur-cur}$ in the momentum space. To this end we make the Fourier transform for currents and shadows,
\be
|\phi(x)\rangle  = \int \frac{d^d p}{ (2\pi)^{d/2} }\ e^{\irm p^a x^a } |\widetilde\phi(p)\rangle \,.
\ee
Now it is easy to check that expressions for $\Gamma^{\rm sh-sh}$ \rf{19052015-man-03}
$\Gamma^{\rm cur-cur}$ \rf{19052015-man-06} take the following respective forms
\beq
\label{20062015-man-01} \widetilde\Gamma^{\rm sh-sh} & = &  \int d^dp \ \widetilde\LL^{\rm sh-sh} \,, \hspace{1.2cm} \widetilde\LL^{\rm sh-sh} \equiv \half \langle\widetilde\phi_\sh(p)|\Box_p^\nu |\widetilde\phi_\sh (p)\rangle \,,
\\
\label{20062015-man-02} \widetilde\Gamma^{\rm cur-cur} & = &  \int d^dp \ \widetilde\LL^{\rm cur-cur} \,,  \hspace{1cm} \widetilde\LL^{\rm cur-cur} \equiv \half \langle\widetilde\phi_\cur(p)|\Box_p^{-\nu} |\widetilde\phi_\cur (p)\rangle \,,
\\
&& \hspace{4cm} \Box_p \equiv  - p^a p^a \,,
\eeq
where the momentum-space bra-vectors $\langle\widetilde\phi_\cur(p)|$ and $\langle\widetilde\phi_\sh(p)|$ are defined according the rule $ \langle\widetilde\phi(p)| = (|\widetilde\phi(p)\rangle)^\dagger$. Note that expressions for $\Gamma^{\rm sh-sh}$ \rf{19052015-man-03} and $\Gamma^{\rm cur-cur}$ \rf{19052015-man-06} are equal to the respective expressions for $\widetilde\Gamma^{\rm sh-sh}$ \rf{20062015-man-01} and $\widetilde\Gamma^{\rm cur-cur}$ \rf{20062015-man-02} up to overall normalization factors.

\noindent {\bf $so(d,2)$ symmetries of light-cone gauge currents and shadows}. In the light-cone approach to currents and shadows, the Lorentz $so(d-1,1)$ algebra symmetries are not realized manifestly. This is to say that, in the framework of the light-cone approach, the complete description of currents and shadows implies that we have to work out an explicit realization of the $so(d-1,1)$ algebra symmetries and the remaining symmetries of the $so(d,2)$ algebra as well. To this end we now  discuss the $so(d,2)$ algebra symmetries of the light-cone gauge 2-point vertices above given.

The $so(d,2)$ algebra transformations of light-cone gauge currents and shadows can be presented as
\be \label{19052015-man-10}
\delta
|\phi_\cur\rangle
=
G_\cur
|\phi_\cur\rangle \,,
\qquad
\delta
|\phi_\sh\rangle
=
G_\sh
|\phi_\sh\rangle,
\ee
where the quantities $G_\cur$ and $G_\sh$ stand for realization of generators of the $so(d,2)$ algebra in terms of differential operators defined on the respective ket-vectors of currents and shadows. Expressions for the differential operators $G_\cur$ and $G_\sh$ we found admit the following representation:
\beq
\label{19052015-man-12}
&&
P^+ = \partial^+\,,  \quad P^- = \partial^-\,,  \hspace{1cm} P^i =  \partial^i\,,
\\
\label{19052015-man-13}
&&
J^{+-}
=
x^+ \partial^-
-
x^-\partial^+\,,
\qquad \ \
J^{+i}
=
x^+\partial^i
- x^i\partial^+\,,
\\
\label{19052015-man-15}
&&
J^{ij}
=
x^i\partial^j
-
x^j\partial^i
+
M^{ij}\,,
\\
\label{19052015-man-16} &&
J^{-i}
=
x^-\partial^i
-
x^i\partial^-
+
M^{-i}\,,
\\
\label{19052015-man-17} &&
D
=
x^+ \partial^-
+
x^-\partial^+
+
x^i\partial^i
+
\Delta\,,
\\
\label{19052015-man-18} &&
K^+
=
K_\Delta^+\,,
\\
\label{19052015-man-19}
&&
K^i
=
K_\Delta^i
+
M^{ij}x^j
+
\half\{ M^{i-},x^+\}
+
M^{\ominussm i}\,,
\\
\label{19052015-man-20}
&&
K^-
=
K_\Delta^-
+
\half \{M^{-i}, x^i\}
-
M^{\ominussm i}
\frac{\partial^i}{\partial^+}
+
\frac{1}{\partial^+}B\,,
\\
\label{19052015-man-21}
&&
\hspace{1cm}
K_\Delta^a
\equiv
-
\half
(2x^+x^-+x^jx^j) \partial^a
+
x^a D\,,
\qquad
a=\pm,i\,,
\\
\label{19052015-man-22}
&& \hspace{1cm}
M^{-i}
\equiv
M^{ij}
\frac{\partial^j}{\partial^+}
+
\frac{1}{\partial^+}
M^{\oplussm i}\,,
\qquad M^{i-}
=
-M^{-i}\,.
\eeq
Expressions for generators given in \rf{19052015-man-12}-\rf{19052015-man-22} are valid for both currents and shadows. Note however that explicit expressions for the operators $\Delta$, $M^{\ominussm i}$, $M^{\oplussm i}$ corresponding to the currents and shadows are different. Using the subscript `$\cur$' and `$\sh$' we note that the operators $\Delta$, $M^{\ominussm i}$, $M^{\oplussm i}$ corresponding to the currents and shadows are given by
\beq
\label{19052015-man-23} && \Delta_\cur = \frac{d}{2} + \nu\,,
\\
\label{19052015-man-24} && M_\cur^{\oplussm i} =  W^i\Box + \Wb^i\,,
\\
\label{19052015-man-25} && M_\cur^{\ominussm i} =  - W^i(2\nu+1)\,,
\\
\label{19052015-man-26} && B_\cur = \half \Bigl( \nu^2  - \langle \CC_{so(d,2)} \rangle - \half M^{ij}M^{ij} - \frac{d^2}{4} \Bigr)\,,
\\[10pt]
\label{19052015-man-27} && \Delta_\sh = \frac{d}{2} - \nu\,,
\\
\label{19052015-man-28} && M_\sh^{\oplussm i} =  W^i + \Wb^i \Box\,,
\\
\label{19052015-man-29} && M_\sh^{\ominussm i} =  (2\nu+1)\Wb^i\,,
\\
\label{19052015-man-30} && B_\sh = \half \Bigl( \nu^2  - \langle \CC_{so(d,2)} \rangle - \half M^{ij}M^{ij} - \frac{d^2}{4} \Bigr)\,.
\eeq
The D'Alembertian operator $\Box$ in $R^{d-1,1}$ appearing in \rf{19052015-man-24}, \rf{19052015-man-28} is given in \rf{18052015-man-04}.

The following remarks are in order.

\noindent \ibf) The representation for the generators of conformal symmetries we obtained is local, i.e., generators \rf{19052015-man-12}-\rf{19052015-man-30} are polynomial in the derivatives $\partial^i$, $\partial^-$. In fact, the main advantage of the light-cone approach to currents (shadows) in this paper as compared the one in Ref.\cite{Metsaev:2005ws} is that the representation for generators of conformal symmetries obtained in this paper is local.

\noindent \iibf) Operators $\nu$, $W^i$, $\Wb^i$, $M^{ij}$, which enter 2-point functions in \rf{19052015-man-03}, \rf{19052015-man-06} and generators of $so(d,2)$ algebra symmetries of currents and shadows in \rf{19052015-man-12}-\rf{19052015-man-30}, satisfy the same defining equations as the ones in light-cone gauge AdS field theory in \rf{04032015-00}-\rf{04032015-05}. In other words, in our light-cone approach, {\it the $so(d,2)$ algebra symmetries of bulk AdS field theory and the $so(d,2)$ algebra symmetries of boundary CFT are governed by one and the same operators} $\nu$, $W^i$, $\Wb^i$, $M^{ij}$. Also, in our light-cone gauge approach, {\it the Lagrangian of bulk AdS fields and 2-point functions of boundary currents and shadows are governed by the same operator} $\nu$. To summarize, finding a solution to defining equations given in \rf{04032015-00}-\rf{04032015-05} leads immediately to the complete description of light-cone gauge AdS field theory and light-cone gauge currents and shadows in CFT.

\noindent \iiibf) The quantity $\langle \CC_{so(d,2)} \rangle$ appearing in \rf{19052015-man-26}, \rf{19052015-man-30} stands for the second order Casimir operator of the $so(d,2)$ algebra. For current labelled by $E_0,\hbf$ and for shadow labelled by $d-E_0,\hbf$, the $\langle \CC_{so(d,2)}\rangle$ takes the form given in \rf{19062015-man-01}.

\noindent \ivbf) Operators $B_\cur$ \rf{19052015-man-26} and $B_\sh$ \rf{19052015-man-30} coincide and take the same form as in light-cone gauge AdS field theory \rf{19052015-man-49}.

\noindent \vbf) Realization for generators of $so(d,2)$ algebra symmetries on space of  currents and shadows given in \rf{19052015-man-12}-\rf{19052015-man-30} can be derived by using the general light-cone gauge approach to CFT developed in Ref.\cite{Metsaev:2005ws}. For some comments, see Appendix B.

\newsection{AdS/CFT correspondence} \label{section-4}

In the framework of AdS/CFT, boundary currents are related to normalizable solution of equations of motion for bulk AdS fields, while boundary shadows are related to non-normalizable solution of equations of motion for bulk AdS fields.
In Sec.\ref{section-02}, we obtained the light-cone gauge formulation for bulk AdS fields, while,  in Sec.\ref{section-03}, we obtained the light-cone gauge formulation for boundary  currents and shadows. We ready therefore to demonstrate the AdS/CFT explicitly. In the light-cone approach, the study of AdS/CFT implies, firstly, the matching of bulk and boundary symmetries of the $so(d,2)$ algebra and, secondly, the matching of effective action of bulk AdS field and boundary 2-point function of shadow. As a side remark we note that, in the framework of our light-cone gauge approach, the study of AdS/CFT is essentially simplified in view of the following two reasons.

\noindent \ibf) In the light-cone gauge approach, we deal with unconstrained fields on both the AdS and CFT sides. This implies that on both the AdS and CFT sides we deal with the same spin degrees of freedom. As we have already said, it is the matching of bulk and boundary spin degrees of freedom that explains why the generators of the $so(d,2)$ algebra in the bulk AdS theory and on the boundary CFT are governed by one and the same operators $\nu$, $W^i$, $\Wb^i$, $M^{ij}$ which we discussed in Sections \ref{section-02},\ref{section-03}.

\noindent \iibf) Light-cone gauge bulk action turns out be surprisingly simple and leads to the decoupled equations of motion.%
\footnote{ It will be interesting to find Lagrangian formulation of massive fields in non-conformal gravitational backgrounds admitting decoupled equations of motion and apply such formulation for the study of holography along the lines in Refs.\cite{Barvinsky:2005ms}. For recent discussion of massive fields in gravitational backgrounds, see, e.g., Refs.\cite{Cortese:2013lda}  (see also Refs.\cite{Buchbinder:2000fy}).
}
Therefore, finding a solution to the Dirichlet problem and the computation of the bulk action on solution of the Dirichlet problem is considerably simplified.

We now study the AdS/CFT duality for normalizable and non-normalizable modes of bulk AdS fields and corresponding boundary currents and shadows in turn.

\subsection{ AdS/CFT for normalizable
modes of AdS field and boundary current}

In this section, we study the AdS/CFT for bulk AdS field and
boundary current. As we have already said, in the framework of AdS/CFT, boundary current is related to normalizable solution of equations of motion for bulk AdS field. Therefore we start with analysis of the normalizable solution. To this end we note that the light-cone gauge action in \rf{18052015-man-02} leads to the following equations of motion for AdS field
\be \label{22062015-man-01}
\Bigl(\Box
+  \partial_z^2
- \frac{1}{z^2}(\nu^2
- \frac{1}{4})
\Bigr)|\phi\rangle = 0 \,.
\ee
The normalizable solution of Eq.\rf{22062015-man-01} is well known
\beq
\label{22062015-man-02}
|\phi(x,z)\rangle
& = & U_\nu  |\phi_\cur(x)\rangle \,,
\\
\label{22062015-man-03} &&
U_\nu
\equiv  h_\kappa
\sqrt{zq} J_\nu(zq)
q^{-(\nu + \half)}\,, \qquad
h_\kappa
\equiv 2^\kappa\Gamma(\kappa+1)\,,
\qquad
q^2\equiv \Box\,, \qquad
\eeq
where $J_\nu$ in \rf{22062015-man-03} stands for the Bessel function. From relation \rf{22062015-man-02}, we see that the normalizable solution for bulk AdS field $\phik$ is entirely governed by the operator $U_\nu$. To proceed we note the following helpful interesting relations for the operator $U_\nu$ which streamline the study of AdS/CFT:
\beq
\label{22062015-man-05} &&
\TT_{\nu-\half} U_\nu
= U_{\nu-1}\,,
\\
\label{22062015-man-06} &&
\TT_{-\nu-\half}
U_\nu
= - U_{\nu+1}\Box\,,
\\
\label{22062015-man-07} &&
\TT_{\nu+\half}
U_{\nu+1}
= U_\nu\,,
\\
\label{22062015-man-08} &&
\TT_{-\nu+\half}
U_{\nu-1}
= - U_\nu \Box\,,
\\
\label{22062015-man-09} &&
z U_{\nu-1}
+ z \Box U_{\nu+1}
= 2\nu U_\nu\,,
\\
\label{22062015-man-10}&&
x^a U_\nu
= U_\nu x^a
+ z U_{\nu+1} \partial^a\,,
\eeq
where the operator $\TT_\nu$ is defined in \rf{19052015-man-50}.
Relations in \rf{22062015-man-05}-\rf{22062015-man-10} should be understood in weak sense.  Namely, those relations hold true on space of ket-vectors which depend on the boundary coordinate $x^a$ and do not depend on the radial coordinate $z$. We note also that relations \rf{22062015-man-05}-\rf{22062015-man-10} can easily be proved by using the following well known formulas for the Bessel function:
\be \label{26062015-man-01}
\TT_\nu J_{\nu }
= J_{\nu-1}\,,
\qquad
\TT_{-\nu} J_{\nu }
= - J_{\nu + 1}\,.
\ee
In fact, it is concise form of relations for the Bessel function in \rf{26062015-man-01} that motivates us to use the operator $\TT_\nu$. Using textbook formulas for the Bessel function, we find that the asymptotic behavior of the normalizable solution given in  \rf{22062015-man-02} takes the following form
\be \label{14102014-07}
|\phi(x,z)\rangle \ \ \stackrel{z\rightarrow 0}{\longrightarrow} \ \ z^{\nu +
\half} \frac{2^\kappa\Gamma(\kappa+1)}{2^\nu\Gamma(\nu+1)} |\phi_\cur(x)\rangle\,.
\ee
The asymptotic behaviour in \rf{14102014-07} tells us that, up to normalization factor, the ket-vector $|\phi_\cur\rangle$ is nothing but the boundary value of the normalizable solution of equations for bulk AdS field.

From AdS/CFT dictionary, we expect that the ket-vector $|\phi_\cur\rangle$ appearing in \rf{14102014-07} describes boundary current. The fact that $|\phi_\cur\rangle$ \rf{14102014-07} is indeed realized as boundary current can be checked by matching of bulk and boundary symmetries of the $so(d,2)$ algebra. To check that $|\phi_\cur\rangle$ appearing on right hand side \rf{14102014-07} is realized as boundary current we should prove the following statement:

\noindent {\it Representation of the light-cone gauge symmetries of the $so(d,2)$ algebra on space of bulk normalizable solution \rf{22062015-man-02} amounts to representation of the light-cone gauge symmetries of the $so(d,2)$ algebra on space of boundary current}.

To this end let us use the notation $G_\AdSsm$ for the representation of the $so(d,2)$ algebra generators on space of bulk AdS field given in \rf{19052015-man-32}-\rf{19052015-man-46} and the notation  $G_\cur$ for representation of the $so(d,2)$ algebra generators on space of boundary current given in \rf{19052015-man-12}-\rf{19052015-man-26}.  With the use of such notation we note that all that is required is to demonstrate that the following relation
\be \label{27062015-man-01}
G_\AdSsm \phik = U_\nu G_\cur |\phi_\cur\rangle
\ee
holds true, where $\phik$ is the normalizabe solution given in \rf{22062015-man-02}.

The relation in \rf{27062015-man-01} can be proved by following the procedure we described in Section 6.1 in Ref.\cite{Metsaev:2014sfa}. For the reader convenience we present some relations which are helpful for analysis of relation \rf{27062015-man-01}.  We note that the following relations
\be \label{27062015-man-02}
\MM^2 U_\nu =  U_\nu \Box\,, \qquad  M_\AdSsm^{\oplussm i} U_\nu  = U_\nu M_\cur^{\oplussm i}
\ee
are helpful to verify relation \rf{27062015-man-01} for the generators $P^-$ and $J^{-i}$. Other relations given by
\beq
\label{27062015-man-04} &&
\bigl(
K_\Delta^a
- \half z^2 \partial^a
\bigr)\Bigr|_{ \Delta
= z\partial_z
+ \frac{d-1}{2} }
U_\nu
=
U_\nu
K_\Delta^a\Bigr|_{ \Delta
= \frac{d}{2}
+ \nu}\,,
\\
\label{27062015-man-05} &&
K_\Delta^a
\equiv
-
\half x^2
\partial^a
+ x^a
(x\partial
+ \Delta)\,,
\qquad a=\pm,i\,,
\eeq
where $x^2\equiv x^a x^a$, $x \partial \equiv x^a\partial^a$, are helpful to verify relation \rf{27062015-man-01} for the generators $K^a$. Relations \rf{27062015-man-02}-\rf{27062015-man-05} are easily proved by using relations given in  \rf{22062015-man-05}-\rf{22062015-man-10}. Also note that, as in \rf{22062015-man-05}-\rf{22062015-man-10}, relations in \rf{27062015-man-02}-\rf{27062015-man-05} should be understood in weak sense. Namely, relations \rf{27062015-man-02}-\rf{27062015-man-05} hold true on space of ket-vectors which depend on the boundary coordinate $x^a$ and do not depend on the radial coordinate $z$.

Representation of bulk fields in terms of currents \rf{22062015-man-02} is referred sometimes as holographic representation of bulk fields.
Our relation \rf{22062015-man-02} is valid for massless and massive arbitrary spin totally symmetric and mixed-symmetry fields. Thus, we see that our light-cone approach allows us
us to study the holographic representation for all just mentioned fields on an equal footing.%
\footnote{In Lorentz covariant approach, the holographic representation of totally symmetric arbitrary spin massless and massive AdS field was obtained for first time in the respective Ref.\cite{Metsaev:2008fs} and  Ref.\cite{Metsaev:2011uy}. In Lorentz covariant approach, the holographic representation of arbitrary spin mixed-symmetry fields is still to be investigated.}

\subsection{ AdS/CFT for non-normalizable
modes of AdS field and boundary shadow }\label{section4-2}

This section is devoted to the study of AdS/CFT for bulk AdS field and
boundary shadow. As we have already said, in the framework of AdS/CFT, boundary shadow is related to non-normalizable solution of equations of motion for bulk AdS field.%
\footnote{ In earlier literature, shadows and related dualities were discussed in Refs.\cite{Petkou:1994ad}. Recent interesting discussion of shadows may be found in Ref.\cite{SimmonsDuffin:2012uy}.}
Therefore we start with the analysis of the non-normalizable solution of bulk equations of motion.

The bulk equations of motion obtained from light-cone gauge action \rf{18052015-man-02} are given in \rf{22062015-man-01}. Non-normalizable solution to equations in  \rf{22062015-man-01} with the
Dirichlet problem corresponding to shadow is well known,
\beq
\label{27062015-man-06} |\phi(x,z)\rangle
& = &
\sigma_\nu \int d^d y\, G_\nu (x-y,z)
|\phi_\sh(y)\rangle\,,
\\
&& G_\nu(x,z)
= \frac{c_\nu z^{\nu+\half}}{
(z^2+ |x|^2)^{\nu + \frac{d}{2}} }\,,
\\
\label{27062015-man-06a} &&
c_\nu
\equiv
\frac{\Gamma(
\nu
+ \frac{d}{2})}{\pi^{d/2}
\Gamma(\nu)} \,,
\qquad \sigma_\nu
\equiv
\frac{2^\nu\Gamma(\nu)}{
2^\kappa\Gamma(\kappa)}  \,,
\eeq
where $G_\nu$ appearing on right hand side in \rf{27062015-man-06} is the bulk to boundary Green function. The asymptotic behaviors of the Green function and solution in \rf{27062015-man-06} are well known
\beq
&& G_\nu(x,z) \ \ \ \stackrel{z \rightarrow 0}{\longrightarrow} \ \ \ z^{-\nu
+ \half} \delta^d(x)\,,
\\
\label{27062015-man-08} && |\phi(x,z)\rangle  \,\,\, \stackrel{z\rightarrow 0 }{\longrightarrow}\,\,\,
z^{-\nu + \half} \sigma_\nu |\phi_\sh(x)\rangle\,.
\eeq
Relation \rf{27062015-man-08} tells us that, up to normalization factor, the ket-vector $|\phi_\sh\rangle$ is nothing but the boundary value of the non-normalizable solution of the Dirichlet  problem for bulk equations of motion.

Using the AdS/CFT dictionary, we expect that the ket-vector $|\phi_\sh\rangle$ appearing in \rf{27062015-man-06} describes boundary shadow. To make sure that ket-vector $|\phi_\sh\rangle$ \rf{27062015-man-06} is indeed realized as boundary shadow we should prove the following two statements:

\noindent \ibf) {\it Representation of the light-cone gauge symmetries of the $so(d,2)$ algebra on space of bulk non-normalizable solution \rf{27062015-man-06} amounts to the representation of the light-cone gauge symmetries of the $so(d,2)$ algebra on space of boundary shadow}.

\noindent \iibf) {\it The light-cone gauge action of bulk AdS field evaluated on the solution of the Dirichlet problem \rf{27062015-man-06} amounts to the boundary 2-point vertex for the shadow}.

We outline proof of these statements.

\noindent {\bf Matching of bulk and boundary representations of the $so(d,2)$ algebra}.
We use the notation $G_\AdSsm$ for the representation of the $so(d,2)$ algebra generators on space of bulk AdS field given in \rf{19052015-man-32}-\rf{19052015-man-46} and the notation  $G_\sh$ for the representation of the $so(d,2)$ algebra generators on space of boundary shadow given in \rf{19052015-man-12}-\rf{19052015-man-22}, \rf{19052015-man-27}-\rf{19052015-man-30}.  With the use of such notation we note that all that is required is to demonstrate that  $G_\AdSsm$ and $G_\sh$  respect the relation
\be \label{27062015-man-09}
G_\AdSsm  |\phi(x,z)\rangle   =   \sigma_\nu \int d^d y\, G_\nu (x-y,z)
G_\sh |\phi_\sh(y)\rangle\,,
\ee
where $\phik$ is given in \rf{27062015-man-06}. Relation given in \rf{27062015-man-09} can be proved by following the procedure we described in Section 6.1 in Ref.\cite{Metsaev:2014sfa}. For the reader convenience, we present the list of relations which are helpful for the analysis of relation \rf{27062015-man-09}
\beq
\label{27062015-man-10} &&
\TT_{\nu-\half}
(\sigma_\nu G_\nu )
=
\Box \sigma_{\nu-1} G_{\nu-1}\,,
\\
\label{27062015-man-11} &&
\TT_{-\nu - \half}
(\sigma_\nu G_\nu )
=
- \sigma_{\nu+1}
G_{\nu+1}\,,
\\
\label{27062015-man-12} && \TT_{\nu+\half}
(\sigma_{\nu+1}
G_{\nu+1})
=
\Box \sigma_\nu
G_\nu\,,
\\
\label{27062015-man-13} &&
\TT_{-\nu+\half}
(\sigma_{\nu-1}
G_{\nu-1})
=
- \sigma_\nu
G_\nu\,,
\\
\label{27062015-man-14} &&
z \Box \sigma_{\nu-1}
G_{\nu-1}
+ z \sigma_{\nu+1}
G_{\nu+1}
= 2\nu G_\nu\,,
\\
\label{27062015-man-15} && x^a \sigma_\nu
G_\nu
=
\sigma_\nu
G_\nu y^a
- z \partial^a
( \sigma_{\nu-1}
G_{\nu-1})\,,
\\
\label{27062015-man-16} && \MM^2 (\sigma_\nu G_\nu) =  \sigma_\nu G_\nu \Box\,, \qquad  M_\AdSsm^{\oplussm i} (\sigma_\nu G_\nu) = \sigma_\nu
G_\nu M_\sh^{\oplussm i}\,,
\\[3pt]
\label{27062015-man-18} &&
\bigl(
K_\Delta^a
- \half z^2 \partial^a
\bigr)\Bigr|_{ \Delta
= z\partial_z
+ \frac{d-1}{2} }
(\sigma_\nu G_\nu)
=
(\sigma_\nu G_\nu)
K_\Delta^a\Bigr|_{ \Delta
= \frac{d}{2}
- \nu}\,,
\\
\label{27062015-man-19} &&
K_\Delta^a
\equiv
-
\half x^2
\partial^a
+ x^a
(x\partial
+ \Delta)\,,
\qquad a=\pm,i\,,
\eeq
where $x^2\equiv x^a x^a$, $x \partial \equiv x^a\partial^a$, and operator $\TT_\nu$ is given in \rf{19052015-man-50}. Relations in \rf{27062015-man-10}-\rf{27062015-man-18} should be understood in weak sense, i.e., relations \rf{27062015-man-10}-\rf{27062015-man-18} hold true on space of ket-vectors which depend on the boundary coordinate $x^a$ and do not depend on the radial coordinate $z$. We note that relations \rf{27062015-man-16} are helpful to verify relation \rf{27062015-man-09} for the generators $P^-$ and $J^{-i}$, while relation \rf{27062015-man-18} is helpful to verify relation \rf{27062015-man-09} for the generators $K^a$. In turn,  relations \rf{27062015-man-16}-\rf{27062015-man-19} can be easily proved by using relations in \rf{27062015-man-10}-\rf{27062015-man-15}.

\noindent {\bf Matching of effective action and 2-point vertex for shadow}. Action of AdS field  evaluated on non-normalizable solution of Dirichlet problem \rf{27062015-man-06} we refer to as effective action. The effective action is found by plugging solution \rf{27062015-man-06} into  action \rf{18052015-man-02}, where action \rf{18052015-man-02} should be supplemented with appropriate boundary term. Using the method in Ref.\cite{Arutyunov:1998ve}, we verify that a light-cone gauge Lagrangian which involves an appropriate boundary term takes the form
\be \label{27062015-man-20}
\LL
=
\half  \langle
\partial^a
\phi|
| \partial^a
\phi\rangle
+
\half  \langle
\TT_{\nu-\half}
\phi|
| \TT_{\nu-\half}
\phi\rangle\,.
\ee
Note also that, to respect the commonly used Euclidean signature, the overall sign of Lagrangian  \rf{18052015-man-03} has been changed, $\LL\rightarrow - \LL$, when passing from expression \rf{18052015-man-03} to expression \rf{27062015-man-20}.

Now it is easy to check that action \rf{18052015-man-02}, \rf{27062015-man-20} considered on the solution of the Dirichlet problem admits the following general representation:
\be   \label{27062015-man-21}
- S_\eff
=
\int d^d x\,
\LL_\eff\Bigr|_{z\rightarrow
0} \,, \qquad  \ \ \
\LL_\eff
\equiv
\half \phibr
\TT_{\nu -\half }
\phik \,.
\ee
Plugging solution to the Dirichlet problem \rf{27062015-man-06} into  \rf{27062015-man-21}, we obtain the effective action
\be  \label{27062015-man-22}
-S_\eff  =   2\kappa c_\kappa \Gamma^{\rm sh-sh} \,,
\ee
where $c_\kappa$ is given in \rf{27062015-man-06a}, while the 2-point vertex of shadow $\Gamma^{\rm sh-sh}$ takes the form given in \rf{19052015-man-03}, \rf{19052015-man-04}. We recall also that $\kappa$ is defined in \rf{18052015-man-07}. Relation \rf{27062015-man-22} tells that the effective action and the 2-point vertex of shadow indeed match. By product, we find the normalization factor $2\kappa c_\kappa$ entering \rf{27062015-man-22}. This normalization factor might be important for the systematical investigation of the AdS/CFT duality.

Our relation for effective action \rf{27062015-man-22} is valid for massless and massive arbitrary spin totally symmetric and mixed-symmetry fields. Thus we see that the light-cone gauge approach allows us to study the AdS/CFT correspondence for all just mentioned fields on an equal footing.%
\footnote{ In Lorentz covariant approach, the computation of effective action for massless mixed-symmetry field in $AdS_{d+1}$ with $d$-arbitrary and $h_1=2$, $h_2=1$ may be found in Ref.\cite{Alkalaev:2012ic}, while the computation of effective action for massive self-dual field in $AdS_5$ with $\kappa$-arbitrary and $h_1=1$, $h_2=1$ may be found in Ref.\cite{Arutyunov:1998xt}. In Lorentz covariant approach, the problem of the effective action for arbitrary spin mixed-symmetry AdS fields is still to be investigated. The realization of AdS/CFT duality in terms of intertwining operator was investigated in Refs.\cite{Dobrev:1998md}.}

To summarize, we used new light-cone gauge formulation of bulk AdS field theory and boundary CFT for the investigation of the AdS/CFT correspondence of AdS fields and boundary currents (shadows). We believe that our new light-cone gauge approach to AdS field theory and CFT might have other interesting applications along the lines in Refs.\cite{Koch:2010cy,Florakis:2014kfa}.

\newsection{ Light-cone gauge approach to conformal fields in $R^{d-1,1}$} \label{sec-05}

Our expression for 2-point vertex of shadow given in \rf{19052015-man-03} provides interesting opportunity for finding a light-cone gauge Lagrangian of conformal field. To explain what has just been said we note that a kernel of the 2-point vertex of shadow in \rf{19052015-man-03}, \rf{19052015-man-04} is not well-defined for the cases when $\nu$ given in \rf{18052015-man-07} takes integer values (see, e.g., Ref.\cite{Aref'eva:1998nn}), while for the cases when $\nu$ takes non-integer values the kernel turns out to be well-defined. Therefore 2-point vertex of shadow with integer $\nu$ can be regularized by using non-integer $\nu$. Removing the regularization, we are left with a logarithmic divergence of the 2-point vertex for shadow which turns out to be light-cone gauge action of conformal field.

Let us describe the regularization procedure we are going to use. Taking into account \rf{18052015-man-07} and recalling that eigenvalues of the operator $M^z$ take integer values, we see that non-integer values of $\kappa$ imply non-integer values of $\nu$. Denoting the integer part of $\kappa$ by $\kappa_\intrm $, we define the regularization parameter $\varepsilon$ by the following relation:
\be \label{27062015-man-01a}
\kappa
- \kappa_\intrm
= - \varepsilon\,,
\qquad
\qquad
\kappa_\intrm
-\hbox{ integer}.
\ee
Taking into account \rf{27062015-man-01a} and \rf{18052015-man-07}, we use the textbook asymptotic behavior for the kernel:
\be \label{27062015-man-03a}
\frac{1}{|x|^{2\nu
+ d}}\,\,\,
\stackrel{\varepsilon \sim
0}{\mbox{\Large$\sim$}}\,\,\,
\frac{1}{\varepsilon}
\varrho_{\nu_\intrm} \Box^{\nu_{\intrm}}
\delta^{(d)}(x)\,,
\qquad
\nu_\intrm
\equiv
\kappa_\intrm
+ M^z,
\qquad
\varrho_\nu
\equiv
\frac{\pi^{d/2}}{4^\nu \Gamma(\nu
+ 1)
\Gamma(\nu
+ \frac{d}{2})}\,.
\ee
Plugging expression \rf{27062015-man-03a} into $\Gamma^{\rm sh-sh}$ given in \rf{19052015-man-03}, we get
\beq
&& \Gamma^{\sh-\sh}\,\,\,
\stackrel{ \varepsilon
\sim
0}{\mbox{\Large$\sim$}}\,\,\,
\frac{1}{\varepsilon}
\varrho_{\kappa_\intrm} \int
d^d x\,\,
\LL\,,
\hspace{1cm}
\nonumber\\
\label{27062015-man-04a}
&&
\LL
=
\half  \phibr
\Box^{\nu_\intrm}
\phik \,, \hspace{1cm} \nu_\intrm = \kappa_\intrm + M^z\,,
\eeq
where in expression for $\LL$ we use  the  identification $\phik\equiv |\phi_\sh\rangle$. Lagrangian \rf{27062015-man-04a} describes light-cone gauge dynamics of conformal field. In general, Lagrangian \rf{27062015-man-04a} describes higher-derivative theory of conformal field.

\noindent {\bf Long and short conformal fields}. In Sec.\ref{section4-2}, we demonstrated that, in the framework of AdS/CFT, the 2-point function $\Gamma^{\sh-\sh}$ is realized as effective action of bulk AdS field \rf{27062015-man-22}. Relation \rf{27062015-man-04a} tells us then that, for integer $\kappa$, the logarithmic divergence of the effective action of AdS field turns out to be the action of conformal field, i.e., AdS field theory leads to higher-derivative theory of conformal field only when $\kappa$ takes integer values. Also note that, in the framework AdS/CFT, the $\kappa$ is connected with $d$ and lowest eigenvalue of the energy operator $E_0$ of AdS field by relation \rf{18052015-man-07}.
From relation \rf{18052015-man-07}, we see that requiring the $\kappa$ be integer leads to restrictions on the $E_0$
and $d$. This is to say that not all AdS fields are related to conformal fields. Namely, as we have already said, AdS field is related to conformal field only when $\kappa$  \rf{18052015-man-07} takes integer values.
Let us now describe AdS fields which are related to conformal fields. We consider massless and massive AdS fields in turn. Requiring the $\kappa$ \rf{18052015-man-07} to be integer we are led to the following statements.

\medskip
\noindent \ibf) {\it Massless field in $AdS_{d+1}$ is related to conformal field in $R^{d-1,1}$ only for even $d\geq 4$}. This can easily been seen by noticing that, for the massless AdS field, the $E_0$ takes integer values (see \rf{27062015-man-05a}).  Therefore relation \rf{18052015-man-07} and requirement the $\kappa$ to be integer imply that $d$ should be even. Conformal field in $R^{d-1,1}$ related to massless field in $AdS_{d+1}$  is referred to as short conformal field. Thus we see that {\it short conformal field propagates in space-time of even dimension $d$}.%
\footnote{
The fact that short totally (anti)symmetric conformal fields propagate in $R^{d-1,1}$ only with even $d$ is well-known. It was also expected that short mixed-symmetry conformal fields propagate in $R^{d-1,1}$ only with even $d$. Our discussion demonstrates that the restriction the $d$ to be even integer for all short conformal fields can be explained by the fact that dependence of $E_0$ on $d$ in \rf{27062015-man-05a} is the same for all totally symmetric and mixed-symmetry massless AdS fields.
}

\medskip
\noindent \iibf) {\it Massive field in $AdS_{d+1}$ is related to conformal field in $R^{d-1,1}$ for arbitrary $d\geq 4$}. For even $d$, the $E_0$ takes integer values, while, for odd $d$, the $E_0$ takes half-integer values.
Allowed values of the $E_0$, denoted by $E_{0,\longrm}$ are given by
\beq
\label{27062015-man-09a} &&  E_{0,\longrm} = \kappa_\intrm + \frac{d}{2}\,, \qquad
\\
\label{27062015-man-09b} &&  \kappa_\intrm >  E_0^{\mrm =0} - \frac{d}{2}\,,
\eeq
where $\kappa_\intrm$ in \rf{27062015-man-09a} is arbitrary integer which satisfies \rf{27062015-man-09b}.
Conformal field which is related to massive AdS field is referred to as long conformal field.
As there are no restrictions on space-time dimension $d\geq 4$, we conclude that {\it the long conformal field propagates in space-time of arbitrary dimension $d\geq 4$}. Conformal dimension of the long conformal field is given by $\Delta = d- E_{0,\longrm}$, where $E_{0,\longrm}$ is given in \rf{27062015-man-09a}, \rf{27062015-man-09b}.

Thus, the short conformal fields propagate in space-time of even dimension $d\geq 4$ and these fields are related to massless fields, while the long conformal fields propagate in space-time of arbitrary dimension $d\geq 4$ and are related to massive AdS fields having $E_0$ defined by \rf{27062015-man-09a}, \rf{27062015-man-09b}.

Light-cone gauge action of conformal field \rf{27062015-man-04a} is invariant under symmetries of the $so(d,2)$ algebra. Realization of light-cone gauge symmetries of the $so(d,2)$ algebra on space of conformal field $\phik$ is described by the same relations as for shadow (see \rf{19052015-man-10}, \rf{19052015-man-12}-\rf{19052015-man-22} \rf{19052015-man-27}-\rf{19052015-man-30}).
Operators $\nu$, $W^i$, $\Wb^i$, $M^{ij}$, which enter light-cone gauge action \rf{27062015-man-04a} and generators of the $so(d,2)$ algebra symmetries of conformal fields  \rf{19052015-man-12}-\rf{19052015-man-30}, satisfy the same defining equations as the ones in  light-cone gauge AdS field theory in \rf{04032015-00}-\rf{04032015-05}. In other words, in our light-cone approach, {\it the $so(d,2)$ algebra symmetries of massless (massive) fields in $AdS_{d+1}$ and the $so(d,2)$ algebra symmetries of short (long) conformal fields in $R^{d-1,1}$ are governed by one and the same operators} $\nu$, $W^i$, $\Wb^i$, $M^{ij}$. Also, in our approach, {\it the Lagrangian of massless (massive) fields in $AdS_{d+1}$ and Lagrangian of conformal short (long) fields in $R^{d-1,1}$ are governed by the same operator} $\nu$. To summarize, finding a solution to defining equations given in \rf{04032015-00}-\rf{04032015-05} leads immediately to the complete description of light-cone gauge theory of massless (massive) fields in $AdS_{d+1}$ and light-cone gauge theory of short (long) conformal fields in $R^{d-1,1}$. We note also that irreps of the $so(d-2)$ algebra which enter the light-cone gauge formulation of massless (massive) field in $AdS_{d+1}$ coincide with irreps of the $so(d-2)$ algebra which enter the light-cone gauge formulation of short (long) conformal field in $R^{d-1,1}$. In fact, it is  matching of irreps of the $so(d-2)$ algebra entering the massless (massive) light-cone gauge fields in $AdS_{d+1}$ and the irreps of the $so(d-2)$ algebra entering the light-cone gauge short (long) conformal fields that explains why the light-cone gauge formulations of massless (massive) AdS fields and conformal short (long) fields are governed by one and the same operators $\nu$, $W^i$, $\Wb^i$, $M^{ij}$.

For the illustration purposes, we now discuss totally symmetric conformal fields. The light-cone gauge Lagrangian of totally symmetric short conformal fields has already been presented in Ref.\cite{Metsaev:2009ym}, while light-cone gauge Lagrangian of totally symmetric long conformal fields has not been discussed in the earlier literature. For the reader convenience we discuss the both short and long conformal fields.

\medskip
\noindent {\bf spin-1 short conformal field in $R^{d-1,1}$}. Field content we use for the light-cone gauge description of spin-1 short conformal field in $R^{d-1,1}$ involves vector field $\phi^i$ and scalar field $\phi$ of the $so(d-2)$ algebra. These fields can be collected into the following ket-vector:%
\footnote{ To build ket-vectors, we use oscillators with commutation relations $[\alphab^i,\alpha^j]=\delta^{ij}$,  $[\alphab^z,\alpha^z]=1$,  $[\zetab,\zeta]=1$ and hermitian conjugation rules, $\alphab^i = \alpha^{i\dagger}$, $\alphab^z = \alpha^{z\dagger}$, $\zetab = \zeta^\dagger$. The vacuum is defined as $\alphab^i|0\rangle=0$, $\alphab^z|0\rangle=0$, $\zetab|0\rangle=0$.
}
\be
\label{29062015-man-01} |\phi\rangle  =  \bigl( \phi^i \alpha^i  + \phi \alpha^z\bigr) |0\rangle\,.
\ee
Realization of the operator $\nu$ on space of ket-vector \rf{29062015-man-01} takes the form
\be \label{29062015-man-01a}
\nu = \kappa - N_z\,, \qquad N_z=\alpha^z\alphab^z, \qquad  \kappa = \frac{d-2}{2}\,, \qquad d-\hbox{ even }\,, \qquad d \geq 4\,.\qquad
\ee
Plugging \rf{29062015-man-01} and \rf{29062015-man-01a} into \rf{27062015-man-04a}, we get Lagrangian for spin-1 short conformal field
\be \label{29062015-man-01b}
\LL  =    \half \phi^i \Box^\kappa \phi^i + \half \phi \Box^{\kappa - 1} \phi \,, \qquad  \kappa = \frac{d-2}{2}\,.
\ee
The number of propagating D.o.F described by the Lagrangian \rf{29062015-man-01b} is given by
\be
\nbf =  \half d(d-3).
\ee

\noindent {\bf spin-1 long conformal field in $R^{d-1,1}$}.  Field content we use for the light-cone gauge description of spin-1 long conformal field in $R^{d-1,1}$ involves one vector fields $\phi^i$ and two scalar fields $\phi_{-1}$, $\phi_1$ of the $so(d-2)$ algebra,
\beq
& \phi^i \ \ &
\nonumber\\[-9pt]
\label{30062015-man-01} &&
\\[-9pt]
& \phi_{-1}\qquad \quad \ \phi_1  &
\nonumber
\eeq
We collect fields \rf{30062015-man-01} into the ket-vector defined by
\be
\label{29062015-man-02}  |\phi\rangle   = \bigl( \phi^i \alpha^i  + \phi_{-1} \alpha^z + \phi_1  \zeta \bigr) |0\rangle\,.
\ee
Realization of the operator $\nu$ on space of ket-vector \rf{29062015-man-02} takes the form
\be \label{29062015-man-02a}
\nu = \kappa_\intrm + N_\zeta - N_z\,, \qquad \kappa_\intrm > \frac{d-2}{2}\,, \qquad \kappa_\intrm - \hbox{ integer }\,, \qquad d \geq 4\,,
\ee
where $N_z=\alpha^z\alphab^z$, $N_\zeta=\zeta\zetab$. Plugging \rf{29062015-man-02} and \rf{29062015-man-02a} into \rf{27062015-man-04a}, we get Lagrangian for spin-1 long conformal field,

\beq
\label{29062015-man-03} \LL_{\kappa_\intrm} &  = &   \half \phi^i \Box^{\kappa_\intrm} \phi^i  +  \half \phi_{-1} \Box^{\kappa_\intrm-1} \phi_{-1} + \half \phi_1 \Box^{\kappa_\intrm+1} \phi_1 \,,
\\
&& \kappa_\intrm = [\frac{d}{2}] - 1 + N\,, \qquad N = 1,2,\ldots
\eeq
Using the notation $\nbf$ for the number of propagating D.o.F described by the Lagrangian \rf{29062015-man-03} we note the relation
\be \label{29062015-man-03a}
\nbf =  \kappa_\intrm n_1^{so(d)}\,,  \qquad n_1^{so(d)} = d,
\ee
where $n_1^{so(d)}$ in \rf{29062015-man-03a} stands for dimension of the spin-1 irreps of the $so(d)$ algebra.

\noindent {\bf spin-2 short conformal field in $R^{d-1,1}$}.  Field content we use for the light-cone gauge description of spin-2 short conformal field in $R^{d-1,1}$ involves traceless tensor field $\phi^{ij}$,   vector field $\phi^i$ and scalar field $\phi$. All these fields transform as irreps of the $so(d-2)$ algebra. We introduce ket-vector by the relation

\be \label{29062015-man-04}
|\phi\rangle  = \Bigl( \half \phi^{ij} \alpha^i \alpha^j +
\phi^i \alpha^i \alpha^z + \frac{1}{\sqrt{2}} \phi\alpha^z
\alpha^z \Bigr) |0\rangle\,.
\ee
Realization of the operator $\nu$ on space of ket-vector \rf{29062015-man-04} takes the form
\be   \label{29062015-man-04a}
\nu = \kappa - N_z\,, \qquad \kappa = \frac{d}{2}\,, \qquad d-\hbox{ even }\,, \qquad d \geq 4\,.
\ee
Plugging \rf{29062015-man-04} and \rf{29062015-man-04a} into \rf{27062015-man-04a}, we get Lagrangian for spin-2 short conformal field

\be  \label{29062015-man-05}
\LL  =  \frac{1}{4} \phi^{ij} \Box^\kappa \phi^{ij} +   \half \phi^i \Box^{\kappa-1} \phi^i
+  \half \phi \Box^{\kappa-2} \phi\,,
\ee
where $\kappa$ is given in \rf{29062015-man-04a}. Note that number of propagating D.o.F described by the Lagrangian \rf{29062015-man-05} is given by
\be \label{29062015-man-06}
\nbf = \frac{1}{4} (d-3)d(d+2).
\ee

\bigskip

\noindent {\bf spin-2 long conformal field in $R^{d-1,1}$}.  Field content we use for the light-cone gauge description of spin-2 long conformal field in $R^{d-1,1}$ involves one traceless rank-2 tensor field, two vector fields, and three scalar fields
\beq
& \phi^{ij} \ \ &
\nonumber\\[5pt]
\label{29062015-man-07} &\! \phi_{-1}^i\qquad \ \phi_1^i  &
\\[5pt]
& \phi_{-2} \qquad \phi_0  \ \qquad \phi_2 &
\nonumber
\eeq
All fields in \rf{29062015-man-07} transform as irreps of the $so(d-2)$ algebra. We collect fields \rf{29062015-man-07} into ket-vector defined by
\be \label{29062015-man-08}
|\phi\rangle  = \Bigl( \half \phi^{ij} \alpha^i \alpha^j +
\phi_{-1}^i \alpha^i \alpha^z  + \phi_1^i \alpha^i \zeta + \frac{1}{\sqrt{2}} \phi_{-2} \alpha^z \alpha^z+
\phi_0 \alpha^z\zeta  + \frac{1}{\sqrt{2}} \phi_2 \zeta^2  \Bigr) |0\rangle\,.
\ee
Realization of the operator $\nu$ on space of ket-vector \rf{29062015-man-08} takes the form
\be \label{29062015-man-09}
\nu = \kappa_\intrm + N_\zeta - N_z\,, \qquad \kappa_\intrm > \frac{d}{2}\,, \qquad \kappa_\intrm - \hbox{ integer }\,, \qquad d \geq 4\,.
\ee
Plugging \rf{29062015-man-08} and \rf{29062015-man-09} into \rf{27062015-man-04a}, we get Lagrangian for spin-2 long conformal field
\beq
\label{29062015-man-10} \LL_{\kappa_\intrm}  & = &  \frac{1}{4} \phi^{ij} \Box^{\kappa_\intrm} \phi^{ij}
\nonumber\\
& + & \half \phi_{-1}^i \Box^{\kappa_\intrm - 1} \phi_{-1}^i + \half \phi_1^i \Box^{\kappa_\intrm + 1} \phi_1^i \,,
\nonumber\\
& + & \half \phi_{-2} \Box^{\kappa_\intrm - 2} \phi_{-2} + \half \phi_0 \Box^{\kappa_\intrm} \phi_0 + \half \phi_2 \Box^{\kappa_\intrm + 2} \phi_2 \,,
\\
&& \kappa_\intrm = [\frac{d}{2}] + N\,, \qquad N = 1,2,\ldots
\eeq
Using the notation $\nbf$ for the number of propagating D.o.F described by the Lagrangian \rf{29062015-man-10} we note the relation
\be \label{29062015-man-10a}
\nbf = \kappa_\intrm  n_2^{{so(d)}}\,, \qquad    n_2^{{so(d)}} = \half (d-1)(d+2)\,,
\ee
where $n_2^{so(d)}$ in \rf{29062015-man-10a} stands for dimension of the totally symmetric spin-2 irrep of $so(d)$ algebra.

\noindent {\bf spin-$s$ totally symmetric short conformal field in $R^{d-1,1}$}.  Field content we use for the light-cone gauge description of spin-$s$ totally symmetric short conformal field in $R^{d-1,1}$ involves scalar, vector, and traceless tensor fields of the $so(d-2)$ algebra, $\phi^{i_1\ldots i_{s'}}$, $s'=0,1,\ldots,s$.   We collect these fields  into ket-vector defined by
\be  \label{29062015-man-11}
\phik = \sum\limits_{s'=0}^s \frac{\alpha_z^{s-s'} \alpha^{i_1} \ldots
\alpha^{i_{s'}}}{s'!\sqrt{(s - s')!}} \, \phi^{i_1\ldots i_{s'}} |0\rangle\,.
\ee
Realization of the operator $\nu$ on space of ket-vector \rf{29062015-man-11} takes the form
\be \label{29062015-man-12}
\nu = \kappa - N_z\,, \qquad N_z =\alpha^z \alphab^z, \qquad \kappa = s+ \frac{d-4}{2}\,, \qquad d-\hbox{ even }\,, \qquad d \geq 4\,.
\ee
Plugging \rf{29062015-man-11} and \rf{29062015-man-12} into \rf{27062015-man-04a}, we get Lagrangian for spin-$s$ short conformal field
\be \label{29062015-man-12-ext}
\LL  = \sum_{s'=0}^s \frac{1}{2 s'!} \phi^{i_1\ldots i_{s'}} \Box^{\nu_{s'}} \phi^{i_1\ldots
i_{s'}}\,, \qquad\quad  \nu_{s'} \equiv s' + \frac{d-4}{2}\,.
\ee
Note that number of propagating D.o.F described by the Lagrangian \rf{29062015-man-12-ext} is given by
\be \label{29062015-man-12a}
\nbf = \half (d-3)(2s+d-2)(2s+d-4) \frac{(s+d-4)!}{s!(d-2)!} \,.
\ee

\noindent {\bf spin-$s$ totally symmetric long conformal field in $R^{d-1,1}$}.  Field content we use for the light-cone gauge description of spin-$s$ totally symmetric long conformal field in $R^{d-1,1}$ involves scalar, vector and traceless tensor fields of the $so(d-2)$ algebra, $\phi_\lambda^{i_1\ldots i_{s'}}$, $s'=0,1,\ldots,s$, $\lambda \in [s-s']_2$. Here and below the notation $\lambda \in [n]_2$
implies that $\lambda =-n,-n+2,\ldots, n-2,n$. We collect the fields into ket-vector defined by
\be \label{29062015-man-14}
\phik = \sum\limits_{s'=0}^s\,\,\sum\limits_{\lambda\in [s-s']_2}
\frac{\zeta_{\phantom{z}}^{\frac{s-s' + \lambda}{2}}
\alpha_z^{\frac{s-s'- \lambda}{2}}\alpha^{i_1}\ldots
\alpha^{i_{s'}}}{s'!\sqrt{(\frac{s-s'+ \lambda}{2})! (\frac{s-s'- \lambda}{2})!}}
\, \phi_\lambda^{i_1\ldots i_{s'}} |0\rangle\,.
\ee
Realization of the operator $\nu$ on space of ket-vector \rf{29062015-man-14} takes the form
\be \label{29062015-man-15}
\nu = \kappa_\intrm + N_\zeta - N_z\,, \qquad \kappa_\intrm > s+ \frac{d-4}{2}\,, \qquad \kappa_\intrm - \hbox{ integer}\,, \qquad d \geq 4\,,
\ee
where $N_\zeta=\zeta\zetab$, $N_z = \alpha^z\alphab^z$. Plugging \rf{29062015-man-14} and \rf{29062015-man-15} into \rf{27062015-man-04a}, we get Lagrangian for spin-$s$ long conformal field
\beq
\label{01072015-man-01} \LL  & = &  \sum\limits_{s'=0}^s\,\,\sum\limits_{\lambda\in [s-s']_2} \frac{1}{2 s'!} \phi_\lambda^{i_1\ldots i_{s'}} \Box^{\kappa_\intrm + \lambda} \phi_\lambda^{i_1\ldots
i_{s'}}\,,
\\
&& \kappa_\intrm = s-2 + [\frac{d}{2}] + N\,, \qquad N = 1,2,\ldots
\eeq
Using the notation $\nbf$ for the number of propagating D.o.F described by the Lagrangian \rf{01072015-man-01} we note the relation
\be \label{01072015-man-02}
\nbf = \kappa_\intrm n_s^{so(d)}\,,  \qquad n_s^{{so(d)}} = (2s+d-2)\frac{(s+d-3)!}{(d-2)!s!}\,,
\ee
where $n_s^{so(d)}$ in \rf{01072015-man-02} stands for dimension of the totally symmetric spin-$s$ irrep of $so(d)$ algebra. As is well known, the $n_s^{so(d)}$ describes a number of propagating D.o.F of totally symmetric massive spin-$s$ field in $(d+1)$ dimensions. Thus we find the following rule. {\it Number of propagating D.o.F of the totally symmetric spin-$s$ long conformal field in $d$ dimensions is equal to $\kappa_\intrm$ times the number of propagating D.o.F for totally symmetric spin-$s$ massive field in $(d+1)$ dimensions.}

For the case of short conformal fields, the expression for $\nbf$ given in \rf{29062015-man-12a} agrees with expression for $\nbf$ when $d=4$ in Ref.\cite{Fradkin:1985am} and $d\geq 4$ in Ref.\cite{Metsaev:2007rw}.%
\footnote{ For $d\geq 4$, derivation of $\nbf$ \rf{29062015-man-12a} based on the computation of partition function of conformal field may be found in Ref.\cite{Tseytlin:2013fca}.}
To our knowledge, for the case of long conformal fields, expression for $\nbf$ given in \rf{01072015-man-02} has not been discussed in the earlier literature.

\newsection{ Antisymmetric (one-column) fields}\label{s6}

In the framework of Lorentz covariant approach, Lagrangian formulation of antisymmetric massless and massive fields in AdS is well known. Using such formulation, the AdS/CFT for totally antisymmetric AdS fields and related boundary shadows was studied in Ref.\cite{l'Yi:1998pi}. To our knowledge, the AdS/CFT for totally antisymmetric AdS fields and related boundary currents has not been considered in the literature.

In this section, we discuss light-cone gauge formulation of massless and massive antisymmetric AdS fields. In the framework of our approach, equations of motion turn out to be decoupled and this considerably simplifies the study of AdS/CFT. Also, we recall that finding a solution to the defining equations for the spin operators \rf{04032015-00}-\rf{04032015-05} leads immediately to the complete light-cone gauge description of AdS field theory and boundary currents, shadows, and conformal fields. This is to say that using the antisymmetric fields we demonstrate how our general approach works.

\subsection{ Antisymmetric (one-column) massive fields in $AdS_{d+1}$ and
long currents, shadows, and conformal fields in $R^{d-1,1}$ }

{\bf Massive field in $AdS_{d+1}$}. Antisymmetric rank-$s$ massive field in $AdS_{d+1}$ is associated with unitary irreps $D(E_0,\hbf)$ of the $so(d,2)$ algebra, where $E_0$, $\hbf$ are given by
\beq
\label{12072015-man-01} && \hbf = (\underbrace{1,1,\cdots ,1}_{s \ \scriptstyle{ {\rm times} }}, \underbrace{0,0,\cdots,0}_{r - s \ \scriptstyle{ {\rm times} }})\,, \qquad r \equiv [ \frac{d}{2}]\,, \qquad
\left\{\begin{array}{l}
1 \leq s \leq r\,, \hspace{0.8cm} \hbox{ for odd } d
\\[5pt]
1 \leq s \leq r-1\,, \ \hbox{ for even } d
\end{array}\right. \qquad
\\
\label{12072015-man-02} && E_0 > d - s \,.
\eeq
Restriction on $E_0$ \rf{12072015-man-02} is the standard unitarity constraint \rf{05072015-man-04} represented in terms of the label $s$. For even $d$ and $s=d/2$, we deal with a sum of self-dual and anti self-dual massive fields associated with the respective unitary irreps $D(E_0,\hbf_+)$ and $D(E_0,\hbf_-)$, where
\beq
\label{12072015-man-01a} && \hbf_\pm = (\underbrace{1,1,\cdots ,1}_{r-1 \ \scriptstyle{ {\rm times} }}, \pm 1)\,, \qquad s=r \,, \qquad r  \equiv \frac{d}{2}\,, \qquad \hbox{ for even } d\,,
\\
&& E_0 > \frac{d}{2} \,.
\eeq

\noindent {\bf Field content of massive field}. In order to describe field content entering dynamics of antisymmetric massive field it is convenient to introduce anti-commuting oscillators $\alpha^i$, $\alpha^z$, $\zeta$. The oscillators $\alpha^i$, $i=1,\ldots,d-2$, transform as vector of the $so(d-2)$ algebra, while the oscillators $\alpha^z$ and $\zeta$ transform as scalars of the $so(d-2)$ algebra.
Anti-commutation relations for the oscillators, the vacuum $|0\rangle$, and hermitian conjugation rules, are defined as
\beq
&& \{\alphab^i,\alpha^j\} = \delta^{ij}\,, \qquad \{\alphab^z,\alpha^z\} = 1\,, \qquad \{\zetab,\zeta\} = 1 \,,
\\
&& \alphab^i |0\rangle = 0 \,, \hspace{1.7cm}  \alphab^z|0\rangle = 0\,, \hspace{1.4cm} \zetab |0\rangle =  0\,,
\\
&& \alpha^{i\dagger} = \alphab^i\,, \hspace{1.9cm} \alpha^{z\dagger} = \alphab^z\,,
\hspace{1.6cm} \zeta^\dagger = \zetab\,.
\eeq
All the remaining anti-commutators for the oscillators are equal to zero.
Using such oscillators we introduce the following
ket-vector to discuss the antisymmetric rank-$s$ massive fields:
\be \label{25052015-man-01}
\phik = \phi(\alpha^i,\alpha^z,\zeta)|0\rangle\,,
\ee
where the ket-vector, by definition, satisfies the following constraint
\be
( N_\alpha + N_{\alpha^z} + N_\zeta -s)\phik  = 0\,, \qquad N_\alpha \equiv \alpha^i\alphab^i\,, \quad \qquad N_z\equiv \alpha^z\alphab^z\,, \quad N_\zeta\equiv \zeta\zetab\,.
\ee
Finite number of ordinary light-cone gauge fields depending of space-time coordinates $x^a,z$ are obtained by expanding ket-vector \rf{25052015-man-01} into
the oscillators $\alpha^i$, $\alpha^z$, $\zeta$,
\beq
\label{20052015-man-01} \phik &  = &|\phi_1\rangle + |\phi_0\rangle +  |\phi_{-1}\rangle\,,
\\[3pt]
\label{20052015-man-02} && |\phi_1\rangle = \zeta |\phi_{1}^{s-1}\rangle\,,
\\[3pt]
\label{20052015-man-03} && |\phi_0\rangle = |\phi_0^s\rangle  + \zeta \alpha^z|\phi_0^{s-2}\rangle\,,
\\[3pt]
\label{20052015-man-05} && |\phi_{-1}\rangle  = \alpha^z |\phi_{-1}^{s-1}\rangle\,,
\\[3pt]
\label{20052015-man-06} && |\phi_\lambda^\Ksm\rangle  = \frac{1}{K!}\alpha^{i_1}\ldots \alpha^{i_\Ksm} \phi_\lambda^{i_1\ldots i_\Ksm} |0\rangle\,,
\eeq
where field $\phi_\lambda^{i_1\ldots i_\Ksm}$ in \rf{20052015-man-06} is antisymmetric rank-$K$ tensor field of the $so(d-2)$ algebra. Thus we see that ket-vector $\phik$ \rf{20052015-man-01} is expanded into finite set of antisymmetric tensor fields of the $so(d-2)$ algebra.%
\footnote{ In this paper, we use vector-like oscillators to build ket-vectors $\phik$. In recent literature, the use of spinor-like and vector like oscillators entering ket-vectors of massive fields may be found in the respective Refs.\cite{Metsaev:2014sfa,deAzcarraga:2014hda} and Ref.\cite{Buchbinder:2015apa}.
}
To label ket-vectors \rf{20052015-man-02}-\rf{20052015-man-05} we use the subscript $\lambda$ which is eigenvalue of the operator $M^z$ (see \rf{20052015-man-07}).
For the illustration purposes, we use the shortcut $\phi_\lambda^\Ksm$ for rank-$K$ antisymmetric tensor field $\phi_\lambda^{i_1\ldots i_\Ksm}$ to represent
the tensor fields entering ket-vector in \rf{20052015-man-01} as
\beq
&& \hspace{1.8cm} \phi_0^{s}
\nonumber\\[5pt]
&& \phi_{-1}^{s-1} \hspace{3cm} \phi_1^{s-1}
\\[5pt]
&& \hspace{1.8cm} \phi_0^{s-2}
\nonumber
\eeq

\noindent {\bf Lagrangian of massive field}. From \rf{18052015-man-03}, \rf{18052015-man-06}, we see that the Lagrangian is determined by the operator $\nu$. We find the following realization of the operator $\nu$ on space of ket-vector \rf{20052015-man-01}:
\be \label{20052015-man-07}
\nu = \kappa + M^z, \qquad \kappa = E_0 - \frac{d}{2}\,, \qquad M^z = N_\zeta - N_z\,, \qquad  N_\zeta\equiv \zeta\zetab\,, \qquad N_z\equiv \alpha^z\alphab^z\,.
\ee
Plugging then ket-vector \rf{20052015-man-01} into \rf{18052015-man-03}, we see that Lagrangian takes the form
\be
\LL  =  \sum_{\lambda=-1,0,1}\langle \phi_\lambda | \left( \Box + \partial_z^2 - \frac{1}{z^2}\bigl( \nu_\lambda^2 -\frac{1}{4}\bigr)\right) |\phi_\lambda \rangle\,, \qquad \nu_\lambda \equiv \kappa + \lambda\,,
\ee
where ket-vectors $|\phi_\lambda\rangle$, $\lambda=0,\pm1$, are defined in \rf{20052015-man-02}-\rf{20052015-man-05}.

According to the general setup in Sec.\ref{section-02}, to complete the light-cone gauge description of the antisymmetric field we should provide a realization of the operators $M^{ij}$, $W^i$, $\Wb^i$, $B$ on space of ket-vector \rf{20052015-man-01}. The realization of these operators on space of ket-vector $\phik$ \rf{20052015-man-01} is given by
\beq
\label{01072015-man-03}  M^{ij} & = &  \alpha^i \alphab^j - \alpha^j\alphab^i\,,
\\
\label{07032015-01} W^i & = &  X \alphab^i + \alpha^i \Yb\,,
\\
\label{07032015-02} \Wb^i & = &  Y \alphab^i + \alpha^i \Xb\,,
\\
\label{07032015-03} && X = \zeta f_\zeta \,, \qquad \Xb = f_\zeta \zetab\,,  \qquad \ \ f_\zeta = \Bigl(\frac{2\kappa - d + 2s}{4\kappa}\Bigr)^{1/2}\,,
\\
\label{07032015-04} && Y = \alpha^z f_z \,, \qquad \Yb = f_z \alphab^z\,, \qquad f_z = \Bigl(\frac{2\kappa + d-2s}{4\kappa}\Bigr)^{1/2}\,,
\\
\label{01072015-man-04} B & = &  \kappa M^z - s  - \half (d-2-2s) (N_\zeta + N_z) - 2 N_\zeta N_z \,.
\eeq
We note that the realization for the spin operator of the $so(d-2)$ algebra $M^{ij}$ given in \rf{01072015-man-03} is well known, while the realization of the operators $\nu$, $W^i$, $\Wb^i$ in \rf{20052015-man-07}, \rf{07032015-01}, \rf{07032015-02} is found by solving the defining equations given in \rf{04032015-01}-\rf{04032015-05}. Expression for the operator $B$ in \rf{01072015-man-04} can be obtained by using relation \rf{18052015-man-08a} and the eigenvalue of the second order Casimir operator of the $so(d)$ algebra for antisymmetric rank-$s$ irrep,
\be
\langle \CC_{so(d)}\rangle = s (d- s) \,.
\ee

\noindent {\bf Long currents and shadows in $R^{d-1,1}$ }. Antisymmetric rank-$s$ long current in $R^{d-1,1}$ is associated with the unitary representation $D(\Delta_\cur,\hbf)$ of the $so(d,2)$ algebra, where $\Delta_\cur = E_0$, and $E_0$, $\hbf$ are given in \rf{12072015-man-01},\rf{12072015-man-02}.
Accordingly, antisymmetric rank-$s$ long shadow in $R^{d-1,1}$ is associated with non-unitary  representation of the $so(d,2)$ algebra with labels $\Delta_\sh= d-E_0$, $\hbf$, where $E_0$, $\hbf$ are given in \rf{12072015-man-01},\rf{12072015-man-02}. For $s=d/2$, $d$-even, we deal with a sum of self-dual and anti self-dual currents (shadows) having $\hbf_\pm$ as in   \rf{12072015-man-01a}.

For long currents and shadows in $R^{d-1,1}$, ket-vectors $|\phi_\cur\rangle$, $|\phi_\sh\rangle$ and the operators $\nu$, $W^i$, $\Wb^i$, $M^{ij}$ take the same form as the ones for massive field in $AdS_{d+1}$ (see \rf{20052015-man-01}, \rf{01072015-man-03}-\rf{01072015-man-04}). Plugging the ket-vectors $|\phi_\cur\rangle$, $|\phi_\sh\rangle$ and the operators  $\nu$, $W^i$, $\Wb^i$, $M^{ij}$ into expressions for 2-point functions and generators of $so(d,2)$ algebra presented in Sec.\ref{section-03}, we get the complete light-cone gauge description of long currents and shadows.
For example, plugging the ket-vector $|\phi_\sh\rangle$ into expression for 2-point function in \rf{19052015-man-04}, we get
\beq
\LL_{12}^{\rm sh-sh}  & = &  \sum_{\lambda=-1,0,1} \LL_{12,\lambda}^{\rm sh-sh}\,, \qquad
\LL_{12,\lambda}^{\rm sh-sh} \equiv \half \langle\phi_{\sh,\lambda}(x_1)|
\frac{f_{\nu_\lambda}^\sh}{ |x_{12}|^{2\nu_\lambda + d }} |\phi_{\sh,\lambda} (x_2)\rangle \,,
\\
&&  f_{\nu_\lambda}^\sh \equiv \frac{4^{\nu_\lambda} \Gamma(\nu_{\lambda} + \frac{d}{2})\Gamma(\nu_{\lambda} + 1)}{4^\kappa
\Gamma(\kappa + \frac{d}{2})\Gamma(\kappa + 1)} \,, \qquad  \nu_\lambda \equiv \kappa + \lambda\,, \qquad \kappa = E_0 - \frac{d}{2}\,,\qquad
\eeq
where the ket-vectors $|\phi_{\sh,\lambda}\rangle$ take the same form as in \rf{20052015-man-02}-\rf{20052015-man-06}.

\noindent {\bf Long conformal field in $R^{d-1,1}$ }. Antisymmetric rank-$s$ long conformal field in $R^{d-1,1}$ is associated with non-unitary representation of the $so(d,2)$ algebra with labels $\Delta$, $\hbf$, where $\hbf$ is given in \rf{12072015-man-01}, while the conformal dimension $\Delta$ is given by
\be \label{01072015-man-04a}
\Delta = d - E_{0,\longrm}\,,  \qquad E_{0,\longrm} = \kappa_\intrm + \frac{d}{2}\,, \qquad \kappa_\intrm > \frac{d}{2} - s \qquad \kappa_\intrm - \hbox{ integer}.\qquad
\ee
For $s=d/2$, $d$-even, we deal with a sum of self-dual and anti self-dual conformal fields having $\hbf_\pm$ as in   \rf{12072015-man-01a}.

Ket-vector of long conformal field $|\phi\rangle$ in $R^{d-1,1}$ takes the same form as the one for massive field in $AdS_{d+1}$ \rf{20052015-man-01}. Relations in  \rf{01072015-man-03}-\rf{01072015-man-04} with the substitution $\kappa \rightarrow \kappa_\intrm$  provide solution for the operators $\nu$, $W^i$, $\Wb^i$, $M^{ij}$ entering the light-cone gauge description of the long conformal field. Plugging such operators  $\nu$, $W^i$, $\Wb^i$, $M^{ij}$  into \rf{19052015-man-12}-\rf{19052015-man-22}, \rf{19052015-man-27}-\rf{19052015-man-30} provides the realization of the conformal $so(d,2)$ symmetries on space of the ket-vector $\phik$, while
plugging the ket-vector $|\phi\rangle$ into \rf{27062015-man-04a}, we get Lagrangian for the long conformal field
\be \label{12072015-man-03}
\LL = \half  \sum_{\lambda=-1,0,1} \langle \phi_\lambda| \Box^{\nu_{\intrm,\lambda} } |\phi_\lambda\rangle\,, \qquad \nu_{\intrm,\lambda} = \kappa_\intrm + \lambda\,,
\ee
where the ket-vectors $|\phi_\lambda\rangle$ take the same form as in  \rf{20052015-man-02}-\rf{20052015-man-06}. In terms of tensor fields \rf{20052015-man-06}, Lagrangian \rf{12072015-man-03} takes the form
\beq
\label{12072015-man-04} \LL & = &  \frac{1}{2s!} \phi_0^{i_1\ldots i_s} \Box^{\kappa_\intrm }\phi_0^{i_1\ldots i_s}
+ \frac{1}{2(s-2)!} \phi_0^{i_1\ldots i_{s-2}} \Box^{\kappa_\intrm }\phi_0^{i_1\ldots i_{s-2}}
\nonumber\\
& + & \frac{1}{2(s-1)!} \phi_{-1}^{i_1\ldots i_{s-1}} \Box^{\kappa_\intrm-1}\phi_{-1}^{i_1\ldots i_{s-1}} + \frac{1}{2(s-1)!} \phi_1^{i_1\ldots i_{s-1}} \Box^{\kappa_\intrm + 1}\phi_1^{i_1\ldots i_{s-1}}\,,
\eeq
where $s$ and $\kappa_\intrm$ satisfy restrictions in \rf{12072015-man-01}, \rf{01072015-man-04a}. Using the notation $\nbf$ for the number of propagating D.o.F described by the Lagrangian \rf{12072015-man-04}, we note the relation
\be \label{12072015-man-05}
\nbf = \kappa_\intrm n_s^{so(d)}\,,  \qquad n_s^{{so(d)}} = \frac{d!}{s!(d-s)!}\,,
\ee
where $n_s^{so(d)}$ in \rf{12072015-man-05} stands for dimension of the rank-$s$ antisymmetric irrep of $so(d)$ algebra. As is well known, the $n_s^{so(d)}$ describes a number of propagating D.o.F of  rank-$s$ antisymmetric massive spin-$s$ field in $(d+1)$ dimensions.

\subsection{ Antisymmetric (one-column) massless fields in $AdS_{d+1}$ and
short currents, shadows, and conformal fields in $R^{d-1,1}$ }

{\bf Massless field in $AdS_{d+1}$}. From general relations \rf{05072015-man-03}, \rf{27062015-man-05a}, we see that in order to realize the limit of antisymmetric massless field we have to consider the limit%
\footnote{ As well known, in massless limit \rf{13072015-man-01}, unitary irreps $D(E_0,\hbf_\pm)$ in \rf{12072015-man-01a} are not associated with local AdS fields propagating in $AdS_{d+1}$. Therefore we ignore the case \rf{12072015-man-01a} when considering the massless limit. As we ignore the case $s=d/2$ for even $d$, then the constraints on $s$ in \rf{12072015-man-01} can be represented on equal footing as in \rf{13072015-man-01}.
}
\beq
\label{13072015-man-01} && E_0 \rightarrow d - s \,,   \qquad  s  \leq \frac{d-1}{2}\,.
\eeq
In the framework of Lorentz covariant approach, it is well-known that, in the massless limit, antisymmetric rank-$s$ massive AdS field is decomposed into rank-$s$ and rank-$(s-1)$ massless AdS fields. We now demonstrate how this result is obtained in the framework of light-cone gauge approach. To this end we note that the ket-vector of massive field in \rf{20052015-man-01} can be represented as
\beq
 \label{13072015-man-02} \phik &  = & |\phibf\rangle + \zeta |\varphibf\rangle\,,
\\
&& \label{13072015-man-03} |\phibf\rangle   =  |\phibf_0\rangle + |\phibf_{-1}\rangle\,, \hspace{1cm} |\phibf_0\rangle \equiv  |\phi_0^s\rangle\,, \hspace{1.6cm} |\phibf_{-1}\rangle \equiv  \alpha^z |\phi_{-1}^{s-1}\rangle\,,
\\
&& \label{13072015-man-04} |\varphibf\rangle  = |\varphibf_0\rangle + |\varphibf_{-1}\rangle\,,  \hspace{1cm} |\varphibf_{-1}\rangle  \equiv  |\phi_{1}^{s-1}\rangle\,, \hspace{1cm} |\varphibf_0\rangle  \equiv \alpha^z|\phi_0^{s-2}\rangle\,,\qquad
\eeq
where the ket-vectors $|\phi_\lambda^\Ksm\rangle$ appearing in \rf{13072015-man-03},\rf{13072015-man-04} are defined in \rf{20052015-man-06}.  It turns out that, in the massless limit, it is the ket-vectors $|\phibf\rangle$ and $|\varphibf\rangle$ that describe the respective rank-$s$ and rank-$(s-1)$ massless AdS fields.
Namely, considering the spin operators \rf{20052015-man-07}, \rf{01072015-man-03}-\rf{01072015-man-04} in the limit \rf{13072015-man-01}, we verify that ket-vectors $|\phibf\rangle$, $|\varphibf\rangle$ \rf{13072015-man-03}, \rf{13072015-man-04} are  invariant under action of the $so(d,2)$ algebra generators given in \rf{19052015-man-32}-\rf{19052015-man-41}.%
\footnote{ It is easy to see that invariance of the ket-vectors under action of $so(d,2)$ algebra generators given in \rf{19052015-man-32}-\rf{19052015-man-41} amounts to the invariance of the ket-vectors under action of the spin operators $W^i$, $\Wb^i$.
}
In other words, in limit \rf{13072015-man-01}, the ket-vector $\phik$ is decomposed into two invariant sub-spaces described by the ket-vectors $|\phibf\rangle$, $|\varphibf\rangle$. The ket-vectors $|\phibf\rangle$ and $|\varphibf\rangle$ describe the respective antisymmetric rank-$s$ and rank-$(s-1)$ irreps of $so(d-1)$ algebra. The ket-vectors $|\phibf\rangle$ and $|\varphibf\rangle$ are associated with the unitary irreps of the $so(d,2)$ algebra  with $E_0=d-s$ and $E_0=d+1-s$ respectively. This implies that, in the massless limit, the antisymmetric rank-$s$ massive field is decomposed into antisymmetric rank-$s$ and rank-$(s-1)$ massless fields described by the respective ket-vectors $|\phibf\rangle$ and $|\varphibf\rangle$ .

Plugging ket-vector \rf{13072015-man-03} into \rf{18052015-man-03}, we get Lagrangian of rank-$s$ antisymmetric massless field
\be \label{15072015-man-01}
\LL   =    \sum_{\lambda=-1,0}\langle \phibf_\lambda | \left( \Box + \partial_z^2 - \frac{1}{z^2}\bigl( \nu_\lambda^2 -\frac{1}{4}\bigr)\right) |\phibf_\lambda \rangle\,, \qquad  \nu_\lambda \equiv  \frac{d}{2} - s  + \lambda\,,\qquad s  \leq \frac{d-1}{2}\,.
\ee
To complete the light-cone gauge description of the antisymmetric massless field, we provide the realization of the operators $M^{ij}$, $W^i$, $\Wb^i$, $B$ on space of ket-vector $|\phibf\rangle$ \rf{13072015-man-03},
\beq
\label{15072015-man-02} && \nu  =  \frac{d}{2} - s  - N_z\,, \qquad  M^{ij}  =  \alpha^i \alphab^j - \alpha^j\alphab^i\,,
\\
\label{15072015-man-03} && W^i  =   \alpha^i \alphab^z\,, \hspace{1.7cm}  \Wb^i  =  \alpha^z \alphab^i \,,
\\
\label{15072015-man-04} && B  =   -s - (d-1-2s) N_z\,, \qquad N_z \equiv \alpha^z\alphab^z\,.
\eeq
We note that the spin operators in \rf{15072015-man-02}-\rf{15072015-man-04} are simply obtained by taking the limit \rf{13072015-man-01} in expressions for spin operators in \rf{01072015-man-03}-\rf{01072015-man-04}.

\noindent {\bf Short currents and shadows in $R^{d-1,1}$ }. Antisymmetric rank-$s$ short current in $R^{d-1,1}$ is related to unitary irrep $D(\Delta_\cur,\hbf)$ of the $so(d,2)$ algebra, where $\Delta_\cur = d-s$, while $\hbf$ is given in \rf{12072015-man-01} and $s\leq (d-1)/2$.
Antisymmetric rank-$s$ short shadow in $R^{d-1,1}$ is associated with representation of the $so(d,2)$ with $\Delta_\sh= s$ and $\hbf$ as in \rf{12072015-man-01}, where $s\leq (d-1)/2$.

For short currents and shadows in $R^{d-1,1}$, ket-vectors $|\phibf_\cur\rangle$, $|\phibf_\sh\rangle$ and the operators $\nu$, $W^i$, $\Wb^i$, $M^{ij}$ take the same form as the ones for massless field in $AdS_{d+1}$ (see \rf{13072015-man-03}, \rf{15072015-man-02}-\rf{15072015-man-04}). Plugging the  ket-vectors $|\phibf_\cur\rangle$, $|\phibf_\sh\rangle$ and the operators  $\nu$, $W^i$, $\Wb^i$, $M^{ij}$ into expressions for 2-point functions and generators of $so(d,2)$ algebra presented in Sec.\ref{section-03}, we get the complete light-cone gauge description of the short currents and shadows.

\noindent {\bf Short conformal field in $R^{d-1,1}$ }. Antisymmetric rank-$s$ short conformal field in $R^{d-1,1}$ is associated with representation of the $so(d,2)$ algebra with labels $\Delta$, $\hbf$, where $\hbf$ is given in \rf{12072015-man-01}, while the conformal dimension $\Delta$ is given by
\be
\Delta = s\,,   \qquad 1 \leq s \leq \frac{d}{2}-1\,, \qquad d-\hbox{ even}.
\ee

For rank-$s$ antisymmetric short conformal field in $R^{d-1,1}$, ket-vector $|\phibf\rangle$ and operators $\nu$, $W^i$, $\Wb^i$, $M^{ij}$ take the same form as the ones for massless field in $AdS_{d+1}$ (see \rf{13072015-man-03}, \rf{15072015-man-02}-\rf{15072015-man-04}). Plugging the operators  $\nu$, $W^i$, $\Wb^i$, $M^{ij}$  into \rf{19052015-man-12}-\rf{19052015-man-22}, \rf{19052015-man-27}-\rf{19052015-man-30} provides realization of the conformal $so(d,2)$ symmetries on space of the ket-vector $|\phibf\rangle$, while
plugging the ket-vector $|\phibf\rangle$ into \rf{27062015-man-04a}, we get Lagrangian for the short antisymmetric conformal field
\be \label{15072015-man-05}
\LL = \half  \sum_{\lambda=-1,0} \langle \phibf_\lambda| \Box^{\nu_{\intrm,\lambda} } |\phibf_\lambda\rangle\,, \qquad \nu_{\intrm,\lambda} =\frac{d}{2} - s + \lambda\,.
\ee
In terms of tensor fields \rf{20052015-man-06}, Lagrangian \rf{15072015-man-05} takes the form
\beq
\label{15072015-man-06} && \LL  =  \frac{1}{2s!} \phi_0^{i_1\ldots i_s} \Box^{\kappa  }\phi_0^{i_1\ldots i_s}
+  \frac{1}{2(s-1)!} \phi_{-1}^{i_1\ldots i_{s-1}} \Box^{\kappa -1}\phi_{-1}^{i_1\ldots i_{s-1}}\,,
\\
&& \kappa \equiv \frac{d}{2} - s \,, \qquad 1 \leq s \leq \frac{d}{2}-1\,, \qquad d-\hbox{ even}\,.
\eeq
Using the notation $\nbf$ for the number of propagating D.o.F described by the Lagrangian \rf{15072015-man-06} we note the relation
\be
\nbf = \frac{d(d-1-2s)(d-2)!}{2s!(d-1-s)!}\,.
\ee
For $s<(d-2)/2$, Lagrangian \rf{15072015-man-06} describes non unitary conformal fields, while, for $s=(d-2)/2$, Lagrangian \rf{15072015-man-06} descries unitary conformal fields.%
\footnote{ For $s=(d-2)/2$, the field $\phi_{-1}^{i_1\ldots i_{s-1}}$ in \rf{15072015-man-06} is non-dynamical, while the field $\phi_0^{i_1\ldots i_{s-1}}$ can be decomposed into self-dual and anti self-dual irreps of the $so(d-2)$ algebra.}

\newsection{ Mixed-symmetry (two-column) fields } \label{sec-07}

In the framework of Lorentz covariant approach, Lagrangian formulation of two-column massless and massive fields in AdS was developed in Ref.\cite{Alkalaev:2003hc} and Ref.\cite{deMedeiros:2003px} respectively. AdS/CFT for such fields and related boundary currents (shadows) has not been considered in the literature. Lorentz covariant description of two-column short conformal field can be obtained from Ref.\cite{Vasiliev:2009ck}. Long conformal fields have not been discussed in the framework of Lorentz covariant approach.

In this section, we discuss the light-cone gauge description of two-column massless and massive AdS fields. We recall that, in our approach, the complete light-cone description of {\it massless} AdS fields leads immediately to the complete light-cone description of {\it short} currents, shadows, and conformal fields, while the complete light-cone description of {\it massive} AdS fields leads immediately to the complete light-cone description of {\it long} currents, shadows, and conformal fields. We discuss AdS fields and CFT in turn.

\subsection{ Mixed-symmetry (two-column) massive fields in $AdS_{d+1}$ and
long currents, shadows, and conformal fields in $R^{d-1,1}$ }

\noindent {\bf Massive field in $AdS_{d+1}$}. We now consider mixed-symmetry (two-column) massive fields in $AdS_{d+1}$. Physical D.o.F of such fields are described by irreducible tensor fields of the $so(d)$ algebra whose $so(d)$ space tensor indices have the structure of two-column Young tableaux. We use integers $s_1$, $s_2$, $s_1\geq s_2$, to indicate of height of the first and second columns of the two-column Young tableaux. Two-column field described by such Young tableaux will be referred to as type $[s_1,s_2]$ field. The type $[s_1,s_2]$ massive field in $AdS_{d+1}$ is associated with the unitary representation $D(E_0,\hbf)$ of the $so(d,2)$ algebra, where $E_0$, $\hbf$ are given by
\beq
\label{17072015-man-01} && \hbf = (\underbrace{2,2,\cdots 2, 2}_{s_2 \ \scriptstyle{ {\rm times} }}, \underbrace{1,1,\cdots 1, 1}_{s_1-s_2 \ \scriptstyle{ {\rm times} }}, \underbrace{0,0,\cdots 0, 0}_{r - s_1 \ \scriptstyle{ {\rm times} }})
\qquad
\left\{\begin{array}{l}
1 \leq s_1 \leq r\,, \hspace{0.8cm} \hbox{ for odd } d
\\[5pt]
1 \leq s_1 \leq r-1\,, \ \hbox{ for even } d
\end{array}\right. \qquad
\\[5pt]
\label{17072015-man-02} && E_0  \geq d + 1 - s_2 \hspace{2cm} \hbox{ for } s_2\geq 1\,,
\\
\label{17072015-man-03} && E_0 \geq d - s_1 \hspace{2.8cm} \hbox{ for } s_2 = 0 \,,
\eeq

\noindent {\bf Field content of massive field with $s_1>s_2\geq 1$}. In order to describe field content entering dynamics of two-column massive field we introduce anti-commuting oscillators $\alpha_n^i$, $\alpha_n^z$, $\zeta_n$, $n=1,2$. The oscillators $\alpha_n^i$, $i=1,\ldots,d-2$, transform as vector of the $so(d-2)$ algebra, while the oscillators $\alpha_n^z$ and $\zeta_n$ transform as scalars of the $so(d-2)$ algebra. Anti-commutation relations for the oscillators, the vacuum $|0\rangle$, and hermitian conjugation rules are defined as
\beq
\label{17072015-man-03-a1} && \{\alphab_m^i,\alpha_n^j\} = \delta_{mn}\delta^{ij}\,, \qquad \{\alphab_m^z,\alpha_n^z\} = \delta_{mn}\,, \qquad \{\zetab_m,\zeta_n\} = \delta_{mn} \,,
\\
\label{17072015-man-03-a2} && \alphab_n^i |0\rangle = 0 \,, \hspace{2.3cm}  \alphab_n^z|0\rangle = 0\,, \hspace{1.8cm} \zetab_n |0\rangle =  0\,,
\\
\label{17072015-man-03-a3} && \alpha_n^{i\dagger} = \alphab_n^i\,, \hspace{2.5cm} \alpha_n^{z\dagger} = \alphab_n^z\,,
\hspace{2cm} \zeta_n^\dagger = \zetab_n\,.
\eeq
We note also that all remaining anti-commutators for the oscillators are equal to zero.
Using such oscillators we introduce the ket-vector $\phik$,
\be \label{17072015-man-04}
\phik = \phi(\alpha_1^i,\alpha_2^i,\alpha_1^z,\alpha_2^z,\zeta_1,\zeta_2)|0\rangle\,,
\ee
which, by definition, satisfies the following algebraic constraints:
\beq
\label{17072015-man-05} && ( N_{\alpha_n} + N_{\alpha_n^z} + N_{\zeta_n} - s_n)\phik  = 0\,, \qquad n =1,2\,,
\\
\label{17072015-man-06} && N_{\alpha_{12}}\phik = 0 \,,
\\
\label{17072015-man-07} && \alphab_{12}\phik = 0 \,,
\\
\label{17072015-man-08} && N_{\alpha_{mn}} \equiv \alpha_m^i\alphab_n^i\,, \qquad \alphab_{mn} \equiv \alphab_m^i\alphab_n^i\,, \qquad m \ne n\,,
\\
\label{17072015-man-09} && N_{\alpha_n} \equiv  \alpha_n^i\alphab_n^i\,, \qquad  \ N_{\alpha_n^z} \equiv  \alpha_n^z\alphab_n^z\,, \qquad  N_{\zeta_n} \equiv   \zeta_n\zetab_n\,.
\eeq
Constraints \rf{17072015-man-05} tell us that the ket-vector $\phik$ is degree-$s_n$ homogeneous polynomial in the oscillators $\alpha_n^i$, $\alpha_n^z$, $\zeta_n$. Constraints \rf{17072015-man-06} and \rf{17072015-man-07} are the respective Young symmetry constraint and tracelessness constraint.
Finite number of ordinary light-cone gauge fields depending of space-time coordinates $x^a,z$ are obtained by expanding ket-vector \rf{17072015-man-04} into
the oscillators $\alpha_n^i$, $\alpha_n^z$, $\zeta_n$,
\beq
\label{17072015-man-10} && \hspace{-1cm} \phik  =  |\phi_2\rangle  + |\phi_1\rangle  + |\phi_0\rangle +  |\phi_{-1}\rangle +  |\phi_{-2}\rangle \,,
\\
\label{17072015-man-11} && |\phi_2\rangle  =  \zeta_1 \zeta_2 |\phi_2^{s_1-1,s_2-1}\rangle \,,
\\[5pt]
\label{17072015-man-11-a1} && |\phi_1\rangle  =  \zeta_1 |\phi_1^{s_1-1,s_2}\rangle + \zeta_2 |\phi_1^{s_1,s_2-1}\rangle + \zeta_1\zeta_2 \alpha_1^z |\phi_1^{s_1-2,s_2-1}\rangle+ \zeta_1\zeta_2 \alpha_2^z |\phi_1^{s_1-1,s_2-2}\rangle\,,
\\[5pt]
\label{17072015-man-11-a2} && |\phi_0\rangle  =    |\phi_0^{s_1,s_2}\rangle + \zeta_1 \alpha_1^z |\phi_0^{s_1-2,s_2}\rangle + \zeta_1 \alpha_2^z |\phi_{0'}^{s_1-1,s_2-1}\rangle
\nonumber\\[5pt]
&& \hspace{2.6cm} + \ \zeta_2 \alpha_1^z|\phi_0^{s_1-1,s_2-1}\rangle + \zeta_2\alpha_2^z  |\phi_0^{s_1,s_2-2}\rangle+ \zeta_1\zeta_2 \alpha_1^z\alpha_2^z |\phi_0^{s_1-2,s_2-2}\rangle,\qquad
\\[5pt]
\label{17072015-man-11-a3} && |\phi_{-1}\rangle  =   \alpha_1^z |\phi_{-1}^{s_1-1,s_2}\rangle + \alpha_2^z |\phi_{-1}^{s_1,s_2-1}\rangle + \zeta_1 \alpha_1^z  \alpha_2^z |\phi_{-1}^{s_1-2,s_2-1}\rangle + \zeta_2 \alpha_1^z \alpha_2^z |\phi_{-1}^{s_1-1,s_2-2}\rangle\,,
\\[5pt]
\label{17072015-man-11-a4} && |\phi_{-2}\rangle  =  \alpha_1^z \alpha_2^z |\phi_{-2}^{s_1-1,s_2-1}\rangle
\,,
\\
\label{17072015-man-12} && |\phi_\lambda^{\Ksm_1 \Ksm_2}\rangle  = \frac{1}{K_1!}\frac{1}{K_2!}\alpha_1^{i_1}\ldots \alpha_1^{i_{\Ksm_1} } \alpha_2^{j_1}\ldots \alpha_2^{j_{\Ksm_2}} \phi_\lambda^{i_1\ldots i_{\Ksm_1},j_1\ldots j_{\Ksm_2}} |0\rangle\,,
\eeq
where field $\phi_\lambda^{i_1\ldots i_{\Ksm_1},j_1\ldots j_{\Ksm_2}}$, $K_1\geq K_2$, appearing in  \rf{17072015-man-12} is mixed-symmetry (two-column) irreducible tensor field of the $so(d-2)$ algebra. By definition, ket-vector \rf{17072015-man-12} satisfies constraints \rf{17072015-man-06},\rf{17072015-man-07}. Note that we assume the convention $|\phi_\lambda^{\Ksm_1 \Ksm_2}\rangle =  0 $ for $K_1 < K_2$. To label ket-vectors \rf{17072015-man-11}-\rf{17072015-man-11-a4} we use the subscript $\lambda$ which is eigenvalue of the operator $M^z$ (see \rf{17072015-man-14}).
For the illustration purposes, we use the shortcut $\phi_\lambda^{\Ksm_1\Ksm_2}$ for two-column tensor field $\phi_\lambda^{i_1\ldots i_{\Ksm_1},j_1\ldots j_{\Ksm_2}}$
to represent the tensor fields entering ket-vector in \rf{17072015-man-10} as
\beq
&& \hspace{4cm} \phi_0^{s_1,s_2}
\nonumber\\[10pt]
&& \phi_{-1}^{s_1-1,s_2}, \ \phi_{-1}^{s_1,s_2-1}  \hspace{3cm} \phi_1^{s_1-1,s_2}, \ \phi_1^{s_1,s_2-1}
\nonumber\\[10pt]
&& \hspace{-2cm} \phi_{-2}^{s_1-1,s_2-1} \hspace{2cm}  \phi_0^{s_1-2,s_2}, \ \phi_0^{s_1-1,s_2-1}, \ \phi_{0'}^{s_1-1,s_2-1}, \ \phi_0^{s_1,s_2-2} \hspace{2cm} \phi_2^{s_1-1,s_2-1} \qquad
\\[10pt]
&& \phi_{-1}^{s_1-2,s_2-1}, \ \phi_{-1}^{s_1-1,s_2-2}  \hspace{2.5cm}  \phi_1^{s_1-2,s_2-1}, \        \phi_1^{s_1-1,s_2-2}
\nonumber\\[10pt]
&&  \hspace{3.8cm} \phi_0^{s_1-2,s_2-2}
\nonumber
\eeq

\noindent {\bf Lagrangian of massive field}.  We note the following realization of the operator $\nu$ on space of ket-vector \rf{17072015-man-10}:
\be \label{17072015-man-14}
\nu = \kappa + M^z\,, \qquad  \kappa = E_0 - \frac{d}{2}\,, \qquad M^z = N_{\zeta_1} + N_{\zeta_2} - N_{\alpha_1^z} - N_{\alpha_2^z}\,,
\ee
where the operators $N_{\alpha_n^z}$, $N_{\zeta_n}$ are defined in \rf{17072015-man-09}.
Plugging ket-vector \rf{17072015-man-10} into \rf{18052015-man-03}, we see that Lagrangian takes the form
\be
\label{17072015-man-15} \LL  =  \sum_{\lambda=-2,-1,0,1,2}\langle \phi_\lambda | \left( \Box + \partial_z^2 - \frac{1}{z^2}\bigl( \nu_\lambda^2 -\frac{1}{4}\bigr)\right) |\phi_\lambda \rangle\,,
\qquad  \nu_\lambda \equiv \kappa + \lambda\,,
\ee
where ket-vectors $|\phi_\lambda\rangle$, $\lambda=0,\pm1,\pm2$, are defined in \rf{17072015-man-11}-\rf{17072015-man-11-a4}.
To complete the light-cone gauge description of the two-column massive field we provide a realization of the operators $M^{ij}$, $W^i$, $\Wb^i$, $B$. The realization of these operators on space of ket-vector \rf{17072015-man-10} is given by
\beq
\label{17072015-man-16} && M^{ij}  =   \sum_{n=1,2} M_n^{ij} \,, \qquad M_n^{ij} \equiv \alpha_n^i \alphab_n^j - \alpha_n^j \alphab_n^i\,,
\\
\label{17072015-man-17} && W^i = X_1 \Ab_1^i + X_2 \Ab_2^i + A_1^i \Yb_1 + A_2^i \Yb_2\,,
\\
\label{17072015-man-18} && \Wb^i = Y_1 \Ab_1^i + Y_2 \Ab_2^i + A_1^i \Xb_1 + A_2^i \Xb_2\,,
\\
\label{17072015-man-19} && \hspace{1cm} A_1^i  = \alpha_1^i - \alpha_{12} \alphab_2^i T\,,
\\
\label{17072015-man-20} &&  \hspace{1cm} A_2^i  =  \alpha_2^i + \alpha_{12} \alphab_1^i T  -  N_{\alpha_{21}} A_1^i H\,,
\\
\label{17072015-man-21} &&  \hspace{1cm} \Ab_1^i = \alphab_1^i + N_{\alpha_{21}}  \alphab_2^i H\,,
\\
\label{17072015-man-22} &&  \hspace{1cm} \Ab_2^i = \alphab_2^i\,,
\\
\label{17072015-man-23} &&  \hspace{1cm} T  \equiv (d-1-N_{\alpha_1}- N_{\alpha_2})^{-1}\,,
\\
\label{17072015-man-24} &&  \hspace{1cm} H \equiv \left( N_{\alpha_1}- N_{\alpha_2} +1 \right)^{-1}\,,
\\
\label{17072015-man-25} && B  =  \kappa M^z - s_1 - s_2   - \frac{t-h}{2} (N_{\zeta_1} + N_{\alpha_1^z}) - \frac{t+h}{2} (N_{\zeta_2} + N_{\alpha_2^z})
\nonumber\\
&& \hspace{1cm} -  2 N_{\zeta_1} N_{\alpha_1^z} - 2 N_{\zeta_2} N_{\alpha_2^z}
+ ( N_{\zeta_1} - N_{\alpha_1^z})(N_{\zeta_2} - N_{\alpha_2^z})\,,
\\
\label{17072015-man-26} && \hspace{1cm} t = d-1 - s_1 - s_2\,, \qquad h = s_1-s_2+1\,,
\eeq
where the operator $M^z$ appearing in \rf{17072015-man-25} is defined in \rf{17072015-man-14}. The spin operator of the $so(d-2)$ algebra $M^{ij}$ given in \rf{17072015-man-16} is well known, while the realization of the operators $\nu$, $W^i$, $\Wb^i$ in \rf{17072015-man-17}, \rf{17072015-man-18} is found by solving the defining equations given in \rf{04032015-01}-\rf{04032015-05}. Explicit expressions for the operators $W^i$, $\Wb^i$ may be found in Appendix C. Expression for the operator $B$ in \rf{17072015-man-25} can be obtained by using \rf{18052015-man-08a} and well know expression for eigenvalue of the second order Casimir operator of the $so(d)$ algebra for two-column  irrep,
\be
\langle \CC_{so(d)}\rangle = s_1(d - s_1) + s_2(d +2 - s_2)\,.
\ee

\noindent {\bf Long currents and shadows in $R^{d-1,1}$ }. Type $[s_1,s_2]$ long current in $R^{d-1,1}$ is associated with the representation $D(\Delta_\cur,\hbf)$ of the $so(d,2)$ algebra, where $\Delta_\cur = E_0$,  and $E_0$, $\hbf$ are given in \rf{17072015-man-01},\rf{17072015-man-02}.
Type $[s_1,s_2]$ long shadow in $R^{d-1,1}$ is associated with non-unitary  representation of the $so(d,2)$ algebra with labels $\Delta_\sh= d-E_0$, $\hbf$, where $E_0$, $\hbf$ are given in \rf{17072015-man-01},\rf{17072015-man-02}.

For two-column long currents and shadows in $R^{d-1,1}$, ket-vectors $|\phi_\cur\rangle$, $|\phi_\sh\rangle$ and the operators $\nu$, $W^i$, $\Wb^i$, $M^{ij}$ take the same form as the ones for two-column massive field in $AdS_{d+1}$ (see \rf{17072015-man-14}, \rf{17072015-man-16}-\rf{17072015-man-25}). Plugging the ket-vectors $|\phi_\cur\rangle$, $|\phi_\sh\rangle$ and the operators  $\nu$, $W^i$, $\Wb^i$, $M^{ij}$ into expressions for 2-point functions and generators of the $so(d,2)$ algebra presented in Sec.\ref{section-03}, we get the complete light-cone gauge description of the two-column long currents and shadows.

\noindent {\bf Long conformal field in $R^{d-1,1}$ }. Type $[s_1,s_2]$ long conformal field in $R^{d-1,1}$ is associated with non-unitary representation of the $so(d,2)$ algebra with labels $\Delta$, $\hbf$, where $\hbf$ is given in \rf{17072015-man-01}, while the conformal dimension $\Delta$ is given by
\be \label{17072015-man-27}
\Delta = d - E_{0,\longrm}\,,  \qquad E_{0,\longrm} = \kappa_\intrm + \frac{d}{2}\,, \qquad \kappa_\intrm > \frac{d}{2} + 1 - s_2\,, \qquad \kappa_\intrm - \hbox{ integer }.\qquad
\ee
Ket-vector of type $[s_1,s_2]$ long conformal field $|\phi\rangle$ in $R^{d-1,1}$ takes the same form as the one for massive field in $AdS_{d+1}$ \rf{17072015-man-10}. Relations in \rf{17072015-man-16}-\rf{17072015-man-25} with substitution $\kappa \rightarrow \kappa_\intrm$  provide solution for the operators $\nu$, $W^i$, $\Wb^i$, $M^{ij}$ entering the light-cone gauge description of the long conformal field. Plugging such operators  $\nu$, $W^i$, $\Wb^i$, $M^{ij}$  into \rf{19052015-man-12}-\rf{19052015-man-22}, \rf{19052015-man-27}-\rf{19052015-man-30}, we get the realization of the conformal $so(d,2)$ symmetries on space of ket-vector $\phik$, while
plugging the ket-vector $|\phi\rangle$ into \rf{27062015-man-04a}, we get Lagrangian for the type $[s_1,s_2]$ long conformal field,
\be   \label{17072015-man-28}
\LL = \half  \sum_{\lambda=-2,-1,0,1,2} \langle \phi_\lambda| \Box^{\nu_{\intrm,\lambda} } |\phi_\lambda\rangle\,, \qquad \nu_{\intrm,\lambda} = \kappa_\intrm + \lambda\,,
\ee
where the ket-vectors $|\phi_\lambda\rangle$ take the same form as in  \rf{17072015-man-11}-\rf{17072015-man-11-a4}.

\subsection{ Mixed-symmetry (two-column) massless fields in $AdS_{d+1}$ and
short currents, shadows, and conformal fields in $R^{d-1,1}$ }

{\bf Massless field in $AdS_{d+1}$}. From \rf{05072015-man-03}, \rf{27062015-man-05a}, we see that in order to realize the limit of two-column massless field we have to consider the limit
\be  \label{19072015-man-02}
E_0  \rightarrow  d + 1 - s_2\,.
\ee
We note that the ket-vectors of massive field in \rf{17072015-man-10} can be presented in terms of two new ket-vectors $\phibfk$, $\varphibfk$ in the following way
\beq
\label{19072015-man-02a-ext} && \hspace{-1cm} \phik  =  \phibfk + \zeta_2 \varphibfk
\\
\label{19072015-man-02a} && \phibfk  =   |\phibf_1\rangle  + |\phibf_0\rangle +  |\phibf_{-1}\rangle +  |\phibf_{-2}\rangle \,,
\\
\label{19072015-man-02b} && \varphibfk  =   |\varphibf_1\rangle  + |\varphibf_0\rangle +  |\varphibf_{-1}\rangle +  |\varphibf_{-2}\rangle \,,
\\
 \label{19072015-man-03} && \hspace{1cm} |\phibf_1\rangle  =  \zeta_1 |\phi_1^{s_1-1,s_2}\rangle\,,
\\
&& \hspace{1cm} |\phibf_0\rangle  =    |\phi_0^{s_1,s_2}\rangle + \zeta_1 \alpha_1^z |\phi_0^{s_1-2,s_2}\rangle + \zeta_1 \alpha_2^z |\phi_{0'}^{s_1-1,s_2-1}\rangle\,,
\\
&& \hspace{1cm} |\phibf_{-1}\rangle  =   \alpha_1^z |\phi_{-1}^{s_1-1,s_2}\rangle + \alpha_2^z |\phi_{-1}^{s_1,s_2-1}\rangle + \zeta_1 \alpha_1^z  \alpha_2^z |\phi_{-1}^{s_1-2,s_2-1}\rangle\,,
\\
&& \hspace{1cm} |\phibf_{-2}\rangle  =  \alpha_1^z \alpha_2^z |\phi_{-2}^{s_1-1,s_2-1}\rangle
\,,
\\
&& \hspace{1cm} |\varphibf_1\rangle  =  - \zeta_1 |\phi_2^{s_1-1,s_2-1}\rangle \,,
\\
&& \hspace{1cm} |\varphibf_0\rangle  =  |\phi_1^{s_1,s_2-1}\rangle - \zeta_1 \alpha_1^z |\phi_1^{s_1-2,s_2-1}\rangle - \zeta_1 \alpha_2^z |\phi_1^{s_1-1,s_2-2}\rangle\,,
\\
&& \hspace{1cm} |\varphibf_{-1}\rangle  =    \alpha_1^z|\phi_0^{s_1-1,s_2-1}\rangle + \alpha_2^z  |\phi_0^{s_1,s_2-2}\rangle - \zeta_1 \alpha_1^z\alpha_2^z |\phi_0^{s_1-2,s_2-2}\rangle,\qquad
\\
 \label{19072015-man-04} && \hspace{1cm} |\varphibf_{-2}\rangle  =  \alpha_1^z \alpha_2^z |\phi_{-1}^{s_1-1,s_2-2}\rangle\,,
\eeq
where ket-vectors $|\phi_\lambda^{\Ksm_1\Ksm_2}\rangle$ appearing in \rf{19072015-man-03}-\rf{19072015-man-04} are defined in \rf{17072015-man-12}.  Now, considering the massless limit \rf{19072015-man-02}, we are led to the following conclusion:
for $s_2>1$, the ket-vectors $|\phibf\rangle$ and $|\varphibf\rangle$ describe the respective type $[s_1,s_2]$ and type $[s_1,s_2-1]$ massless AdS fields, while, for $s_2=1$, the ket-vectors $|\phibf\rangle$ and $|\varphibf\rangle$ describe the respective type $[s_1,1]$ massless and type $[s_1,0]$ massive AdS fields (see also Refs.\cite{Metsaev:1997nj,deMedeiros:2003px}). The type  $[s_1,0]$ massive field is associated with unitary irreps of the $so(d,2)$ algebra  with $E_0=d+1$.
On space of type $[s_1,s_2]$ massless AdS field $\phibfk$ \rf{19072015-man-02a}, the operator $\nu$ takes the form as in \rf{18052015-man-07} with $M^z = N_{\zeta_1} - N_{\alpha_1^z} - N_{\alpha_2^z}$ and $E_0$ as in \rf{19072015-man-02}.

Plugging ket-vector $\phibfk$ \rf{19072015-man-02a} into \rf{18052015-man-03}, we get Lagrangian of type $[s_1,s_2]$ massless AdS field
\be \label{19072015-man-05}
\LL   =   \half \sum_{\lambda=-2,-1,0,1}\langle \phibf_\lambda | \left( \Box + \partial_z^2 - \frac{1}{z^2}\bigl( \nu_\lambda^2 -\frac{1}{4}\bigr)\right) |\phibf_\lambda \rangle\,, \qquad  \nu_\lambda \equiv  \frac{d}{2} + 1  - s_2  + \lambda\,.
\ee
For massless AdS field, the spin operators $W^i$, $\Wb^i$ are obtained by taking limit \rf{19072015-man-02} in the respective spin operators of the massive field (see Appendix C).

As a side remark we note that the ket-vector of type $[s_1,s_2]$ massless AdS field $\phibfk$ \rf{19072015-man-02a} can be represented as
\beq
\label{19072015-man-06} && \hspace{-1.2cm} \phibfk = |\phibf^{s_1,s_2}\rangle + \zeta_1 |\phibf^{s_1-1,s_2}\rangle\,,
\\
\label{19072015-man-07} && |\phibf^{s_1,s_2}\rangle  \equiv     |\phi_0^{s_1,s_2}\rangle + \alpha_1^z |\phi_{-1}^{s_1-1,s_2}\rangle + \alpha_2^z |\phi_{-1}^{s_1,s_2-1}\rangle +   \alpha_1^z \alpha_2^z |\phi_{-2}^{s_1-1,s_2-1}\rangle
\,,
\\
\label{19072015-man-08} && |\phibf^{s_1-1,s_2}\rangle  \equiv    |\phi_1^{s_1-1,s_2}\rangle +  \alpha_1^z |\phi_0^{s_1-2,s_2}\rangle +  \alpha_2^z |\phi_0^{s_1-1,s_2-1}\rangle +    \alpha_1^z  \alpha_2^z |\phi_{-1}^{s_1-2,s_2-1}\rangle\,.
\eeq
Ket-vectors $|\phibf^{s_1,s_2}\rangle$ \rf{19072015-man-07} and $|\phibf^{s_1-1,s_2}\rangle$ \rf{19072015-man-08} describe the respective type $[s_1,s_2]$ and type $[s_1-1,s_2]$ irreps of the $so(d-1)$ algebra. In other words, in flat space limit, the ket-vector of type $[s_1,s_2]$ massless AdS field $\phibfk$ \rf{19072015-man-06} is decomposed into type $[s_1,s_2]$ and type $[s_1-1,s_2]$ massless fields in $R^{d,1}$. This result is in agreement with the conjecture in Ref.\cite{Brink:2000ag}. For general mixed-symmetry fields, the proof of the conjecture in Ref.\cite{Brink:2000ag} may be found in Ref.\cite{Boulanger:2008up}.

\noindent {\bf Short currents and shadows in $R^{d-1,1}$ }. Type $[s_1,s_2]$ short current in $R^{d-1,1}$ is related to the unitary representation $D(\Delta_\cur,\hbf)$ of the $so(d,2)$ algebra, where $\Delta_\cur = d+1-s_2$, while $\hbf$ is given in \rf{17072015-man-01}.
Type $[s_1,s_2]$ short shadow in $R^{d-1,1}$ is associated with representation of the $so(d,2)$ algebra with labels $\Delta_\sh= s_2-1$ and $\hbf$ \rf{17072015-man-01}. For short currents and shadows in $R^{d-1,1}$, ket-vectors $|\phibf_\cur\rangle$, $|\phibf_\sh\rangle$ take the same form as the ket-vector $\phibfk$ \rf{19072015-man-02a} for the massless field in $AdS_{d+1}$, while the operators $\nu$, $W^i$, $\Wb^i$, $M^{ij}$ are obtained by taking the limit \rf{19072015-man-02} in expressions for the respective operators of the massive field in $AdS_{d+1}$. Plugging the ket-vectors $|\phibf_\cur\rangle$, $|\phibf_\sh\rangle$ and the operators  $\nu$, $W^i$, $\Wb^i$, $M^{ij}$ into expressions for 2-point functions and generators of the $so(d,2)$ algebra given in Sec.\ref{section-03}, we get the complete light-cone gauge description of the type $[s_1,s_2]$ short currents and shadows.

\noindent {\bf Short conformal field in $R^{d-1,1}$ }. Type $[s_1,s_2]$ short conformal field in $R^{d-1,1}$ is associated with representation of the $so(d,2)$ algebra with labels $\Delta$, $\hbf$, where $\hbf$ is given in \rf{17072015-man-01}, while the conformal dimension $\Delta$ is given by
\be
\Delta = s_2-1\,,   \qquad s_2 \geq 1\,, \qquad d-\hbox{ even}.
\ee
For the type $[s_1,s_2]$ short conformal field in $R^{d-1,1}$, ket-vector $|\phibf\rangle$ takes the same form as the one for massless field in $AdS_{d+1}$ \rf{19072015-man-02a}, while the operators $\nu$, $W^i$, $\Wb^i$, $M^{ij}$ can be obtained by considering the massless limit \rf{19072015-man-02} in expressions for the respective operators of the massive field in $AdS_{d+1}$.
Plugging the operators  $\nu$, $W^i$, $\Wb^i$, $M^{ij}$  into \rf{19052015-man-12}-\rf{19052015-man-22}, \rf{19052015-man-27}-\rf{19052015-man-30} provides realization of the conformal $so(d,2)$ symmetries on space of the ket-vector $|\phibf\rangle$, while
plugging the ket-vector $|\phibf\rangle$ into \rf{27062015-man-04a}, we get Lagrangian for the type $[s_1,s_2]$ short conformal field
\be
\LL = \half  \sum_{\lambda=-2,-1,0,1} \langle \phibf_\lambda| \Box^{\nu_{\intrm,\lambda} } |\phibf_\lambda\rangle\,, \qquad \nu_{\intrm,\lambda} =\frac{d}{2} + 1 - s_2 + \lambda\,,\qquad d-\hbox{even}.
\ee

\noindent {\bf Conclusion}. In conclusion, we make the following comment. Our presentation for the (one)column and (two)column AdS fields demonstrates how the field content and explicit form of the operator $M^z$ which are required for the light-cone gauge formulation of  mixed-symmetry arbitrary spin AdS field can be specified. This is to say that, by analogy with the light-cone gauge formulation of massive field in $R^{d,1}$, in order to develop the light-cone gauge formulation of massive field in $AdS_{d+1}$ we can use field content which is realized as irreducible representation of the $so(d)$ algebra. Note that, in this respect, the light-cone gauge formulation of massive field is similar to the Lorentz covariant formulation of massive field. Namely, we recall that the field content entering the Lorentz covariant formulation of massive field in $R^{d,1}$ coincides with  the field content entering the Lorentz covariant formulation of massive field in $AdS_{d+1}$. Use of the oscillators provides the easy way to describe the field content and the operator $M^z$.
For example, in order to describe field content entering the light-cone gauge formulation of  mixed-symmetry arbitrary spin massive AdS field \rf{05072015-man-04}, we introduce a finite set of anti-commuting oscillators $\alpha_n^i$, $\alpha_n^z$, $\zeta_n$, $n=1,\ldots,h_k$, which satisfy the same relations as in \rf{17072015-man-03-a1}-\rf{17072015-man-03-a3}. Also, by analogy with \rf{17072015-man-04}, we can introduce a ket-vector $\phik$ which, by definition, depends on all just mentioned oscillators and satisfies the well-known algebraic constraints needed to select the irreducible representation of the $so(d)$ algebra. Those constraints for mixed-symmetry arbitrary spin field are the counterparts of the ones for two-column field in \rf{17072015-man-05}-\rf{17072015-man-07}. For such oscillator realization of $\phik$, the operator $M^z$ takes the form $M^z = \sum_{n=1}^{h_k} M_n^z$, where we use $M_n^z\equiv N_{\zeta_n} - N_{\alpha_n^z}$ and notation as in \rf{17072015-man-09}. On space of $\phik$, eigenvalues of such operator $M^z$ take values $-h_k,-h_k+1, \ldots, h_k$, as it should be. For massless AdS field, the field content is realized as an invariant subspace in space of massive AdS field.

\bigskip
\bigskip

{\bf Acknowledgments}. This work was supported by the Russian Science Foundation grant 14-42-00047.

\setcounter{section}{0} \setcounter{subsection}{0}

\appendix{Notation and conventions}

The Cartesian coordinates in $R^{d-1,1}$ are denoted by $x^a$, $a=0,1,\ldots d-1$. Our flat metric tensor $\eta^{ab}$ is mostly positive. To simplify our expressions we drop the flat  metric in scalar product, i.e., we use the convention $X^a Y^a \equiv \eta_{ab} X^a Y^b$.

We use the Poincar\'e parametrization of $AdS_{d+1}$ space,
\be
ds^2= \frac{1}{z^2}(- dx^0 dx^0 + dx^idx^i + dx^{d-1} dx^{d-1}
+ dz dz)\,,  \qquad i=1,\ldots , d-2\,.
\ee

In the light-cone frame, we use the decomposition $x^a= x^+,x^-,x^i$, $i=1,\ldots,d-2$, where the  coordinates $x^\pm$ are defined by the relations
\be
x^\pm \equiv  \frac{1}{\sqrt{2}} (x^{d-1} \pm x^0)\,,
\ee
and $x^+$ is considered as the light-cone evolution parameter. In the frame of coordinates $x^\pm$, $x^i$, non-zero values of the flat metric are given by
\be \label{21062015-man-04}
\eta^{ij} = \delta^{ij}\,, \qquad   \eta^{+-}=1\,,
\ee
where $\delta^{ij}$ is the Kronecker delta. This implies that a scalar product for two vectors $X^a$, $Y^a$ of the Lorentz $so(d-1,1)$ algebra is decomposed as
\be
X^a Y^a  =  X^+ Y^- + X^- Y^+ + X^i Y^i \,.
\ee
Derivative with respect to $z$ is defined as $\partial_z = \partial/\partial_z$. For derivatives $\partial^\pm$, $\partial^i$, and the D'Alembertian operator $\Box$ in $R^{d-1,1}$ we adopt the following conventions:
\be
\partial_\pm \equiv \partial/\partial x^\pm\,, \quad  \partial_i \equiv \partial/\partial x^i\,,\quad \partial^+=\partial_-\,, \quad \partial^-=\partial_+\,, \quad \partial^i = \partial_i \,,\quad\Box = 2\partial^+\partial^- + 2 \partial^i \partial^i.
\ee

In a basis of the algebra $so(d-1,1)$, the algebra $so(d,2)$ is decomposed into the translation generators $P^a$, the conformal boost generators $K^a$, the dilatation generator $D$, and generators of the $so(d-1,1)$ algebra $J^{ab}$. The commutations relations of the $so(d,2)$ algebra and the hermitian conjugation rules we use in this paper take the form
\beq
\label{21062015-man-01} && {}[D,P^a]=-P^a\,, \hspace{2.5cm}  [P^a,J^{bc}]=\eta^{ab}P^c -\eta^{ac}P^b
\,,
\nonumber\\
&& [D,K^a]=K^a\,, \hspace{2.7cm} [K^a,J^{bc}]=\eta^{ab}K^c - \eta^{ac}K^b\,,
\\
&& [P^a,K^b]=\eta^{ab}D - J^{ab}\,, \hspace{1.2cm}  [J^{ab},J^{ce}]=\eta^{bc}J^{ae}+3\hbox{ terms} \,,
\nonumber\\
&& P^{a \dagger} = - P^a\,,  \quad K^{a \dagger} = - K^a\,, \quad J^{ab\dagger} = -J^{ab}\,, \quad D^\dagger = - D\,.
\eeq
In the basis of the $so(d-2)$ algebra, the generators $P^a$, $K^a$, $J^{ab}$ are decomposed as
\be
\label{21062015-man-02} P^a = P^\pm, \ P^i\,, \quad K^a = K^\pm\,, K^i\,, \qquad J^{ab} = J^{+-}\,, J^{\pm i}\,, J^{ij}\,.
\ee
Commutators for the generators given in \rf{21062015-man-02} are obtained from the ones in \rf{21062015-man-01} by using the non-zero values of the flat metric $\eta^{ab}$ given in \rf{21062015-man-04}.

\appendix{ Derivation of defining equations for operators $\nu$, $W$, $\Wb^i$, $M^{ij}$.}

In this Appendix, we outline the derivation of defining equations \rf{04032015-00}-\rf{04032015-05}  for the operators $\nu$, $W$, $\Wb^i$, $M^{ij}$ by using equations for the operators $A$, $M^{zi}$, $M^{ij}$ obtained in Ref.\cite{Metsaev:1999ui}. Also we comment on the derivation of the generators of the $so(d,2)$ algebra in CFT.

\noindent {\bf Derivation of defining equations \rf{04032015-00}-\rf{04032015-05}}. Complete system of equations for the basic operators $A$, $M^{zi}$, $M^{ij}$ obtained in Ref.\cite{Metsaev:1999ui} takes the form%
\footnote{ Relations \rf{19052015-man-32}-\rf{19052015-man-46} and \rf{31012015-01}-\rf{31012015-03b} were obtained in the following way. First, in Sec.3 in Ref.\cite{Metsaev:1999ui}, using the Lorentz covariant formulation of totally symmetric AdS fields and applying the standard method for the derivation of the light-cone gauge formulation from the Lorentz covariant formulation, we obtained  \rf{19052015-man-32}-\rf{19052015-man-46} and the operators $A$, $B$, $M^{ij}$ which satisfy \rf{31012015-01}-\rf{31012015-03b}. Second, in Secs.4,5 in Ref.\cite{Metsaev:1999ui}, using the group theoretical description of representations of the $so(d,2)$ algebra, we generalized our results to the case of mixed-symmetry fields. In the same way, we obtained the light-cone gauge formulation of CFT.  Namely, in Sec.6 in Ref.\cite{Metsaev:1999ui}, using the Lorentz covariant formulation of totally symmetric currents (shadows) and applying the standard method for the derivation of the light-cone gauge formulation from the Lorentz covariant formulation, we obtained the light-cone gauge formulation of totally symmetric currents (shadows), while, in Ref.\cite{Metsaev:2005ws}, we generalized our results to the case of mixed-symmetry currents (shadows). Derivation of relations \rf{19052015-man-12}-\rf{19052015-man-30} in this paper from the results in Ref.\cite{Metsaev:2005ws} is outlined in this Appendix.}
\beq
\label{31012015-01} && 2\{ M^{zi},A\} - [[M^{zi} ,A],A] = 0 \,,
\\
\label{31012015-02} && [M^{zi},[M^{zj},A]] + \{ M^{il},M^{lj}\} - \{ M^{zi},M^{zj}\} = - 2\delta^{ij} B \,,
\\
\label{31012015-02a} && -A + 2B + \half M^{ij}M^{ij} + \frac{d^2-1}{4}+
\langle \CC_{so(d,2)}\rangle = 0\,,
\\
\label{31012015-02b} && [A, M^{ij}] = 0\,,
\\
\label{31012015-03} && [M^{zi},M^{zj}]=-M^{ij}\,,
\\
&& [M^{ij},M^{kl}]= \delta^{jk}M^{il} + 3 \hbox{ terms} \,, \quad \qquad M^{ij} = -M^{ji}\,,
\\
\label{31012015-03b} && [M^{ij}, M^{zk}]=\delta^{jk} M^{zi} - \delta^{ik} M^{zj}\,.
\eeq
From the results in Section 5.2 in Ref.\cite{Metsaev:1999ui}, we learn that the operator $A$ can be presented as in \rf{18052015-man-06}, \rf{18052015-man-07}. Note however that for the derivation of defining equations \rf{04032015-00}-\rf{04032015-05} we do not need to use the particular representation for the operator $\nu$ given in \rf{18052015-man-07}. In other words, all that is required for the derivation of defining equations \rf{04032015-00}-\rf{04032015-05} is to use the representation for the operator $A$ given in \rf{18052015-man-06} and equations \rf{31012015-01}-\rf{31012015-03b} We now outline proof of the following

\noindent {\bf Statement}. Equations \rf{31012015-01}-\rf{31012015-03b} with the operator $A$ as in \rf{18052015-man-06} amount to the defining equations given in \rf{04032015-00}-\rf{04032015-05}.

\noindent {\bf Proof of the statement}. We prove the statement in the following three steps.

\noindent {\bf Step 1}. Introducing new operators $M^i$, $\Mwt^i$ by the following commutators
\be
\label{19022015-16}  M^i \equiv [\nu,M^{zi}]\,, \qquad \Mwt^i \equiv [\nu,M^i]\,,
\ee
we find the relations
\beq
\label{19022015-07} && [[ M^{zi},A],A] = \{\nu ,\{\nu,\Mwt^i\}\}\,,
\qquad \ 2 \{ A,M^{zi} \} = \{\nu,\{\nu, M^{zi}\}\} - M^{zi} + \Mwt^i\,.\qquad
\eeq
Using \rf{19022015-07} in \rf{31012015-01}, we find that Eq.\rf{31012015-01} amounts to the equations
\be \label{19022015-09}
\{\nu,\{\nu, M^{zi} - \Mwt^i \}\} - M^{zi} + \Mwt^i = 0 \,.
\ee
Equations \rf{19022015-09} imply
\be\label{19022015-10}
M^{zi} = \Mwt^i\,.
\ee

\noindent {\bf Step 2}. We introduce new operators $W^i$, $\Wb^i$ by the relations
\be
\label{19022015-14} M^{zi} = W^i - \Wb^i\,, \qquad M^i = W^i + \Wb^i\,,
\ee
and note that \rf{19022015-16}, \rf{19022015-10} lead to commutators in \rf{04032015-01}, \rf{04032015-02}, while the commutator \rf{31012015-03} leads to \rf{04032015-03}-\rf{04032015-04}. Also we note that \rf{31012015-02b} and \rf{18052015-man-06} lead to \rf{04032015-00}.

\noindent {\bf Step 3}. We note that Eqs.\rf{04032015-05} are obtained from \rf{31012015-02} by using the following relation:
\beq
&& [M^{zi},[M^{zj},A]] - \{M^{zi},M^{zj}\}
\nonumber\\
&& \hspace{2cm}  =   2(\nu+1)(\Wb^i W^j + \Wb^j W^i)] - 2(\nu-1) (W^i \Wb^j + W^j \Wb^i)\,.
\eeq

\noindent  {\bf Derivation of generators of $so(d,2)$ algebra in CFT}. We now outline the derivation of generators of the $so(d,2)$ algebra symmetries for currents and shadow given in relations \rf{19052015-man-12}-\rf{19052015-man-30} in this paper. To this end, we use the general light-cone gauge approach to CFT developed in Ref.\cite{Metsaev:2005ws}.
Namely, the defining equations for operators $A$, $M^i$, $M^{ij}$ entering general light-cone gauge approach to CFT are given in Eqs.(3.11), (A1)-(A3) in Ref.\cite{Metsaev:2005ws}. For the study of those defining equations, we apply the above demonstrated procedure we used for the study of Eqs.\rf{31012015-01}-\rf{31012015-03b} in this Appendix.
Doing so, we learn that the operator $A$ in Ref.\cite{Metsaev:2005ws} takes the form as in Eq.\rf{18052015-man-06} in this paper, while the operator $M^i$ in Ref.\cite{Metsaev:2005ws} takes the form $M^i = W^i + \Wb^i$. Also we learn that  Eqs.(3.11), (A1)-(A3) in Ref.\cite{Metsaev:2005ws} lead to the same equations for the operators $\nu$, $W^i$, $\Wb^i$, $M^{ij}$ we derived in the light-cone gauge AdS field theory (see Eqs.\rf{04032015-00}-\rf{04032015-05} in this paper). We note however that the representation for generators of the $so(d,2)$ algebra symmetries on space of conformal field given by relations (3.1)-(3.15) in Ref.\cite{Metsaev:2005ws} is realized as the non-local representation. Let us use the notation $|\OO\rangle$ for a conformal field on which such non-local representation of the $so(d,2)$ algebra is realized. Then introducing a current by the relation $q^{\nu + \half} |\OO\rangle =  |\phi_\cur\rangle$, $q \equiv \sqrt{\Box}$, we verify that the non-local representation on space of the conformal field $|\OO\rangle$ given by relations (3.1)-(3.15) in Ref.\cite{Metsaev:2005ws} is realized as the local representation on the space of $ |\phi_\cur\rangle$ given by relations in \rf{19052015-man-12}-\rf{19052015-man-26} in this paper, while introducing a shadow by the relation $q^{-\nu + \half}|\OO\rangle =  |\phi_\sh\rangle$, $q\equiv \sqrt{\Box}$, we verify that the non-local representation on  space of conformal field $|\OO\rangle$ given by relations (3.1)-(3.15) in Ref.\cite{Metsaev:2005ws} is realized as the local representation on space of $ |\phi_\sh\rangle$ given by relations \rf{19052015-man-12}-\rf{19052015-man-22}, \rf{19052015-man-27}-\rf{19052015-man-30} in this paper.

\appendix{ Operators $W$, $\Wb^i$ for two-column fields}

We start with the description of the explicit form of operators $W^i$, $\Wb^i$ \rf{17072015-man-17}, \rf{17072015-man-18} for the type $[s_1,s_2]$, $s_1>s_2$, massive fields. These operators should satisfy the defining equations \rf{04032015-00}-\rf{04032015-05} and respect constraints \rf{17072015-man-05}-\rf{17072015-man-07}. By using operators $A_n^i$, $\Ab_n^i$ \rf{17072015-man-19}-\rf{17072015-man-22} we respect  constraints \rf{17072015-man-06},\rf{17072015-man-07}. All that is required then is to determine the operators $X_n$ and $Y_n$, $n=1,2$ which depend on the scalar oscillators $\alpha_n^z$, $\zeta_n$ and do not depend on the vector oscillators $\alpha_n^i$. General representation for the operators $X_n$, $Y_n$ which respects constraint \rf{17072015-man-05} and commutation relations \rf{04032015-01},\rf{04032015-02} is given by
\beq
\label{15032015-06} && X_1 = \zeta_1 f_{\zeta_2\alpha_1^z\alpha_2^z} + g_{\zeta_1} \zeta_2 \alpha_1^z \alphab_2^z\,,
\hspace{1.5cm}
Y_1 = \alpha_1^z f_{\alpha_2^z\zeta_1\zeta_2} + g_{\alpha_1^z} \alpha_2^z \zeta_1 \zetab_2\,,
\qquad
\\
&& X_2 = \zeta_2 f_{\zeta_1\alpha_2^z\alpha_1^z} + g_{\zeta_2} \zeta_1 \alpha_2^z \alphab_1^z\,,
\hspace{1.5cm}
Y_2 = \alpha_2^z f_{\alpha_1^z\zeta_2\zeta_1} + g_{\alpha_2^z} \alpha_1^z \zeta_2 \zetab_1\,,
\\
&& \Xb_1 = \zetab_1 f_{\zeta_2\alpha_1^z\alpha_2^z} + g_{\zeta_1}  \alpha_2^z \alphab_1^z \zetab_2 \,, %
\hspace{1.5cm}
\Yb_1 = \alphab_1^z f_{\alpha_2^z\zeta_1\zeta_2} + g_{\alpha_1^z}   \zeta_2 \zetab_1 \alphab_2^z\,,
\\
\label{15032015-06a} && \Xb_2 = \zetab_2 f_{\zeta_1\alpha_2^z\alpha_1^z} + g_{\zeta_2}\alpha_1^z \alphab_2^z  \zetab_1\,,
\hspace{1.5cm}
\Yb_2 = \alphab_2^z f_{\alpha_1^z\zeta_2\zeta_1} + g_{\alpha_2^z}  \zeta_1 \zetab_2 \alphab_1^z \,,
\eeq
where quantities $f_{xyw}$, $g_v$ appearing in \rf{15032015-06}-\rf{15032015-06a} take the form
\beq
\label{17072015-15} && f_{xyw} = f_{xyw}(N_x, N_y, N_w)\,, \hspace{2.7cm} g_v = g_v(N_v)\,,
\\
&& N_x\equiv x\xb\,, \quad N_y \equiv y\yb\,,\quad N_w = w\wb\,, \qquad N_v \equiv v\vb\,.
\eeq
Namely, $f_{xyw}$ \rf{17072015-15} depends on three operators $N_x, N_y, N_w$, while $g_v$ \rf{17072015-15} depends only on one operator $N_v$. For example, the quantity $f_{\zeta_2\alpha_1^z\alpha_2^z}$ in \rf{15032015-06} depends on the three operators $N_{\zeta_2}$, $N_{\alpha_1^z}$, $N_{\alpha_2^z}$, while $g_{\zeta_1}$ in \rf{15032015-06} depends only on the one operator $N_{\zeta_1}$. Using notation $a$, $b$, $c$, $e$ for eigenvalues of the respective operators $N_x$, $N_y$, $N_w$, $N_v$, we introduce the quantities
\be
f_{xyw}(a,b,c)=f_{xyw}(N_x, N_y, N_w)\bigr|_{N_x=a, N_y=b, N_w=c}\,, \qquad g_v(e) = g_v(N_v)\bigr|_{N_v=e}\,.
\ee
On space of ket-vector $\phik$ \rf{17072015-man-10}, the $a,b,c,e$ take the values $a,b,c,e=0,1$. Thus we should determine the $f_{xyw}(a,b,c)$ and $g_v(e)$ for the just mentioned values of $a,b,c,e$.  The $f_{xyw}(a,b,c)$ and $g_v(e)$ are determined from equations \rf{04032015-03}-\rf{04032015-05}. We now describe our results for the $f_{xyw}(a,b,c)$ and $g_v(e)$. To this end we introduce the following notation
\beq
\label{17072015-16} && \hspace{0.7cm} \kappa = E_0-\frac{d}{2}\,, \qquad t = d-1 - s_1 - s_2\,, \qquad h = s_1-s_2+1\,,
\\
\label{17072015-17} && \sigma_{\zeta_1}  = \kappa - \half (t-h+2)\,, \hspace{2cm} \sigma_{\alpha_1^z}  = \kappa + \half (t-h+2)\,,
\\
\label{17072015-18} && \sigma_{\zeta_2}  = \kappa - \half (t+h+2)\,, \hspace{2cm} \sigma_{\alpha_2^z}  = \kappa + \half (t+h+2)\,,
\\
\label{17072015-19} && \rho_{\zeta_1} = \sigma_{\zeta_1} - \frac{(t+h+2)^2}{2\kappa ((t+2)h+1)}\,, \hspace{1cm}  \rho_{\alpha_1^z} = \sigma_{\alpha_1^z} - \frac{(t+h+2)^2}{2\kappa ((t+2)h+1) }\,,
\\
\label{17072015-20} && \rho_{\zeta_2} = \sigma_{\zeta_2} + \frac{(t-h+2)^2}{2\kappa ((t+2)h-1)}\,, \hspace{1cm} \rho_{\alpha_2^z} = \sigma_{\alpha_2^z} + \frac{(t-h+2)^2}{2\kappa ((t+2)h-1) }\,.
\eeq
Using notation in \rf{17072015-16}-\rf{17072015-20}, we find the following expressions for $f_{xyw}(a,b,c)$ and $g_v(e)$:
{\small
\beq
\label{05052015-01} && \hspace{-0.8cm}  f_{\zeta_2\alpha_1^z\alpha_2^z}(0,0,0)  = \Bigl( \frac{\sigma_{\zeta_1}}{2\kappa}\Bigr)^{\scriptscriptstyle 1/2}\!\!,
\hspace{2.7cm}
f_{\alpha_2^z\zeta_1\zeta_2}(0,0,0)  = \Bigl( \frac{\sigma_{\alpha_1^z}}{2\kappa}\Bigr)^{\scriptscriptstyle 1/2}\!\!,
\\
\label{05052015-02} && \hspace{-0.8cm}  f_{\zeta_2\alpha_1^z\alpha_2^z}(1,0,0)  = \Bigl(\frac{h+1}{2h} \frac{\sigma_{\zeta_1}}{\kappa+1}\Bigr)^{\scriptscriptstyle 1/2}\!\!,
\hspace{1.4cm}
f_{\alpha_2^z\zeta_1\zeta_2}(1,0,0)  = \Bigl(\frac{h+1}{2h} \frac{\sigma_{\alpha_1^z}}{\kappa-1}\Bigr)^{\scriptscriptstyle 1/2}\!\!,
\\
\label{05052015-03} && \hspace{-0.8cm} f_{\zeta_2\alpha_1^z\alpha_2^z}(0,1,0)  = \Bigl(  \frac{\sigma_{\zeta_1}}{2\kappa}\Bigr)^{\scriptscriptstyle 1/2}\!\!,
\hspace{2.7cm}
f_{\alpha_2^z\zeta_1\zeta_2}(0,1,0)  = \Bigl(  \frac{\sigma_{\alpha_1^z}}{2\kappa}\Bigr)^{\scriptscriptstyle 1/2}\!\!,
\\
\label{05052015-04} && \hspace{-0.8cm}  f_{\zeta_2\alpha_1^z\alpha_2^z}(0,0,1)  =   r_{\zeta_1}(0) \cos\psi_{\zeta_1}\,,
\hspace{1.9cm}
f_{\alpha_2^z\zeta_1\zeta_2}(0,0,1)  =   r_{\alpha_1^z}(0) \cos\psi_{\alpha_1^z}\,,
\\
\label{05052015-05} && \hspace{-0.8cm}  f_{\zeta_2\alpha_1^z\alpha_2^z}(0,1,1)  = \Bigl(\frac{t+3}{2(t+2)} \frac{\sigma_{\zeta_1}}{\kappa-1}\Bigr)^{\scriptscriptstyle 1/2}\!\!,
\hspace{1cm}
f_{\alpha_2^z\zeta_1\zeta_2}(0,1,1)  = \Bigl(\frac{t+3}{2(t+2)} \frac{\sigma_{\alpha_1^z}}{\kappa+1}\Bigr)^{\scriptscriptstyle 1/2}\!\!,
\\
\label{05052015-06} && \hspace{-0.8cm} f_{\zeta_2\alpha_1^z\alpha_2^z}(\!1,\!0,\!1)\!  = \! \Bigl(\! \frac{(t+3)(h+2)\sigma_{\zeta_1}}{2(t+2)(h+1)\kappa}\!\Bigr)^{\scriptscriptstyle 1/2}\!\!\!,
\hspace{1cm}
f_{\alpha_2^z\zeta_1\zeta_2}(\!1,\!0,\!1)\!  = \!\! \Bigl(\! \frac{(t+3)(h+2)\sigma_{\alpha_1^z}}{2(t+2)(h+1)\kappa}\!\Bigr)^{\scriptscriptstyle 1/2}\!\!\!,
\\
\label{05052015-07} && \hspace{-0.8cm}  f_{\zeta_2\alpha_1^z\alpha_2^z}(1,1,0)  = r_{\zeta_1}(1) \cos\psi_{\zeta_1}\,,
\hspace{1.8cm}
f_{\alpha_2^z\zeta_1\zeta_2}(1,1,0)  = r_{\alpha_1^z}(1) \cos\psi_{\alpha_1^z}\,,
\\
\label{05052015-08} && \hspace{-0.8cm}  f_{\zeta_2\alpha_1^z\alpha_2^z}(\!1,\!1,\!1)\! = \! \Bigl(\!\frac{(t+4)(h+1)\sigma_{\zeta_1}}{2(t+3)h\kappa}\!\Bigr)^{\scriptscriptstyle 1/2}\!\!\!,
\hspace{1cm}
f_{\alpha_2\zeta_1\zeta_2}(\!1,\!1,\!1)\! = \! \Bigl(\!\frac{(t+4)(h+1)\sigma_{\alpha_1^z}}{2(t+3)h\kappa}\!\Bigr)^{\scriptscriptstyle 1/2}\!\!\!,
\\[10pt]
\label{05052015-09} && \hspace{-0.8cm} f_{\zeta_1\alpha_2^z\alpha_1^z}(0,0,0)  = \Bigl(\frac{h+1}{2h} \frac{\sigma_{\zeta_2}}{\kappa}\Bigr)^{\scriptscriptstyle 1/2}\!,
\hspace{1.9cm}
f_{\alpha_1^z\zeta_2\zeta_1}(0,0,0)  = \Bigl(\frac{h+1}{2h} \frac{\sigma_{\alpha_2^z}}{\kappa}\Bigr)^{\scriptscriptstyle 1/2}\!,
\\
\label{05052015-10} && \hspace{-0.8cm}  f_{\zeta_1\alpha_2^z\alpha_1^z}(1,0,0)  = \Bigl(\half \frac{\sigma_{\zeta_2}}{\kappa+1}\Bigr)^{\scriptscriptstyle 1/2}\!,
\hspace{2.2cm}
f_{\alpha_1^z\zeta_2\zeta_1}(1,0,0)  = \Bigl(\half \frac{\sigma_{\alpha_2^z}}{\kappa-1}\Bigr)^{\scriptscriptstyle 1/2}\!,
\\
\label{05052015-11} && \hspace{-0.8cm}  f_{\zeta_1\alpha_2^z\alpha_1^z}(0,1,0)  = \Bigl(\frac{h+2}{2(h+1)} \frac{\sigma_{\zeta_2}}{\kappa}\Bigr)^{\scriptscriptstyle 1/2}\!,
\hspace{1.5cm}
f_{\alpha_1^z\zeta_2\zeta_1}(0,1,0)  = \Bigl(\frac{h+2}{2(h+1)} \frac{\sigma_{\alpha_2^z}}{\kappa}\Bigr)^{\scriptscriptstyle 1/2}\!,
\\
\label{05052015-12} && \hspace{-0.8cm}  f_{\zeta_1\alpha_2^z\alpha_1^z}(0,0,1)  = r_{\zeta_2}(0) \cos\psi_{\zeta_2}\,,
\hspace{2.1cm}
f_{\alpha_1^z\zeta_2\zeta_1}(0,0,1)  = r_{\alpha_2^z}(0) \cos\psi_{\alpha_2^z}\,,
\\
\label{05052015-13} && \hspace{-0.8cm} f_{\zeta_1\alpha_2^z\alpha_1^z}(\!0,\!1,\!1)\!  =\! \Bigl(\!\frac{(t+3)(h+1)\sigma_{\zeta_2}}{2(t+2)h(\kappa-1)}\!\Bigr)^{\scriptscriptstyle 1/2}\!\!\!,
\hspace{1.3cm}
f_{\alpha_1^z\zeta_2\zeta_1}(\!0,\!1,\!1)\!  =\! \Bigl(\!\frac{(t+3)(h+1)\sigma_{\alpha_2^z}}{2(t+2)h(\kappa+1)}\!\Bigr)^{\scriptscriptstyle 1/2}\!\!\!,
\\
\label{05052015-14} && \hspace{-0.8cm} f_{\zeta_1\alpha_2^z\alpha_1^z}(1,0,1)  = \Bigl(\frac{t+3}{2(t+2)} \frac{\sigma_{\zeta_2}}{\kappa}\Bigr)^{\scriptscriptstyle 1/2}\!,
\hspace{1.6cm}
f_{\alpha_1^z\zeta_2\zeta_1}(1,0,1)  = \Bigl(\frac{t+3}{2(t+2)} \frac{\sigma_{\alpha_2^z}}{\kappa}\Bigr)^{\scriptscriptstyle 1/2}\!,
\\
\label{05052015-15} &&  \hspace{-0.8cm} f_{\zeta_1\alpha_2^z\alpha_1^z}(1,1,0)  = r_{\zeta_2}(1) \cos\psi_{\zeta_2}\,,
\hspace{2.2cm}
f_{\alpha_1^z\zeta_2\zeta_1}(1,1,0)  = r_{\alpha_2^z}(1) \cos\psi_{\alpha_2^z}\,,
\\
\label{05052015-16} && \hspace{-0.8cm} f_{\zeta_1\alpha_2^z\alpha_1^z}(1,1,1)  = \Bigl(\frac{t+4}{2(t+3)} \frac{\sigma_{\zeta_2}}{\kappa}\Bigr)^{\scriptscriptstyle 1/2}\!,
\hspace{1.7cm}
f_{\alpha_1^z\zeta_2\zeta_1}(1,1,1)  = \Bigl(\frac{t+4}{2(t+3)} \frac{\sigma_{\alpha_2^z}}{\kappa}\Bigr)^{\scriptscriptstyle 1/2}\!.
\eeq
\beq
\label{17072015-21} && g_{\zeta_1}(0) = r_{\zeta_1}(0) \sin\psi_{\zeta_1}\,,
\hspace{2cm}
g_{\alpha_1^z}(0) = r_{\alpha_1^z}(0) \sin\psi_{\alpha_1^z}\,,
\\
&& g_{\zeta_1}(1) = r_{\zeta_1}(1) \sin\psi_{\zeta_1}\,,
\hspace{2cm}
g_{\alpha_1^z}(1) = r_{\alpha_1^z}(1) \sin\psi_{\alpha_1^z}\,,
\\
&& g_{\zeta_2}(0) = r_{\zeta_2}(0) \sin\psi_{\zeta_2}\,,
\hspace{2cm}
g_{\alpha_2^z}(0) = r_{\alpha_2^z}(0) \sin\psi_{\alpha_2^z}\,,
\\
\label{17072015-22} && g_{\zeta_2}(1) = r_{\zeta_2}(1) \sin\psi_{\zeta_2}\,,
\hspace{2cm}
g_{\alpha_2^z}(1) = r_{\alpha_2^z}(1) \sin\psi_{\alpha_2^z}\,,
\eeq
where radii $r_x(0)$, $r_x(1)$ appearing in \rf{05052015-04},\rf{05052015-07},\rf{05052015-12}, \rf{05052015-15}, \rf{17072015-21}-\rf{17072015-22} are given by the relations
\beq
\label{17072015-22a} && \hspace{-0.5cm} r_{\zeta_1}(0) = \Bigl(\frac{((t+2)h+1)\rho_{\zeta_1}}{2(t+1) h (\kappa-1)}\Bigr)^{\scriptscriptstyle 1/2}\!, \hspace{1.8cm}
r_{\alpha_1^z}(0) = \Bigl(\frac{ ((t+2)h+1) \rho_{\alpha_1^z}}{2(t+1) h (\kappa+1)}\Bigr)^{\scriptscriptstyle 1/2}\!,
\\
&& \hspace{-0.5cm} r_{\zeta_1}(1) = \Bigl(\frac{((t+2)h+1)\rho_{\zeta_1}}{2(t+2) (h -1) (\kappa+1)}\Bigr)^{\scriptscriptstyle 1/2}\!, \hspace{1cm}
r_{\alpha_1^z}(1) = \Bigl(\frac{ ((t+2)h+1) \rho_{\alpha_1^z}}{2(t+2) (h -1) (\kappa-1)}\Bigr)^{\scriptscriptstyle 1/2}\!,
\\
&& \hspace{-0.5cm} r_{\zeta_2}(0) = \Bigl(\frac{((t+2)h-1)\rho_{\zeta_2}}{2(t+1) (h-1) (\kappa-1)}\Bigr)^{\scriptscriptstyle 1/2}\!, \hspace{1cm}
r_{\alpha_2^z}(0) = \Bigl( \frac{ ((t+2)h-1) \rho_{\alpha_2^z}}{2(t+1) (h-1) (\kappa+1)}\Bigr)^{\scriptscriptstyle 1/2}\!,
\\
&& \hspace{-0.5cm} r_{\zeta_2}(1) = \Bigl(\frac{((t+2)h-1)\rho_{\zeta_2}}{2(t+2) h (\kappa+1)}\Bigr)^{\scriptscriptstyle 1/2}\!, \hspace{1.9cm}
\label{17072015-22b} r_{\alpha_2^z}(1) = \Bigl(\frac{ ((t+2)h-1) \rho_{\alpha_2^z}}{2(t+2) h (\kappa-1)}\Bigr)^{\scriptscriptstyle 1/2}\!,
\eeq
while angle variables $\psi_x$ in  \rf{05052015-04},\rf{05052015-07},\rf{05052015-12}, \rf{05052015-15}, \rf{17072015-21}-\rf{17072015-22} should satisfy the equations
\beq
\label{20072015-man-01} && \cos(\psi_{\zeta_1} + \psi_{\zeta_2}) = \Bigl( \frac{(t+1)(t+3)h^2 (\kappa^2-1) \sigma_{\zeta_1} \sigma_{\zeta_2} }{ ((t+2)h+1) ((t+2)h-1) \kappa^2 \rho_{\zeta_1} \rho_{\zeta_2} } \Bigr)^{1/2}\,,
\\[3pt]
&& \cos(\psi_{\zeta_1} + \psi_{\alpha_1^z}) = \Bigl( \frac{(t+1)(t+3)(h^2-1) \sigma_{\zeta_1} \sigma_{\alpha_1^z} }{ ((t+2)h+1)^2\rho_{\zeta_1} \rho_{\alpha_1^z} } \Bigr)^{1/2}\,,
\\[3pt]
&& \cos(\psi_{\zeta_1} - \psi_{\alpha_2^z}) = \Bigl( \frac{(t+2)^2 (h^2-1)(\kappa^2-1) \sigma_{\zeta_1} \sigma_{\alpha_2^z} }{ ((t+2)h+1)((t+2)h-1)  \kappa^2  \rho_{\zeta_1} \rho_{\alpha_2^z}  } \Bigr)^{1/2} \,,
\\[3pt]
&& \cos(\psi_{\zeta_2} - \psi_{\alpha_1^z}) = \Bigl( \frac{(t+2)^2 (h^2-1) (\kappa^2-1) \sigma_{\zeta_2} \sigma_{\alpha_1^z} }{ ((t+2)h+1)((t+2)h-1)  \kappa^2  \rho_{\zeta_2} \rho_{\alpha_1^z} } \Bigr)^{1/2} \,,
\\[3pt]
&& \cos(\psi_{\zeta_2} + \psi_{\alpha_2^z}) = \Bigl( \frac{ (t+1)(t+3)(h^2-1) \sigma_{\zeta_2} \sigma_{\alpha_2^z} }{ ((t+2)h-1)^2\rho_{\zeta_2} \rho_{\alpha_2^z}  } \Bigr)^{1/2} \,,
\\[3pt]
&& \cos(\psi_{\alpha_1^z} + \psi_{\alpha_2^z}) = \Bigl( \frac{(t+1)(t+3)h^2 (\kappa^2-1) \sigma_{\alpha_1^z} \sigma_{\alpha_2^z} }{ ((t+2)h+1)((t+2)h-1)  \kappa^2  \rho_{\alpha_1^z} \rho_{\alpha_2^z}  }  \Bigr)^{1/2}\,,
\\
&& \sin(\psi_{\zeta_1} + \psi_{\zeta_2}) = - \frac{\kappa-t-2}{\kappa} \Bigl( \frac{(h^2-1) \sigma_{\alpha_1^z}\sigma_{\alpha_2^z} }{((t+2)h +1) ((t+2)h  - 1)  \rho_{\zeta_1} \rho_{\zeta_2} } \Bigr)^{1/2}\,,
\\[3pt]
&& \sin(\psi_{\zeta_1} + \psi_{\alpha_1^z}) = - \frac{t+h + 2}{(t+2)h +1}  \Bigl( \frac{ (\kappa^2-1) \sigma_{\zeta_2} \sigma_{\alpha_2^z} }{ \kappa^2 \rho_{\zeta_1} \rho_{\alpha_1^z} } \Bigr)^{1/2}\,,
\\[3pt]
&& \sin(\psi_{\zeta_1} - \psi_{\alpha_2^z}) = - \frac{\kappa+h}{\kappa} \Bigl( \frac{(t+1)(t+3) \sigma_{\zeta_2} \sigma_{\alpha_1^z} }{ ((t+2)h+1)((t+2)h-1) \rho_{\zeta_1} \rho_{\alpha_2^z}  } \Bigr)^{1/2} \,,
\\[3pt]
&& \sin(\psi_{\zeta_2} - \psi_{\alpha_1^z}) =   \frac{\kappa - h}{\kappa} \Bigl( \frac{(t+1)(t+3) \sigma_{\zeta_1} \sigma_{\alpha_2^z} }{ ((t+2)h+1)((t+2)h-1) \rho_{\zeta_2} \rho_{\alpha_1^z}   } \Bigr)^{1/2}\,,
\\[3pt]
&& \sin(\psi_{\zeta_2} + \psi_{\alpha_2^z}) = \frac{t-h+2}{(t+2)h -1} \Bigl( \frac{ (\kappa^2-1) \sigma_{\zeta_1} \sigma_{\alpha_1^z} }{ \kappa^2 \rho_{\zeta_2} \rho_{\alpha_2^z} } \Bigr)^{1/2} \,,
\\[3pt]
\label{20072015-man-02} && \sin(\psi_{\alpha_1^z} + \psi_{\alpha_2^z}) = - \frac{\kappa + t+2}{\kappa} \Bigl( \frac{(h^2-1)  \sigma_{\zeta_1}\sigma_{\zeta_2} }{((t+2)h +1) ((t+2)h  - 1) \rho_{\alpha_1^z} \rho_{\alpha_2^z} } \Bigr)^{1/2}\,.\qquad
\eeq
}

Eqs.\rf{20072015-man-01}-\rf{20072015-man-02} do not determine the angle variables uniquely. These equations leave one-parametric freedom in solution for the angle variables. In other words, the defining equations leave one-parametric freedom in solution for the operators $X_n$, $Y_n$ \rf{15032015-06}-\rf{15032015-06a}. This freedom reflects the fact that ket-vector $\phik$ \rf{17072015-man-10} involves two ket-vectors  $|\phi_0^{s_1-1,s_2-1}\rangle$, $|\phi_{0'}^{s_1-1,s_2-1}\rangle$ having the same $M^z$-charge and labels of the $so(d-2)$ algebra. We note then that there exists the non-trivial unitary transformation of ket-vector $\phik$ \rf{17072015-man-10},
\be  \label{20072015-man-03}
\phik^U = U \phik\,,
\ee
with unitary operator $U$ that commutes with the operators $M^z$ and $M^{ij}$,
\be \label{20072015-man-04}
 U^\dagger U =1\,,\qquad U^\dagger M^z U = M^z\,,   \qquad U^\dagger M^{ij} U = M^{ij}\,.
\ee
This is to say that Eqs.\rf{20072015-man-04} have the following non-trivial unique solution:
\beq
&& U  =  \sum_{a,b,c,e=0,1} \mu_{abce} \Pi_a^{\zeta_1} \Pi_b^{\zeta_2} \Pi_c^{\alpha_1^z} \Pi_e^{\alpha_2^z} + \sin\psi(\zeta_1 \alpha_2^z \alphab_1^z \zetab_2 - \zeta_2 \alpha_1^z \alphab_2^z \zetab_1)\,,
\\
&& \mu_{abce} = 1 \ \ \hbox{ for } \ \ abce \ne 0110, \ 1001\,, \qquad \mu_{0110} =\mu_{1001} = \cos\psi\,,
\\
&& \Pi_0^x = 1 - N_x\,, \qquad \Pi_1^x \equiv N_x\,, \qquad N_x \equiv x\xb\,.
\eeq
From \rf{17072015-man-10},\rf{20072015-man-03}, we get the transformations of the ket-vectors $|\phi_0^{s_1-1,s_2-1}\rangle$, $|\phi_{0'}^{s_1-1,s_2-1}\rangle$,
\beq
\label{20072015-man-05} && |\phi_0^{s_1-1,s_2-1}\rangle^U = \cos\psi |\phi_0^{s_1-1,s_2-1}\rangle - \sin\psi |\phi_{0'}^{s_1-1,s_2-1}\rangle\,,
\\
\label{20072015-man-06} && |\phi_{0'}^{s_1-1,s_2-1}\rangle^U = \cos\psi |\phi_{0'}^{s_1-1,s_2-1}\rangle + \sin\psi |\phi_0^{s_1-1,s_2-1}\rangle\,.
\eeq
Also, according to textbook, transformation of ket-vector \rf{20072015-man-03} implies the transformations of the  operators $X_n$, $Y_n$,
\be \label{20072015-man-07}
X_n^U = U^\dagger X_n U\,, \qquad Y_n^U = U^\dagger Y_n U\,.
\ee
Using \rf{15032015-06}-\rf{15032015-06a} and relations \rf{05052015-04},\rf{05052015-07},\rf{05052015-12}, \rf{05052015-15}, \rf{17072015-21}-\rf{17072015-22}, we verify that transformations \rf{20072015-man-07} imply the following transformations of the angle variables
\be \label{20072015-man-08}
\psi_{\zeta_1}^U = \psi_{\zeta_1} + \psi\,,\quad \psi_{\alpha_1^z}^U = \psi_{\alpha_1^z} - \psi\,,\quad \psi_{\zeta_2}^U = \psi_{\zeta_2} - \psi\,,\quad \psi_{\alpha_2^z}^U = \psi_{\alpha_2^z} + \psi\,.
\ee
We now note that the following combinations of the angle variables
\be  \label{20072015-man-09}
\psi_{\zeta_1} + \psi_{\zeta_2}\,, \quad \psi_{\zeta_1} + \psi_{\alpha_1^z}\,,\quad \psi_{\zeta_1} - \psi_{\alpha_2^z}\,, \quad \psi_{\zeta_2} - \psi_{\alpha_1^z}\,,\quad \psi_{\zeta_2} + \psi_{\alpha_2^z}\,,\quad \psi_{\alpha_1^z} + \psi_{\alpha_2^z}\,.
\ee
are invariant under transformations \rf{20072015-man-08}. It is the invariant combinations \rf{20072015-man-09} that are determined uniquely by Eqs.\rf{20072015-man-01}-\rf{20072015-man-02}. Note however that quantities $f_{xyw}$ and $g_v$ in \rf{05052015-04}, \rf{05052015-07},\rf{05052015-12}, \rf{05052015-15}, \rf{17072015-21}-\rf{17072015-22} are not expressed in terms of the invariant combinations of the angle variables \rf{20072015-man-09}.
The just-mentioned quantities $f_{xyw}$ and $g_v$ are determined uniquely after fixing the one-parametric freedom in Eqs.\rf{20072015-man-01}-\rf{20072015-man-02}. To be flexible, we do not fix the one-parametric freedom of the angles variables when considering massive field.

\noindent {\bf Limit of massless AdS field}. For considering various limits, some particular fixing of the one-parametric freedom in Eqs.\rf{20072015-man-01}-\rf{20072015-man-02} may be convenient. To explain what has just been said let us consider limit of massless AdS field \rf{19072015-man-02}. For this limit, the convenient choice to fix the one-parametric freedom in Eqs.\rf{20072015-man-01}-\rf{20072015-man-02} is given by $\psi_{\alpha_2^z}=0$. We now note that, in terms of $\sigma_{\zeta_2}$ \rf{17072015-18}, the limit of massless AdS field is realized as $\sigma_{\zeta_2}\rightarrow 0$. Using $\psi_{\alpha_2^z}=0$, $\sigma_{\zeta_2}=0$ in Eqs.\rf{20072015-man-01}-\rf{20072015-man-02}, we find the following unique simple solution for the angle variables
\be  \label{20072015-man-09a}
\psi_{\zeta_1} = 0\,, \qquad \psi_{\zeta_2} =  \frac{\pi}{2}\,, \qquad
\psi_{\alpha_1^z} = 0\,, \qquad \psi_{\alpha_2^z} = 0\,.
\ee
Making use of relations \rf{20072015-man-09a} in expressions for the spin operators $W^i$, $\Wb^i$ \rf{17072015-man-17},\rf{17072015-man-18}, we then verify that, in massless limit \rf{19072015-man-02}, ket-vectors $\phibfk$ \rf{19072015-man-02a} and $\varphibfk$ \rf{19072015-man-02b} are realized as two invariant subspaces in space of ket-vector $\phik$ \rf{19072015-man-02a-ext}.

\noindent {\bf Type $[s,s]$ massive field in $AdS_{d+1}$}. Ket-vector of type $[s,s]$ massive field is given by
\beq
\label{20072015-man-10} && \hspace{-1cm} \phik  =  |\phi_2\rangle  + |\phi_1\rangle  + |\phi_0\rangle +  |\phi_{-1}\rangle +  |\phi_{-2}\rangle \,,
\\
\label{20072015-man-10a} && |\phi_2\rangle  =  \zeta_1 \zeta_2 |\phi_2^{s-1,s-1}\rangle \,,
\\
&& |\phi_1\rangle  =    \zeta_2 |\phi_1^{s,s-1}\rangle  + \zeta_1\zeta_2 \alpha_2^z |\phi_1^{s-1,s-2}\rangle\,,
\\
&& |\phi_0\rangle  =    |\phi_0^{s,s}\rangle  +  \zeta_2 \alpha_1^z|\phi_0^{s-1,s-1}\rangle + \zeta_2\alpha_2^z  |\phi_0^{s,s-2}\rangle+ \zeta_1\zeta_2 \alpha_1^z\alpha_2^z |\phi_0^{s-2,s-2}\rangle,\qquad
\\
&& |\phi_{-1}\rangle  =    \alpha_2^z |\phi_{-1}^{s,s-1}\rangle  + \zeta_2 \alpha_1^z \alpha_2^z |\phi_{-1}^{s-1,s-2}\rangle\,,
\\
\label{20072015-man-11} && |\phi_{-2}\rangle  =  \alpha_1^z \alpha_2^z |\phi_{-2}^{s-1,s-1}\rangle
\,.
\eeq
Ket-vector $\phik$ \rf{20072015-man-10} is obtained from the one in \rf{17072015-man-10} by equating to zero the ket-vector $|\phi_{0'}^{s-1,s-1}\rangle$ and the ket-vectors $|\phi_\lambda^{\Ksm_1\Ksm_2}\rangle$ with $K_1< K_2$. The convenient choice to fix the one-parametric freedom in Eqs.\rf{20072015-man-01}-\rf{20072015-man-02} is given by $\psi_{\alpha_2^z}=0$. Using this choice and taking into account that, for type $[s,s]$ field, one has the relation $h=1$ (see \rf{17072015-16}), we note that Eqs.\rf{20072015-man-01}-\rf{20072015-man-02} uniquely determine the angle variables,
\be \label{20072015-man-12}
\psi_{\zeta_1} = - \frac{\pi}{2}\,, \qquad \psi_{\zeta_2} =  \frac{\pi}{2}\,, \qquad
\psi_{\alpha_1^z} = 0\,, \qquad \psi_{\alpha_2^z} = 0\,.
\ee
We now note that the ket-vectors $|\phi_\lambda^{\Ksm\Ksm}\rangle$ appearing in \rf{20072015-man-10a}-\rf{20072015-man-11} satisfy the algebraic constraints
\beq
 \label{20072015-man-14} && \Ab_1^i|\phi_\lambda^{\Ksm\Ksm}\rangle = 0 \,, \quad A_2^i|\phi_\lambda^{\Ksm\Ksm}\rangle = 0 \,, \quad N_{\alpha_{12}} |\phi_\lambda^{\Ksm\Ksm}\rangle = 0\,, \quad N_{\alpha_{21}}
|\phi_\lambda^{\Ksm\Ksm}\rangle = 0\,,
\\
 \label{20072015-man-15} && (N_{\alpha_1} - N_{\alpha_2})|\phi_\lambda^{\Ksm\Ksm}\rangle=0\,, \qquad \alphab_{12}|\phi_\lambda^{\Ksm\Ksm}\rangle=0\,.
\eeq
Using \rf{20072015-man-12}-\rf{20072015-man-15}, we verify that action of operators $W^i$, $\Wb^i$ \rf{17072015-man-17},\rf{17072015-man-18},\rf{15032015-06}-\rf{17072015-22b} is well defined on the space of ket-vector $\phik$ \rf{20072015-man-10}.

\noindent {\bf Type $[s,s]$ massless field in $AdS_{d+1}$}. Ket-vector of massive field $\phik$ \rf{20072015-man-10} can be presented in terms of two new ket-vectors $\phibfk$, $\varphibfk$ in the following way
\beq
&& \hspace{-1cm} \phik = \phibfk + \zeta_2 \varphibfk \,,
\\
\label{20072015-man-16} && \phibfk  =     |\phibf_0\rangle +  |\phibf_{-1}\rangle +  |\phibf_{-2}\rangle \,,
\\
\label{20072015-man-17} && \varphibfk  =     |\varphibf_1\rangle  +  |\varphibf_0\rangle +  |\varphibf_{-1}\rangle +  |\varphibf_{-2}\rangle \,,
\\
&& \hspace{1cm} |\phibf_0\rangle  =    |\phi_0^{s,s}\rangle\,,
\nonumber\\
&& \hspace{1cm} |\phibf_{-1}\rangle  =   \alpha_2^z |\phi_{-1}^{s,s-1}\rangle\,,
\nonumber\\
&& \hspace{1cm} |\phibf_{-2}\rangle  =  \alpha_1^z \alpha_2^z |\phi_{-2}^{s-1,s-1}\rangle
\,,
\\
&& \hspace{1cm}  |\varphibf_1\rangle  = - \zeta_1   |\phi_2^{s-1,s-1}\rangle \,,
\nonumber\\
&& \hspace{1cm} |\varphibf_0\rangle  =     |\phi_1^{s,s-1}\rangle  - \zeta_1  \alpha_2^z |\phi_1^{s-1,s-2}\rangle\,,
\nonumber\\
&& \hspace{1cm}  |\varphibf_{-1}\rangle  =    \alpha_1^z|\phi_0^{s-1,s-1}\rangle +  \alpha_2^z  |\phi_0^{s,s-2}\rangle - \zeta_1 \alpha_1^z\alpha_2^z |\phi_0^{s-2,s-2}\rangle,\qquad
\nonumber\\
&& \hspace{1cm} |\varphibf_{-2}\rangle  =     \alpha_1^z \alpha_2^z |\phi_{-1}^{s-1,s-2}\rangle\,.
\eeq
In the massless limit \rf{19072015-man-02}, we note that, for $s>1$, ket-vector $\phibfk$ \rf{20072015-man-16} describes type $[s,s]$ massless AdS field, while ket-vector $\varphibfk$ \rf{20072015-man-17} describes type $[s,s-1]$ massless AdS field. Note that type $[s,s]$ massless AdS field is realized as irrep of $so(d-1)$ algebra. Therefore, in the flat limit, the ket-vector $\phibfk$ describes irreducible massless field in $R^{d,1}$.

\noindent {\bf Type $[s,s]$ currents, shadows, and conformal fields in $R^{d-1,1}$}.
Ket-vectors of type $[s,s]$ long currents, shadows, and conformal fields in $R^{d-1,1}$ take the same form as the ket-vector $\phibfk$ of type $[s,s]$ massive field in $AdS_{d+1}$ \rf{20072015-man-10},
while ket-vectors of type $[s,s]$ short currents, shadows, and conformal fields in $R^{d-1,1}$ take the same form as the ket-vector $\phibfk$ of type $[s,s]$ massless field in $AdS_{d+1}$ \rf{20072015-man-16}. Using the ket-vectors of currents, shadows, and conformal fields and our result in Sections \ref{section-03},\ref{sec-05}, we get 2-point functions for the type $[s,s]$ currents and shadows and Lagrangian for the type $[s,s]$ conformal fields.

\end{document}